%% file: DilatationOperatorPaper20arXiv.tex
\documentclass[a4paper,11pt]{article}
\usepackage{jheppub}
\usepackage[utf8]{inputenc}
\usepackage{{amsmath,amsfonts,latexsym, amstext, amssymb, amsthm}}
\usepackage{xspace} 
\usepackage{slashed}
\usepackage[bf]{caption}
\usepackage{subcaption}
\usepackage{comment}
\usepackage{dsfont}
\usepackage{placeins}

\numberwithin{equation}{section}

\usepackage{feynmp}
\input{Macros2.tex}

\newcommand{\DOT}[1]{\dot{#1}}

\makeatletter
\DeclareRobustCommand*{\bfseries}{%
  \not@math@alphabet\bfseries\mathbf
  \fontseries\bfdefault\selectfont
  \boldmath
}
\makeatother

\usepackage{xkeyval}
\makeatletter

\define@boolkey{FDinline}{bubble}[true]{}
\define@boolkey{FDinline}{triangle}[true]{}
\define@boolkey{FDinline}{cut}[true]{}
\define@key{FDinline}{toplabel}%
{\def\FDinlinetoplabel{#1}}%
\define@key{FDinline}{topleglabel}%
{\def\FDinlinetopleglabel{#1}}%
\define@key{FDinline}{bottomleglabel}%
{\def\FDinlinebottomleglabel{#1}}%
\presetkeys{FDinline}{%
bubble=false,%
triangle=false,%
cut=false,%
toplabel=$\phantom{0}$,%
topleglabel=$\phantom{0}$,%
bottomleglabel=$\phantom{0}$%
}{}
\newcommand*\FDinline[1][]{%
 \setkeys{FDinline}{#1}%
\settoheight{\eqoff}{$\times$}%
\setlength{\eqoff}{0.5\eqoff}%
\addtolength{\eqoff}{-4\unitlength}%
\raisebox{\eqoff}{%
\fmfframe(1,0)(1,0){%
\scalebox{1}{%
\begin{fmfchar*}(15,8)%
\fmfcmd{%
dash_len := 1.6mm; 
}%
\begin{fmfshrink}{0.5}
\ifKV@FDinline@cut
\fmftop{vtop}
\fmfbottom{vbot}
\fmf{dashes}{vtop,vbot}
\else\fi
\ifKV@FDinline@bubble
\fmfleft{vq}
\fmfright{vp2,vp1}
\fmf{double,tension=2}{vq,v1}
\fmf{plain,tension=0.5,left=0.7,label=\FDinlinetoplabel}{v1,v2}
\fmf{plain,tension=0.5,right=0.7}{v1,v2}
\fmf{plain,tension=1.2}{v2,vp1}
\fmf{plain,tension=1.2}{v2,vp2}
\fmfv{label=\FDinlinetopleglabel,l.a=0}{vp1}
\fmfv{label=\FDinlinebottomleglabel,l.a=0}{vp2}
\else\fi
\ifKV@FDinline@triangle
\fmfleft{vq}
\fmfright{vp2,vp1}
\fmf{double,tension=2}{vq,v1}
\fmf{plain,tension=0.5,left=0.0,label=\FDinlinetoplabel}{v1,v2t}
\fmf{plain,tension=0.5,right=0.0}{v1,v2b}
\fmf{plain,tension=0}{v2t,v2b}
\fmf{plain,tension=1.2}{v2t,vp1}
\fmf{plain,tension=1.2}{v2b,vp2}
\fmfv{label=\FDinlinetopleglabel,l.a=0}{vp1}
\fmfv{label=\FDinlinebottomleglabel,l.a=0}{vp2}
\else\fi
\end{fmfshrink}
\end{fmfchar*}%
}}}%
}

\makeatother

  \usepackage{lipsum}
  \makeatletter
  \if@titlepage
    \renewenvironment{abstract}{%
        \titlepage
        \null\vfil
        \@beginparpenalty\@lowpenalty
        \begin{center}%
          \bfseries \abstractname
          \@endparpenalty\@M
        \end{center}}%
       {\par\vfil\null\endtitlepage}
  \else
    \renewenvironment{abstract}{%
        \if@twocolumn
          \section*{\abstractname}%
        \else
          \small
          \begin{center}%
            {\bfseries \abstractname\vspace{-.5em}\vspace{\z@}}%
          \end{center}%
          \quotation
        \fi}
        {\if@twocolumn\else\endquotation\fi}
  \fi
  \makeatother

\title{Amplitudes, Form Factors and the \\ Dilatation Operator in $\cN=4$ SYM Theory}
\author{Matthias Wilhelm}
 \affiliation{Institut für Mathematik und Institut für Physik,
 Humboldt-Universität zu Berlin\\
 IRIS Gebäude, 
 Zum Großen Windkanal 6, 
 12489 Berlin}
 \emailAdd{mwilhelm@physik.hu-berlin.de}
 \arxivnumber{1410.6309}

\begin{document}


\begin{fmffile}{diagrams}
\fmfcmd{%
thin := 1pt; 
thick := 2thin;
arrow_len := 3mm;
arrow_ang := 15;
curly_len := 3mm;
dash_len := 1.8mm; 
dot_len := 0.75mm; 
wiggly_len := 2mm; 
wiggly_slope := 60;
zigzag_len := 2mm;
zigzag_width := 2thick;
decor_size := 5mm;
dot_size := 2thick;
}
\fmfcmd{%
marksize=7mm;
def draw_cut(expr p,a) =
  begingroup
    save t,tip,dma,dmb; pair tip,dma,dmb;
    t=arctime a of p;
    tip =marksize*unitvector direction t of p;
    dma =marksize*unitvector direction t of p rotated -90;
    dmb =marksize*unitvector direction t of p rotated 90;
    linejoin:=beveled;
    drawoptions(dashed evenly);
    draw ((-.5dma.. -.5dmb) shifted point t of p);
    drawoptions();
  endgroup
enddef;
style_def phantom_cut expr p =
    save amid;
    amid=.5*arclength p;
    draw_cut(p, amid);
    draw p;
enddef;
}
\fmfcmd{%
smallmarksize=5mm;
def draw_smallcut(expr p,a) =
  begingroup
    save t,tip,dma,dmb; pair tip,dma,dmb;
    t=arctime a of p;
    tip =smallmarksize*unitvector direction t of p;
    dma =smallmarksize*unitvector direction t of p rotated -90;
    dmb =smallmarksize*unitvector direction t of p rotated 90;
    linejoin:=beveled;
    drawoptions(dashed evenly);
    draw ((-.5dma.. -.5dmb) shifted point t of p);
    drawoptions();
  endgroup
enddef;
style_def phantom_smallcut expr p =
    save amid;
    amid=.5*arclength p;
    draw_smallcut(p, amid);
    draw p;
enddef;
}

 \begingroup\parindent0pt
 \begin{flushright}\footnotesize
 \texttt{HU-MATH-2014-26}\\
 \texttt{HU-EP-14/38} 
 \end{flushright}
 \vspace*{4em}
 \centering
 \begingroup\LARGE
 \bf
 Amplitudes, Form Factors and the \\ Dilatation Operator  in  $\cN=4$ SYM Theory
 \par\endgroup
 \vspace{2.5em}
 \begingroup\large\bf
 Matthias Wilhelm
 \par\endgroup
 \vspace{1em}
 \begingroup\itshape
 Institut für Mathematik und Institut für Physik\\
 Humboldt-Universität zu Berlin\\
IRIS Gebäude \\
 Zum Großen Windkanal 6 \\
 12489 Berlin
 \par\endgroup
 \vspace{1em}
 \begingroup\ttfamily
 mwilhelm@physik.hu-berlin.de \\
 \par\endgroup
 \vspace{2.5em}
 \endgroup
 
 
 \begin{abstract}
 
 We study the form factor of a generic gauge-invariant local composite operator in \NfSYMt.
 At tree level and for a minimal number of external on-shell super fields, we find that the form factor precisely yields the spin-chain picture of integrability in the language of scattering amplitudes.
 Moreover, we compute the cut-constructible part of the one-loop correction to this minimal form factor via generalised unitarity. 
 From its UV divergence, we obtain the complete one-loop dilatation operator of \NfSYMt. 
 Thus, we provide a field-theoretic derivation of a relation between the 
 one-loop dilatation operator and the four-point tree-level amplitude 
 which was observed earlier. 
 We also comment on the implications of our findings in the context of 
 integrability.
 \end{abstract}
 

 
 \thispagestyle{empty}
 
 \newpage
%
 \hrule
 \tableofcontents
 \afterTocSpace
 \hrule
 \afterTocRuleSpace

\section{Introduction}

The maximally supersymmetric Yang-Mills theory ($\cN=4$ SYM theory) in $3+1$ dimensions is frequently called the hydrogen atom of the $21^{\text{st}}$ century, and it might be the first non-trivial quantum field theory that can be solved exactly.
During the last decade, tremendous progress has been made in understanding this theory, in particular for gauge group $SU(N)$ in the 't Hooft (planar) limit \cite{'tHooft:1973jz}, where $N\to\infty$, $g_\YM\to0$ with $\lambda=Ng_\YM^2$ fixed.

One source for this progress was the discovery of integrability in the spectrum of anomalous dimensions, see \cite{Beisert:2010jr} for a review.
The origin of this discovery was the realisation that at one loop the action of the dilatation operator on composite operators can be mapped to the Hamiltonian of an integrable spin chain. 
The one-loop spectral problem can thus be solved by finding the eigenvalues of the one-loop dilatation operator via Bethe-ansatz type methods \cite{Minahan:2002ve,Beisert:2003yb}.
Moreover, these methods generalise to higher loop orders; see the review \cite{Beisert:2010jr} for references.%
\footnote{While these methods yield the correct anomalous dimensions up to very high loop orders \cite{Beisert:2010jr}, the complete dilatation operator, whose eigenvalues the corresponding Bethe ansätze are giving, is currently known only at one-loop level \cite{Beisert:2003jj}.}

Another source for the increased understanding of \NfSYMt was the development of highly efficient on-shell techniques in the context of scattering amplitudes, see e.g.\ \cite{Elvang:2013cua,Henn:2014yza} for reviews. Exploiting their analytic properties, generic amplitudes can be built from simpler amplitudes with a lower number of external legs and loops via recursion relations \cite{Cachazo:2004kj,Britto:2004ap,Britto:2005fq} and (generalised) unitarity \cite{Bern:1994zx,Bern:1994cg,Britto:2004nc} --- largely eliminating the need to compute Feynman diagrams. 
These methods allowed for the construction of all tree-level amplitudes \cite{Drummond:2008cr} as well as the unregularised integrand of all loop amplitudes \cite{ArkaniHamed:2010kv}. 

Remarkably, the interplay between integrability-based methods and on-shell techniques 
has been relatively limited. The study of the on-shell structure of amplitudes at weak coupling via integrability was initiated in \cite{Ferro:2012xw} and advanced in \cite{Ferro:2013dga,Chicherin:2013ora,Frassek:2013xza,Beisert:2014qba,Kanning:2014maa,Broedel:2014pia,Broedel:2014hca,Bargheer:2014mxa,Ferro:2014gca}.\footnote{A third important source for progress in \NfSYMt has been the strong coupling description via the $\text{AdS}/\text{CFT}$ correspondence \cite{Maldacena:1997re,Gubser:1998bc,Witten:1998qj}. The interplay between the strong coupling description and integrability has been extensive. In particular, the strong coupling description allows for the construction of amplitudes via integrability-based methods at strong coupling \cite{Alday:2010vh}, and in certain kinematic regimes also at finite coupling \cite{Basso:2013vsa}.}
It was motivated by an interesting connection between scattering amplitudes and the dilatation operator, which was observed in \cite{Zwiebel:2011bx}\footnote{For the simplest case of four-point amplitudes and the one-loop dilatation operator, this observation goes back to Niklas Beisert; see footnote 27 in \cite{Zwiebel:2011bx}.} but could so far not be derived by the methods of field theory.
In this paper, we use the powerful on-shell methods from the study of scattering amplitudes to revisit the spectral problem of integrability, and in particular to derive the relation observed in \cite{Zwiebel:2011bx}. 
We propose that the missing link between these two areas of research is given by form factors.

For a given gauge-invariant local composite operator $\cO(x)$, the form factor $\cF_{\cO}$ is defined as the overlap of the off-shell state created by $\cO$ from the vacuum $\ket{0}$ with an on-shell $n$-particle state $\bra{1,\dots,n}$,\footnote{As in the case of amplitudes, this on-shell $n$-particle state is specified by the momenta, helicities and flavours of the $n$ particles.} i.e.\
\begin{equation}\label{eq: form factor intro}
 \cF_{\cO}(1,\dots,n;x)=\bra{1,\dots,n}\cO(x)\ket{0}\eqndot
\end{equation}
Thus, form factors interpolate between the completely on-shell amplitudes and completely off-shell correlation functions.%
\footnote{For the `unit operator' $\cO=\idm$, we recover the scattering amplitude $\cA(1,\dots,n)=\braket{1,\dots,n}{0}$. In analogy to the notation for scattering amplitudes, the form factor with $n$ on-shell particles is also referred to as $n$-point form factor.}
In particular, form factors, their generalisations to several operators and scattering amplitudes can in principle be used to compute correlation functions via generalised unitarity \cite{Engelund:2012re}.
Form factors were first studied in $\cN=4$ SYM theory by van Neerven almost 30 years ago \cite{vanNeerven:1985ja}.
Interest resurged when the authors of \cite{Alday:2007he,Maldacena:2010kp,Gao:2013dza} 
used the $\text{AdS}/\text{CFT}$ correspondence to determine form factors 
at strong coupling. 
Many studies at weak coupling followed \cite{Brandhuber:2010ad,Bork:2010wf,Brandhuber:2011tv,Bork:2011cj,Henn:2011by,Gehrmann:2011xn,Brandhuber:2012vm,Bork:2012tt,Engelund:2012re,Johansson:2012zv,Boels:2012ew,Penante:2014sza,Brandhuber:2014ica,Bork:2014eqa}.
Form factors can be calculated using many of the successful on-shell techniques that were developed in the context of amplitudes.
At tree level, spinor helicity variables \cite{Brandhuber:2010ad}, $\cN=4$ on-shell superspace \cite{Brandhuber:2011tv}, BCFW- \cite{Brandhuber:2010ad} and MHV-recursion relations \cite{Brandhuber:2011tv} as well as twistors \cite{Brandhuber:2010ad} and momentum twistors \cite{Brandhuber:2011tv} are applicable.\footnote{An interpretation in terms of the volume of polytops also exists \cite{Bork:2014eqa}.} At loop level, generalised unitarity \cite{Brandhuber:2010ad,Johansson:2012zv} and symbols \cite{Brandhuber:2012vm} can be used. 
Both at tree and loop level, colour-kinematic duality \cite{Bern:2008qj}
was found to be present in certain examples \cite{Boels:2012ew}.
However, the previous studies have largely focused on the BPS operator $\tr[\phi_{12}\phi_{12}]$ and the chiral stress-tensor super multiplet it is part of, as well as its generalisation to $\tr[(\phi_{12})^k]$. The corresponding form factors are currently known up to three loops for two points \cite{Gehrmann:2011xn}, i.e.\ on-shell fields, and two loops for $k$ points \cite{Brandhuber:2014ica}, respectively.
The only exceptions to this rule are the operators from the $\mathfrak{su}(2)$ and $\mathfrak{sl}(2)$ subsectors, whose tree-level \MHV form factors\footnote{In analogy to the notation for amplitudes, maximally helicity violating (MHV) refers to the minimal fermionic degree in the corresponding super form factor.} were determined in \cite{Engelund:2012re}, and the Konishi operator, whose two-point form factor was calculated at one-loop level in \cite{Bork:2010wf}.

In this paper, we study the form factor of a generic single-trace operator at tree level and one-loop order. In particular, we establish the form factor's role as the missing link between on-shell methods and the integrability of composite operators. 
We focus on the minimal form factor, i.e.\ the form factor for the minimal number of points $n$ that yields a non-vanishing result.
The structure of this paper is as follows. 
In section \ref{sec: form factors and spin chains}, we give a short review of the elementary gauge-invariant local composite operators of \NfSYMt and the spin-chain picture. We then calculate the form factors of these operators in the free theory, which are equal to the minimal form factors at tree level in the interacting theory. We find that they exactly realise the spin chain of \NfSYMt in the language of on-shell super fields.
In section \ref{sec: one-loop form factors}, we employ generalised unitarity to compute the one-loop corrections to these minimal form factors. We obtain the complete one-loop result except for a finite rational part which is not cut-constructible, i.e.\ not constructible via four-dimensional generalised unitarity. 
In section \ref{sec: one-loop dila}, we discuss the IR and UV divergences of our one-loop result. From its UV divergence, we obtain the complete one-loop dilatation operator. In particular, we derive the connection between the one-loop dilatation operator and the four-point amplitude which was observed in \cite{Zwiebel:2011bx}.
A conclusion and outlook is contained in section \ref{sec: conclusion and outlook}; there, we also comment on the implications of our finding in the context of integrability.
We summarise some elementary facts about amplitudes in appendix \ref{app: amplitudes}. In appendix \ref{app: examples}, we calculate some instructive examples of one-loop form factors of operators from the $\mathfrak{su}(2)$ and $\mathfrak{sl}(2)$ subsectors via ordinary unitarity.
We give explicit expressions for Feynman integrals that are required throughout the paper in appendix \ref{app: integrals}. 

In a forthcoming paper \cite{Nandan:2014oga}, it is demonstrated that form factors can also be employed to compute anomalous dimensions at higher loop orders. In particular, we explicitly calculate the anomalous dimension of the Konishi operator at two-loop order via its two-loop minimal form factor.

\section{Form factors and spin chains}
\label{sec: form factors and spin chains}

In section \ref{subsec: operators and spin chain picture}, we give a short review of the elementary gauge-invariant local composite operators of \NfSYMt and the spin-chain picture, cf.\ for example \cite{Beisert:2004ry,Minahan:2010js}. Moreover, we review the action of the one-loop dilatation operator on these spin chains. In section \ref{subsec: form factors}, we then derive the form factors of these operators in the free theory. They are equal to the tree-level minimal form factors in the interacting theory.


\subsection{Composite operators and spin chains}
\label{subsec: operators and spin chain picture}

Gauge-invariant local composite operators can be built as traces of products of gauge-covariant fields at a common spacetime point $x$. These fields are given by the scalars $\phi_{AB}=-\phi_{BA}$, the fermions $\psi^A_\alpha=\epsilon^{ABCD}\psi_{BCD \alpha}$,\footnote{Hence, $\psi_{BCD \alpha}=\frac{1}{3!}\epsilon_{ABCD}\psi^A_\alpha$. Note that throughout this paper Einstein's summation convention applies.} 
the anti-fermions $\bar\psi_{A\alphadot}$ and the field strength $F_{\mu\nu}$, where $\alpha,\beta=1,2$, $\alphadot,\betadot=\DOT1,\DOT2$, $\mu,\nu=0,1,2,3$ and $A,B,C,D=1,2,3,4$.\footnote{Note that the positive- and negative-helicity components of the gauge field, $g^+$ and $g^-$, do not transform covariantly under gauge transformations.} 
Using the Pauli matrices $(\sigma^\mu)_{\alpha\alphadot}$, the spacetime indices $\mu,\nu$ can be exchanged for spinor indices $\alpha,\alphadot$ and the field strength can be split into its self-dual and anti-self-dual parts $\cfstrength_{\alpha\beta}$ and $\cantifstrength_{\alphadot\betadot}$:
\begin{equation}
\label{eq: def field strength slitting}
 F_{\alpha\beta\alphadot\betadot}=F_{\mu\nu}(\sigma^\mu)_{\alpha\alphadot}(\sigma^\nu)_{\beta\betadot}=\sqrt{2}\epsilon_{\alphadot\betadot}\cfstrength_{\alpha\beta}+\sqrt{2}\epsilon_{\alpha\beta}\cantifstrength_{\alphadot\betadot} \eqncom
\end{equation}
where $\epsilon_{\alpha\beta}$ and $\epsilon_{\alphadot\betadot}$ are the two-dimensional antisymmetric tensors.\footnote{We normalise these via $\epsilon^{12}=\epsilon_{21}=\epsilon^{\DOT1\DOT2}=\epsilon_{\DOT2\DOT1}=1$.}
Furthermore, covariant derivatives 
\begin{equation}
\label{eq: def covariant derivative}
 \cder_{\alpha\alphadot}=\cder_\mu(\sigma^\mu)_{\alpha\alphadot}=(\partial_\mu-i g_\YM A_\mu)(\sigma^\mu)_{\alpha\alphadot}
\end{equation}
may act on the above fields, where $\partial_\mu$ is the ordinary derivative and $A_\mu$ the gauge field.
Using the equations of motion, the definition of the field strengths and the Bianchi identity, any antisymmetric combination of spinor indices $\alpha$ and $\alphadot$ at such a field can be reduced to a linear combination of products of totally symmetric terms. The resulting irreducible fields transform in the so-called singleton representation $\cV_S$ of $\mathfrak{psu}(2,2|4)$. Arbitrary gauge-invariant local composite operators can be constructed as products of single-trace operators that obey the correct Bose-Einstein or Fermi-Dirac statistics.\footnote{At finite $N$, there are also matrix relations connecting traces of more than $N$ irreducible fields to sums of products of traces of less irreducible fields.} In the planar limit, the correlation functions of multi-trace operators are, however, completely determined by those of their single-trace factors.\footnote{Interactions that merge or split traces are suppressed by $\frac{1}{N}$.}

The singleton representation $\cV_S$ of $\mathfrak{psu}(2,2|4)$ can be constructed via two sets of bosonic ocsillators $\aosc_\alpha$, $\aoscdag^{\alpha}$ ($\alpha=1,2$) and $\bosc_{\alphadot}$, $\boscdag^{\alphadot}$ ($\alphadot=\DOT1,\DOT2$) as well as one set of fermionic oscillators $\dosc_A$, $\doscdag^A$ ($A=1,2,3,4$). 
These oscillators obey the usual (anti-)commutation relations:
\begin{equation}\label{eq: (anti)commutation relations}
\comm{\aosc_\alpha}{\aoscdag^\beta}=\delta_\alpha^\beta \eqncom \qquad
\comm{\bosc_{\alphadot}}{\boscdag^{\betadot}}=\delta_{\alphadot}^{\betadot} \eqncom \qquad
\acomm{\dosc_A}{\dosc^{\dagger B}}=\delta^B_A\eqncom  
\end{equation}
with all other (anti-)commutators vanishing.
In terms of the oscillators, the irreducible fields read 
\begin{equation}\label{eq: fields}
\begin{aligned}
&\cder^k\cfstrength_{\phantom{ABC}} &\mathrel{\widehat{=}} \quad&
  (\aoscdag)^{k+2} 
  (\boscdag)^{k\phantom{+0}}
  \dosc^{\dagger 1} \dosc^{\dagger 2} \dosc^{\dagger 3} \dosc^{\dagger 4} 
  \vac \eqncom & &\\
&\cder^k\ferm_{ABC} &\mathrel{\widehat{=}}     \quad&
  (\aoscdag)^{k+1} 
  (\boscdag)^{k\phantom{+0}}
  \dosc^{\dagger A} \dosc^{\dagger B} \dosc^{\dagger C}
  \vac \eqncom & &\\
&\cder^k{}\phi_{AB{}\phantom{C}}{} &\mathrel{\widehat{=}}  \quad &  
  (\aoscdag)^{k\phantom{+0}} 
  (\boscdag)^{k\phantom{+0}} 
  \dosc^{\dagger A} \dosc^{\dagger B} 
  \vac \eqncom & &\\
&\cder^k\antiferm_{A\phantom{BC}} &\mathrel{\widehat{=}} \quad&
  (\aoscdag)^{k\phantom{+0}} 
  (\boscdag)^{k+1} 
  \dosc^{\dagger A}    
  \vac \eqncom & &\\
&\cder^k\cantifstrength_{\phantom{ABC}} &\mathrel{\widehat{=}} \quad  &
  (\aoscdag)^{k\phantom{+0}}
  (\boscdag)^{k+2} 
  \vac \eqncom & &
\end{aligned}
\end{equation} 
where we have suppressed all spinor indices $\alpha$, $\alphadot$. We can characterise each such field via the numbers of $\aosc^{\dagger\alpha}$, $\bosc^{\dagger\alphadot}$  and $\dosc^{\dagger A}$ oscillators as 
\begin{equation}\label{eq: occupation number vector}
  \vec{n}=(\akind[1],\akind[2],\bkind[\DOT1],\bkind[\DOT2],\dkind[1],\dkind[2],\dkind[3],\dkind[4])\eqndot                                                                                                                                                                                         \end{equation}
In the free theory, the generators of $\mathfrak{psu}(2,2|4)$ act on these oscillator states as
\begin{equation}\label{eq: oscillator algebra}
 \begin{aligned}
  \mathfrak{L}^\alpha_\beta&=\aosc^{\dagger \alpha}\aosc_\beta-\frac{1}{2}\delta^\alpha_\beta \aosc^{\dagger \gamma}\aosc_\gamma \eqncom &
  \mathfrak{Q}^{\alpha A}&=\aosc^{\dagger\alpha}\dosc^{\dagger A} \eqncom \\
  \dot{\mathfrak{L}}^{\alphadot}_{\betadot}&=\bosc^{\dagger \alphadot}\bosc_{\betadot}-\frac{1}{2}\delta^{\alphadot}_{\betadot} \bosc^{\dagger \gammadot}\bosc_{\gammadot} \eqncom &
  \mathfrak{S}_{\alpha A}&=\aosc_{\alpha}\dosc_{A}  \eqncom\\
  \mathfrak{R}^A_B&=\dosc^{\dagger A}\dosc_B-\frac{1}{4}\delta^A_B \dosc^{\dagger C}\dosc_C \eqncom&
  \dot{\mathfrak{Q}}^{\alphadot}_A&=\bosc^{\dagger\alphadot}\dosc_{A}  \eqncom\\
  \mathfrak{D}&=\frac{1}{2}(\aosc^{\dagger \gamma}\aosc_\gamma+\bosc^{\dagger \gammadot}\bosc_{\gammadot}+2) \eqncom&
  \dot{\mathfrak{S}}^{ A}_{\alphadot}&=\bosc_{\alphadot}\dosc^{\dagger A}  \eqncom\\
  \mathfrak{P}^{\alpha \alphadot}&=\aosc^{\dagger\alpha}\bosc^{\dagger \alphadot} \eqncom &
  \mathfrak{K}_{\alpha \alphadot}&=\aosc_{\alpha}\bosc_{\alphadot} \eqndot
 \end{aligned}
\end{equation}
They can be supplemented by the central charge $\mathfrak{C}$ and the hypercharge $\mathfrak{B}$,
\begin{equation}\label{eq: oscillator algebra extension}
 \begin{aligned}
 \mathfrak{C}&=\frac{1}{2}(\aosc^{\dagger \gamma}\aosc_\gamma-\bosc^{\dagger \gammadot}\bosc_{\gammadot}-\dosc^{\dagger C}\dosc_C+2) \eqncom&
  \mathfrak{B}&=\dosc^{\dagger C}\dosc_C \eqncom&
 \end{aligned}
\end{equation}
to arrive at the action of $\mathfrak{u}(2,2|4)$.\footnote{Some authors define the hypercharge as $\mathfrak{B}=\frac{1}{2}(\aosc^{\dagger \gamma}\aosc_\gamma-\bosc^{\dagger \gammadot}\bosc_{\gammadot}+\dosc^{\dagger C}\dosc_C+2)$. The two definitions are equivalent for $\mathfrak{C}=0$.}
The central charge $\mathfrak{C}$ vanishes on all physical fields \eqref{eq: fields}.
The hypercharge $\mathfrak{B}$ measures the fermionic degree, i.e.\ the number of fermionic oscillators.
Alternatively, we can write the states and algebra generators as in \cite{Beisert:2003jj,Beisert:2004ry} in terms of $\cosc$ instead of $\dosc$ oscillators, which are defined as $\cosc^\dagger=\dosc$, $\cosc=\dosc^\dagger$. 
The vacuum $\vac$ of the $\dosc$ oscillators then has to be replaced by $\cosc^{\dagger 1}\cosc^{\dagger 2}\cosc^{\dagger 3}\cosc^{\dagger 4}\vac$.
Using the $\cosc$ oscillators, conjugation can simply be written as%
\footnote{At the level of the oscillators, a second kind of conjugation appears natural: the exchange of creation and annihilate operators. This conjugation occurs in the radial quantisation, see e.g.\ \cite{Zwiebel:2011bx,Pappadopulo:2012jk}. 
}
\begin{equation}
 \aosc \leftrightarrow \bosc\eqncom \quad \dosc \leftrightarrow \cosc \eqndot
\end{equation}
It transforms a field characterised by $\vec{n}$ to one characterised by
\begin{equation}
\vec{n}^*=(\bkind[\DOT1],\bkind[\DOT2],\akind[1],\akind[2],1-\dkind[1],1-\dkind[2],1-\dkind[3],1-\dkind[4])\eqncom
\end{equation}
apart from a sign.\footnote{This sign can be worked out from the replacements of the oscillators and the vacuum and an additional sign for $\sum_{A=1}^4 \dkind[A]=2$ and $\sum_{A=1}^4 \dkind[A]=3$ which stems from the fact that the canonical order of the $\coscdag$ oscillators is opposite to that of the $\doscdag$ oscillators.}

Single-trace operators containing $L$ irreducible fields correspond to states in the $L$-fold tensor products of the singleton representation, which in addition have to be graded cyclic invariant.
This graded cyclic invariance stems from the fact that a trace is manifestly invariant under the shift of a field from the last position to the first if the field or the rest of the trace is bosonic. In the case that both are fermionic, the operator acquires a sign.
We can construct states in the tensor product by adding an additional site index $i=1,\dots,L$ to the spin-chain oscillators in \eqref{eq: (anti)commutation relations}--\eqref{eq: oscillator algebra extension} and characterise them via  
$\{\vec{n}_{i}\}_{i=1,\dots,L}$.
The generators $\mathfrak{J}$ of \eqref{eq: oscillator algebra} and \eqref{eq: oscillator algebra extension} act on a length-$L$ state as 
\begin{equation}\label{eq: oscillator action on composite operators}
 \mathfrak{J}=\sum_{i=1}^L \mathfrak{J}_i \eqncom
\end{equation}
where $\mathfrak{J}_i$ acts on the $i^{\text{th}}$ site.

In the interacting theory, the generators \eqref{eq: oscillator algebra} receive quantum corrections. 
In the perturbative regime, the dilatation operator can be expanded in the effective planar coupling constant
\begin{equation}
 g=\frac{\sqrt{\lambda}}{4\pi}=\frac{g_\YM\sqrt{N}}{4\pi}
\end{equation}
as
\begin{equation}
 \mathfrak{D}=\mathfrak{D}_0+g^2 \mathfrak{D}_2 +\cO(g^3) \eqndot
\end{equation}
The one-loop dilatation operator $\mathfrak{D}_2$ of \NfSYMt was found in \cite{Beisert:2003jj} by an explicit Feynman diagram calculation in the so-called $\mathfrak{sl}(2)$ subsector of $\mathfrak{psu}(2,2|4)$,\footnote{This subsector is defined by $\vec{n}_i=(n_i,0,n_i,0,0,0,1,1)$ with $n_i\in \NN_0$ for all $i$.} which was then lifted to the complete theory via symmetry. 
It was later found that apart from a global constant it is completely fixed by symmetry \cite{Beisert:2004ry}. 
The one-loop dilatation operator $\mathfrak{D}_2$ acts only on two fields of the spin chain at a time, which have to be neighbouring in the planar limit. It can hence be written in terms of its density $(\mathfrak{D}_2)_{i\, i+1}$. 
On a length-$L$ state, one has
\begin{equation}
 \mathfrak{D}_2=\sum_{i=1}^L (\mathfrak{D}_2)_{i\, i+1} \eqncom
\end{equation}
where $(\mathfrak{D}_2)_{i\,i+1}$ acts on the $i^{\text{th}}$ and $(i+1)^{\text{th}}$ site and cyclic identification $i+L\sim i$ is understood.
Several different formulations of the so-called harmonic action of $(\mathfrak{D}_2)_{i\, i+1}$ exist, either in terms of harmonic numbers and projection operators or in terms of the sum of weighted hopping operations of the oscillators \cite{Beisert:2003jj}. 
The first kind of formulation \cite{Beisert:2003jj} uses the decomposition of the tensor product $\cV_S\otimes\cV_S$ in terms of irreducible representations $\cV_j$:
\begin{equation}
 \cV_S\otimes\cV_S=\bigoplus_{i=0}^\infty \cV_j \eqndot
\end{equation}
Defining the projectors 
\begin{equation}
 \PP_j: \cV_S\otimes\cV_S \rightarrow \cV_j \eqncom
\end{equation}
the dilatation-operator density can be written as 
\begin{equation}
 (\mathfrak{D}_2)_{i\, i+1}= 2\sum_{j=0}^\infty h(j)(\PP_j)_{i\, i+1} \eqncom
\end{equation}
where $h(j)=\sum_{k=1}^j\frac{1}{k}$ is the $j^\text{th}$ harmonic number and $(\PP_j)_{i\, i+1}$ acts on the $i^{\text{th}}$ and $(i+1)^{\text{th}}$ site.
This formulation is equivalent to a second kind of formulation \cite{Beisert:2003jj}, which is more suitable for concrete calculations and uses the oscillators defined above. 
We combine the $\aosc^{\dagger \alpha}_i$, $\bosc^{\dagger \alphadot}_i$ and $\dosc^{\dagger A}_i$ oscillators into one super oscillator 
\begin{equation}\label{eq: def super oscillator}
\Aosc_i^\dagger=(\aosc_i^{\dagger 1},\aosc_i^{\dagger 2},\bosc_i^{\dagger \DOT1},\bosc_i^{\dagger \DOT2},\dosc_i^{\dagger 1},\dosc_i^{\dagger 2},\dosc_i^{\dagger 3},\dosc_i^{\dagger 4}) 
\end{equation} 
and introduce an index $A_i$ to $\Aosc_i^\dagger$ to label the eight different components in \eqref{eq: def super oscillator}.
The dilatation-operator density acts on a two-site state, i.e.\ a state in the tensor product $\cV_S\otimes\cV_S$, as a weighted sum over all re-distributions of the oscillators:
\begin{equation}\label{eq: one-loop dila in oscillators beisert}
 (\mathfrak{D}_2)_{1\, 2} \, \Aosc_{s_1}^{\dagger A_1}\cdots\Aosc_{s_n}^{\dagger A_n} \vac = \sum_{s^\prime_1,\dots,s^\prime_n=1}^2 \delta_{C_2,0} \,c(n,n_{12},n_{21}) \Aosc_{s_1^\prime}^{\dagger A_1}\cdots\Aosc_{s_n^\prime}^{\dagger A_n} \vac \eqncom
\end{equation}
where the $s_k=1,2$ specify the original sites of the oscillators, $n$ is the total number of oscillators and  $n_{12}$ ($n_{21}$) is the number of oscillators hopping from $1$ to $2$ ($2$ to $1$).
The Kronecker delta ensures that the central charge $C_2$ of the final state at the second site vanishes. Together with the condition that the original states are physical, i.e.\ have vanishing central charge, this implies physicality for the final states at both sites.
The coefficient is 
\begin{equation}\label{eq: harmonic action coefficient}
 c(n,n_{12},n_{21})=\begin{cases}
 2 h(\frac12 n) \hfill \text{if } n_{12}=n_{21}=0 \eqncom& \\
 2 (-1)^{1+n_{12}n_{21}}  \beta(\frac12(n_{12}+n_{21}),1+\frac12(n-n_{12}-n_{21})) \qquad \text{else,}&
                     \end{cases}
\end{equation}
where $\beta$ is the Euler beta function.
For our purpose, a particular integral representation of this oscillator formulation is most suitable, which was found in \cite{Zwiebel:2007cpa}.%
\footnote{An operator variant of this formulation was given in \cite{Ferro:2013dga,Frassek:2013xza} and a different integral representation can be found in \cite{Fokken:2014moa}.}
We define
\begin{equation}
 (\Aosc^\dagger_i)^{\vec{n}_i}=(\aosc_i^{\dagger 1})^{\akindsite[1]{i}}(\aosc_i^{\dagger 2})^{\akindsite[2]{i}}(\bosc_i^{\dagger \DOT1})^{\bkindsite[\DOT1]{i}}(\bosc_i^{\dagger \DOT2})^{\bkindsite[\DOT2]{i}}(\dosc_i^{\dagger 1})^{\dkindsite[1]{i}}(\dosc_i^{\dagger 2})^{\dkindsite[2]{i}}(\dosc_i^{\dagger 3})^{\dkindsite[3]{i}}(\dosc_i^{\dagger 4})^{\dkindsite[4]{i}}\eqndot
\end{equation}
Then, the dilatation-operator density acts on a two-site state as
\begin{equation}\label{eq: one-loop dila in oscillators}
 (\mathfrak{D}_2)_{1\, 2} \, (\Aosc_{1}^{\dagger})^{ \vec{n}_{1}}(\Aosc_{2}^{\dagger})^{ \vec{n}_{2}}\vac = 4 \delta_{C_2,0} \int_0^{\frac{\pi}{2}} \de \theta \cot\theta \left((\Aosc_{1}^{\dagger})^{ \vec{n}_{1}}(\Aosc_{2}^{\dagger})^{ \vec{n}_{2}}-(\Aosc_{1}^{\prime\dagger})^{ \vec{n}_{1}}(\Aosc_{2}^{\prime \dagger})^{ \vec{n}_{2}} \right)\vac \eqncom
\end{equation}
where
\begin{equation}
\left( \begin{array}{c}
\Aosc_{1}^{\prime \dagger}  \\
\Aosc_{2}^{\prime \dagger}   \end{array} \right)
=V(\theta)  \left( \begin{array}{c}
\Aosc_{1}^{\dagger}  \\
\Aosc_{2}^{\dagger}   \end{array} \right) \eqncom \qquad
V(\theta)=
\left( \begin{array}{cc}
\cos\theta & -\sin\theta \\
\sin\theta & \cos\theta \end{array} \right) \eqndot
\end{equation}
This formulation can be shown to be equivalent to the above formulation via the following integral representation of the Euler beta function:
\begin{equation}\label{eq: integral for beta function}
 \beta(x,y)=2\int_0^{\frac{\pi}{2}} \de \theta \sin^{2x-1}\theta\cos^{2y-1}\theta \eqndot
\end{equation}
Note that \eqref{eq: integral for beta function} is divergent for $x=0$. This divergence is precisely cancelled by the first term in \eqref{eq: one-loop dila in oscillators}, leading to the first line in \eqref{eq: harmonic action coefficient}.

Beyond the first loop order, the dilatation operator receives corrections that do not preserve the length $L$ and the hypercharge $\mathfrak{B}$.
At leading order, these are completely fixed by symmetry and were determined in \cite{Zwiebel:2011bx}.
Also the generators $\mathfrak{Q}$, $\dot{\mathfrak{P}}$, $\mathfrak{S}$, $\dot{\mathfrak{S}}$, $\mathfrak{P}$ and $\mathfrak{K}$ receive length-changing quantum corrections. These were determined at leading order in \cite{Zwiebel:2007cpa}.

\subsection{Form factors in the free theory}
\label{subsec: form factors}

We can now compute the form factor of a generic single-trace operator built of irreducible fields.%
\footnote{The form factor of a linear combination of operators is given by the respective linear combination of the individual form factors. Hence, we can restrict the discussion to irreducible fields as defined in the last subsection without loss of generality.}
 We start in the free theory.

\begin{figure}[htbp]
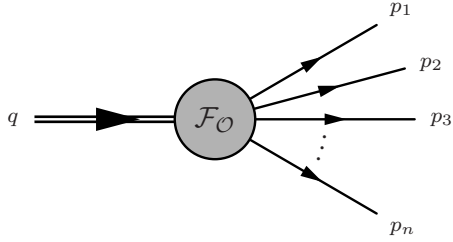

 \centering
\begin{equation*}
 \begin{aligned}
\settoheight{\eqoff}{$\times$}%
\setlength{\eqoff}{0.5\eqoff}%
\addtolength{\eqoff}{-14.5\unitlength}%
\raisebox{\eqoff}{%
\fmfframe(2,2)(6,2){%
\begin{fmfchar*}(50,25)
\fmfleft{vq}
\fmfright{vpL,vp,vp3,vp2,vp1}
\fmf{dbl_plain_arrow,tension=5}{vq,vqa}
\fmf{plain_arrow,tension=1}{vpLa,vpL}
\fmf{plain_arrow,tension=1}{vp3a,vp3}
\fmf{phantom,tension=1}{vpa,vp}
\fmf{plain_arrow}{vp1a,vp1}
\fmf{plain_arrow}{vp2a,vp2}
\fmf{dbl_plain_arrow,tension=25}{vqa,v1}
\fmf{plain_arrow,tension=5}{v1,vpLa}
\fmf{plain_arrow,tension=5}{v1,vp3a}
\fmf{phantom,tension=5}{v1,vpa}
\fmf{plain_arrow,tension=5}{v1,vp1a}
\fmf{plain_arrow,tension=5}{v1,vp2a}
\fmffreeze
\fmfdraw
 \fmfcmd{pair vertq, vertpone, vertptwo, vertpthree, vertpL, vertone, verttwo, vertp; vertone = vloc(__v1); verttwo = vloc(__v2); vertq = vloc(__vq); vertpone = vloc(__vp1); vertptwo = vloc(__vp2); vertpthree = vloc(__vp3);vertp = vloc(__vp);vertpL = vloc(__vpL);}
 \fmfiv{decor.shape=circle,decor.filled=30,decor.size=30}{vertone}
 \fmfiv{label=$ \cF_{\cO}$,l.d=0,l.a=0}{vertone}
 \fmfiv{label=$\scriptstyle q$}{vertq}
 \fmfiv{label=$\scriptstyle p_1$}{vertpone}
 \fmfiv{label=$\scriptstyle p_2$}{vertptwo}
 \fmfiv{label=$\scriptstyle p_3$}{vertpthree}
 \fmfiv{label=$\scriptstyle p_n$}{vertpL}
 \fmfiv{label=$\cdot$,l.d=40,l.a=-10}{vertone}
 \fmfiv{label=$\cdot$,l.d=40,l.a=-16}{vertone}
 \fmfiv{label=$\cdot$,l.d=40,l.a=-22}{vertone}
\end{fmfchar*}%
}}%
\end{aligned}
\end{equation*}
\caption{The momentum-space form factor of an operator $\cO$. The operator has momentum $q$ and is depicted as double line while the $n$ on-shell fields with momenta $p_i$ ($i=1,\dots,n$) are depicted as single lines. The arrows indicate the direction of momentum flow.}
\label{fig: form factor}
\end{figure}

To use the on-shell formalism, we Fourier transform \eqref{eq: form factor intro} to momentum space: 
\begin{equation}\label{eq: form factor momentum space}
\begin{aligned}
 \cF_{\cO}(1,\dots,n;q)&=\int \de^4x \e^{-iqx}\bra{1,\dots,n}\cO(x)\ket{0}\\
 &=\int \de^4x \e^{-iqx}\bra{1,\dots,n}\e^{i x \mathfrak{P}}\cO(0)\e^{-i x \mathfrak{P}}\ket{0}\\
 &=\delta^4\left(q-\sum_{i=1}^n p_i\right)\bra{1,\dots,n}\cO(0)\ket{0} \eqncom
 \end{aligned}
\end{equation}
where $\mathfrak{P}$ is the momentum operator and the delta function guarantees momentum conservation. The resulting momentum-space form factor is depicted in figure \ref{fig: form factor}.

Following the usual approach for amplitudes in \NfSYMt, we use the spinor helicity variables $\lambda^\alpha_{p_i}$, $\lambdat^{\alphadot}_{p_i}$ to write the momenta of the external on-shell particles $i=1,\dots, n$ as $p^{\alpha\alphadot}_i=\lambda^\alpha_{p_i}\lambdat^{\alphadot}_{p_i}$. We also frequently abbreviate these as $\lambda^\alpha_{i}$, $\lambdat^{\alphadot}_{i}$. Contractions of spinor helicity variable are written as $\ab{i j}=\epsilon_{\alpha\beta}\lambda_i^\alpha\lambda_j^\beta$ and $\sb{i j}=-\epsilon_{\alphadot \betadot }\lambdat_i^{\alphadot}\lambdat_j^{\betadot}$.

Furthermore, we use Nair's $\cN=4$ on-shell superspace \cite{Nair:1988bq} to combine the different external fields into a single formal superfield%
\begin{equation}
\label{eq: N=4 superspace}
 \Phi=g^+ +  \eta^A \, \bar\psi_A +\frac{1}{2!}\eta^A\eta^B \, \phi_{AB} +\frac{1}{3!}\epsilon_{ABCD} \eta^A\eta^B\eta^C \, \psi^D+\frac{1}{4!}\epsilon_{ABCD}\eta^A\eta^B\eta^C\eta^D \, g^- \eqndot
\end{equation}
An external field $\Phi_i$ can then be completely characterised by its super momentum $\Lambda_i=(\lambda_i,\lambdat_i,\eta_i)$. 
Alternatively, the superfield can be expanded in fermionic variables $\etat$, which are related to the variables $\eta$ via
\begin{equation}
\begin{aligned}
 1&=\eta^1_i\eta^2_i\eta^3_i\eta^4_i\eqncom \quad& 
 \etat_{iA}&=\frac{1}{3!}\epsilon_{ABCD}\eta^B_i\eta^C_i\eta^D_i\eqncom \quad &
 \etat_{iA}\etat_{iB}&=\frac{1}{2!}\epsilon_{ABCD}\eta^C_i\eta^D_i\eqncom \\ 
 & &\etat_{iA}\etat_{iB}\etat_{iC}&=\epsilon_{ABCD}\eta^D_i\eqncom \quad &
 \etat_{i1}\etat_{i2}\etat_{i3}\etat_{i4}&=1\eqndot&
 \end{aligned}
\end{equation}
In the language of on-shell super fields, conjugation is given by
\begin{equation}
\label{eq: conjugation in on-shell variables}
 \lambda \leftrightarrow \lambdat \eqncom \quad \eta \leftrightarrow \etat \eqndot
\end{equation}
The individual component expressions for the resulting super form factor can be extracted via suitable derivatives with respect to the $\eta$'s.
For example,\footnote{Note the sign which accounts for the anticommuting nature of the $\eta$'s.}
\begin{equation}
 \begin{aligned}
 \cF_{\cO}(1^{g^+},2^{g^-},\dots,n^{\phi_{12}};q)=1\left(\frac{\partial}{\partial \eta_2^1}\frac{\partial}{\partial \eta_2^2}\frac{\partial}{\partial \eta_2^3}\frac{\partial}{\partial \eta_2^4}\right)\cdots\left(-\frac{\partial}{\partial \eta_n^1}\frac{\partial}{\partial \eta_n^2}\right)\cF_{\cO}(1,2,\dots,n;q)\bigg|_{\eta_i^A=0} \eqncom
 \end{aligned}
\end{equation}
where the superscripts specify the helicities and flavours of the respective fields in the component form factor and we have abbreviated the dependence of the (super) form factor on the (super) momenta $p_i$ ($\Lambda_i$) by $i=1,\dots,n$.

In analogy to amplitudes, we introduce colour-ordered form factors $\hat{\cF}_\cO$ via
\begin{equation}\label{eq: def colour-ordered form factor}
 \begin{aligned}
 \cF_{\cO}(1,\dots,n;q)= \sum_{\sigma\in \SS_n/\ZZ_n} \Tr[\T^{a_{\sigma(1)}}\cdots\T^{a_{\sigma(n)}}] \hat{\cF}_{\cO}(\sigma(1),\dots,\sigma(n);q)
 \eqncom
 \end{aligned}
\end{equation}
where $\T^a$ with $a=1,\dots,N^2-1$ are the generators of $SU(N)$ and the sum is over all non-cyclic permutations. 
Note that the insertion of the operator does not take part in the colour ordering as the operator is a colour singlet. 
Starting at one-loop order, also multi-trace terms appear in \eqref{eq: def colour-ordered form factor}, which we have, however, suppressed since they are subleading in the 't Hooft limit.

\begin{figure}[htbp]
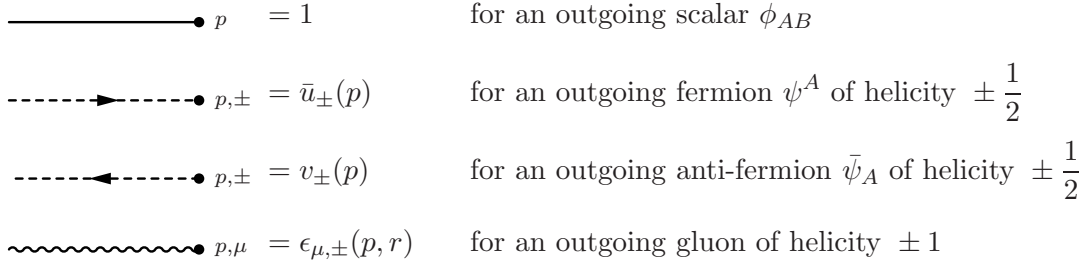
%
\begin{equation*}
 \begin{aligned}
\settoheight{\eqoff}{$\times$}%
\setlength{\eqoff}{0.5\eqoff}%
\addtolength{\eqoff}{-5\unitlength}%
\raisebox{\eqoff}{%
\fmfframe(2,0)(8,0){%
\begin{fmfchar*}(25,8)
\fmfleft{vq}
\fmfright{vp}
\fmf{plain,tension=1}{vq,vp}
\fmffreeze
\fmfdraw
 \fmfcmd{pair vertq, vertp;  vertq = vloc(__vq); vertp = vloc(__vp);}
 \fmfiv{decor.shape=circle,decor.filled=100,decor.size=3}{vertp}
 \fmfiv{label=$\scriptstyle p$}{vertp}
\end{fmfchar*}%
}}%
&=
1 \quad & &\text{for an outgoing scalar }\phi_{AB} \\
\settoheight{\eqoff}{$\times$}%
\setlength{\eqoff}{0.5\eqoff}%
\addtolength{\eqoff}{-5\unitlength}%
\raisebox{\eqoff}{%
\fmfframe(2,0)(8,0){%
\begin{fmfchar*}(25,8)
\fmfleft{vq}
\fmfright{vp}
\fmf{dashes_arrow,tension=1}{vq,vp}
\fmffreeze
\fmfdraw
 \fmfcmd{pair vertq, vertp;  vertq = vloc(__vq); vertp = vloc(__vp);}
 \fmfiv{decor.shape=circle,decor.filled=100,decor.size=3}{vertp}
 \fmfiv{label=$\scriptstyle p,,\pm$}{vertp}
\end{fmfchar*}%
}}%
&=
\bar{u}_\pm(p) \quad & & \text{for an outgoing fermion }\psi^A\text{ of helicity }\pm\frac{1}{2} \\
\settoheight{\eqoff}{$\times$}%
\setlength{\eqoff}{0.5\eqoff}%
\addtolength{\eqoff}{-5\unitlength}%
\raisebox{\eqoff}{%
\fmfframe(2,0)(8,0){%
\begin{fmfchar*}(25,8)
\fmfleft{vq}
\fmfright{vp}
\fmf{dashes_arrow,tension=1}{vp,vq}
\fmffreeze
\fmfdraw
 \fmfcmd{pair vertq, vertp;  vertq = vloc(__vq); vertp = vloc(__vp);}
 \fmfiv{decor.shape=circle,decor.filled=100,decor.size=3}{vertp}
 \fmfiv{label=$\scriptstyle p,,\pm$}{vertp}
\end{fmfchar*}%
}}%
&=
v_\pm(p) \quad & & \text{for an outgoing anti-fermion }\bar{\psi}_A\text{ of helicity }\pm\frac{1}{2}  \\
\settoheight{\eqoff}{$\times$}%
\setlength{\eqoff}{0.5\eqoff}%
\addtolength{\eqoff}{-5\unitlength}%
\raisebox{\eqoff}{%
\fmfframe(2,0)(8,0){%
\begin{fmfchar*}(25,8)
\fmfleft{vq}
\fmfright{vp}
\fmf{photon,tension=1}{vq,vp}
\fmffreeze
\fmfdraw
 \fmfcmd{pair vertq, vertp;  vertq = vloc(__vq); vertp = vloc(__vp);}
 \fmfiv{decor.shape=circle,decor.filled=100,decor.size=3}{vertp}
 \fmfiv{label=$\scriptstyle p,,\mu$}{vertp}
\end{fmfchar*}%
}}%
&=
\epsilon_{\mu,\pm}(p,r) \quad & & \text{for an outgoing gluon of helicity }\pm 1
\end{aligned}
\end{equation*}
\caption{The momentum-space Feynman rules for outgoing scalars, fermions, anti-fermions and gluons; cf.\ for instance \cite{Henn:2014yza}.
In all cases, the momentum $p$ is flowing out of the diagram; the arrow distinguishes fermions from anti-fermions.
As usual, $\bar{u}_\pm(p)$ and $v_\pm(p)$ are solutions of the massless Dirac equation and $\epsilon_{\mu, \pm}(p,r)$ are the polarisation vectors.
The reference vector $r$ in $\epsilon_{\mu, \pm}(p,r)$ can be chosen independently for each gluon and has to drop out of all gauge-invariant quantities.}%
\label{fig: Feynman rules}%
\end{figure}

The form factor can now easily be computed via Feynman rules. 
In the free theory, no interactions can occur and the form factor vanishes unless the number of the external fields $n$ equals the number of the irreducible fields $L$ in $\cO$.\footnote{Moreover, their types have to match.}
In the interacting theory, each occurrence of the Yang-Mills coupling constant $g_\YM$ either increases the number of external fields or the number of loops. Hence, the form factor of the free theory equals the minimal tree-level form factor in the interacting theory.  
The required Feynman rules for the outgoing fields are depicted in figure \ref{fig: Feynman rules}. %

For an outgoing scalar $\phi_{AB}$, the Feynman rules give simply $1$.
This has to be dressed with $\eta^A\eta^B$ to obtain the corresponding super form factor. 
For an outgoing fermion $\psi^A_{\alpha}$ of negative helicity,
the Feynman rules give $\bar{u}_-(p)=(\lambda^\alpha,0)$  and hence $\lambda^\alpha$, which has to be dressed with $\frac{1}{3!}\epsilon_{ABCD}\eta^B\eta^C\eta^D$.
For an outgoing antifermion $\bar\psi_{A \alphadot}$ of positive helicity, we have $v_+(p)=(0,\lambdat^\alphadot)^T$ and hence $\lambdat^{\alphadot}$ dressed with $\eta^A$. 
For vanishing coupling, the covariant derivative $\cder_{\alpha\alphadot}$ reduces to the ordinary derivative $\partial_{\alpha\alphadot}$.
In momentum space, it simply gives the momentum $p^{\alpha\alphadot}=\lambda^\alpha\lambdat^{\alphadot}$ of the external field associated to the field in the operator on which $\cder_{\alpha\alphadot}$ acts.%
\footnote{We have absorbed a factor of the imaginary unit $i$, which one would expect from the Fourier transformation, into the definition of the (covariant) derivative.
} 
For an outgoing gauge field of positive or negative helicity, the Feynman rules yield the polarisation vectors  $\epsilon_{\mu,\pm}(p,r)$, where $r$ is an arbitrary reference vector. In spinor helicity variables, these read 
\begin{equation}\label{eq: def polarisation vectors}
 \epsilon_+^{\alpha\alphadot}(p;r)=\sqrt{2}\frac{\lambda_r^\alpha \lambdat_p^\alphadot}{\ab{ r p}}\eqncom \quad
 \epsilon_-^{\alpha\alphadot}(p;r)=\sqrt{2}\frac{\lambda_p^\alpha\lambdat_r^\alphadot}{\sb{p r}} \eqndot
\end{equation}
As mentioned in section \ref{subsec: operators and spin chain picture}, the operators $\cO$ contain gauge fields only in the gauge-invariant and irreducible combinations of the self-dual and anti-self-dual field strengths \eqref{eq: def field strength slitting}. For vanishing coupling, these read 
\begin{equation}\label{eq: field stengths definition}
 F_{\alpha\beta}=-\frac{1}{2\sqrt{2}}\epsilon^{\alphadot\betadot}(\partial_{\alpha\alphadot}A_{\beta\betadot}-\partial_{\beta\betadot}A_{\alpha\alphadot})\eqncom \quad  \bar{F}_{\alphadot\betadot}=-\frac{1}{2\sqrt{2}} \epsilon^{\alpha\beta}(\partial_{\alpha\alphadot}A_{\beta\betadot}-\partial_{\beta\betadot}A_{\alpha\alphadot})\eqndot
\end{equation}
Inserting the polarisation vectors \eqref{eq: def polarisation vectors} and the momenta in terms of spinor helicity variables into \eqref{eq: field stengths definition}, we obtain for outgoing gluons of the specified helicities
\begin{equation}
 \begin{aligned}
   F_{\alpha\beta}&\overset{\epsilon_+}{\longrightarrow}
   -\frac{1}{2\sqrt{2}}\epsilon_{\alphadot\betadot}(\lambda_p^{\alpha}\lambdat_p^{\alphadot}\epsilon_+^{\beta\betadot}-\lambda_p^{\beta}\lambdat_p^{\betadot}\epsilon_+^{\alpha\alphadot})
   =0\eqncom\\
   F_{\alpha\beta}&\overset{\epsilon_-}{\longrightarrow}
   -\frac{1}{2\sqrt{2}}\epsilon_{\alphadot\betadot}(\lambda_p^{\alpha}\lambdat_p^{\alphadot}\epsilon_-^{\beta\betadot}-\lambda_p^{\beta}\lambdat_p^{\betadot}\epsilon_-^{\alpha\alphadot})
   =\lambda_p^\alpha\lambda_p^\beta\eqncom\\
   F_{\alphadot\betadot}&\overset{\epsilon_+}{\longrightarrow}
   -\frac{1}{2\sqrt{2}}\epsilon_{\alpha\beta}(\lambda_p^{\alpha}\lambdat_p^{\alphadot}\epsilon_+^{\beta\betadot}-\lambda_p^{\beta}\lambdat_p^{\betadot}\epsilon_+^{\alpha\alphadot})
   =\lambdat_p^\alphadot\lambdat_p^\betadot\eqncom\\
   F_{\alphadot\betadot}&\overset{\epsilon_-}{\longrightarrow}
   -\frac{1}{2\sqrt{2}}\epsilon_{\alpha\beta}(\lambda_p^{\alpha}\lambdat_p^{\alphadot}\epsilon_-^{\beta\betadot}-\lambda_p^{\beta}\lambdat_p^{\betadot}\epsilon_-^{\alpha\alphadot})
   =0\eqnsem
 \end{aligned}
\end{equation}
cf.\ \cite{Witten:2003nn}.
In the super form factor, we thus have $\lambda^\alpha\lambda^\beta\eta^1\eta^2\eta^3\eta^4$ and $\lambdat^{\alphadot}\lambdat^{\betadot}$ for self-dual and anti-self-dual field strengths, respectively.%
\footnote{Note that the vertical position of the indices in the operator is opposite to the one in the super momentum variables. For the $\eta$'s, this is a consequence of \eqref{eq: N=4 superspace}. For the $\lambda$'s and $\lambdat$'s, it arises since the Feynman rules for the operators are obtained by taking the functional derivatives w.r.t.\ the fields in the operator.}

Assembling the above pieces, we obtain the colour-ordered form factor of a composite operator $\cO$ with $\{\vec{n}_i\}_{i=1,\dots,L}=\{(\akindsite[1]{i},\akindsite[2]{i},\bkindsite[1]{i},\bkindsite[2]{i},\dkindsite[1]{i},\dkindsite[2]{i},\dkindsite[3]{i},\dkindsite[4]{i})\}_{i=1,\dots,L}$ as defined in \eqref{eq: occupation number vector} in the free theory. It is given by 
\begin{multline}\label{eq: form factor from spin chain}
 \hat{\cF}_{\cO}(\Lambda_1,\dots,\Lambda_L;q)=\delta^4\left(q-\sum_{i=1}^L p_i\right) \sum_{\sigma\in\ZZ_L}\\ \prod_{i=1}^L (\lambda_{\sigma(i)}^1)^{\akindsite[1]{i}}(\lambda_{\sigma(i)}^2)^{\akindsite[2]{i}}(\lambdat_{\sigma(i)}^{\DOT1})^{\bkindsite[\DOT1]{i}}(\lambdat_{\sigma(i)}^{\DOT2})^{\bkindsite[\DOT2]{i}}(\eta_{\sigma(i)}^1)^{\dkindsite[1]{i}}(\eta_{\sigma(i)}^2)^{\dkindsite[2]{i}}(\eta_{\sigma(i)}^3)^{\dkindsite[3]{i}}(\eta_{\sigma(i)}^4)^{\dkindsite[4]{i}} \eqndot
 \end{multline}
The sum over all cyclic permutations stems from the (graded) cyclic invariance of the single-trace operator. It is manifestly invariant under the shift of a field from the last position in the trace to the first if the field or the rest of the trace is bosonic. In the case that both are fermionic, the operator acquires a sign.
This is reflected in the product $\prod$ in \eqref{eq: form factor from spin chain}; it inserts the super spinor helicity variables in the permuted order given by $\sigma$, which in the case of odd fermionic degree leads to signs when restoring the canonical order by commuting the $\lambda^\alpha_i$'s and $\lambdat^\alphadot_i$'s but anticommuting the $\eta_i^A$'s.
The colour-ordered form factor of a generic single-trace operator, which is characterised by a linear combination of $\{\vec{n}_i\}_{i=1,\dots,L}=\{(\akindsite[1]{i},\akindsite[2]{i},\bkindsite[1]{i},\bkindsite[2]{i},\dkindsite[1]{i},\dkindsite[2]{i},\dkindsite[3]{i},\dkindsite[4]{i})\}_{i=1,\dots,L}$, is simply given by the corresponding linear combinations of \eqref{eq: form factor from spin chain}.\footnote{%
Note that a subtlety arises in the normalisation of the operator, which is connected to symmetry factors and the number of different Wick contractions.
The operator $\tr[\phi_{12}\phi_{13}\phi_{12}\phi_{14}]$ allows for one planar contraction with its conjugate. 
The operator $\tr[\phi_{12}\phi_{13}\phi_{12}\phi_{13}]$, however, admits two planar contractions with its conjugate.
If one wants the two-point functions to be normalised to unity, this has to be taken into account.
}

Note that, apart from the momentum-conserving delta function and a normalisation factor of $L$, \eqref{eq: form factor from spin chain} precisely agrees with the result of replacing all oscillators of a normalised and graded cyclically invariant state in the spin-chain picture \eqref{eq: fields} according to%
\footnote{Alternatively, we can replace 
\begin{equation}
    \cosc_i^{\dagger A}\to\etat_i^A\eqncom\quad  \cosc_{i,A}\to\frac{\partial}{\partial \etat_i^A}\eqndot
\end{equation}
}
\begin{equation}\label{eq: oscillator replacements}
 \begin{aligned}
  \aosc_i^{\dagger \alpha} &\to \lambda_i^\alpha \eqncom & \bosc_i^{\dagger \alphadot}&\to \lambdat_i^{\alphadot}\eqncom & \dosc_i^{\dagger A}&\to\eta_i^A\eqncom\\
  \aosc_{i,\alpha} &\to \partial_{i,\alpha}=\frac{\partial}{\partial\lambda_i^\alpha} \eqncom & 
  \bosc_{i,\alphadot}&\to \partial_{i,\alphadot}=\frac{\partial}{\partial\lambdat_i^{\alphadot}}\eqncom & \dosc_{i,A}&\to\partial_{i,A}=\frac{\partial}{\partial \eta_i^A}\eqndot
 \end{aligned}
\end{equation}
If we replace the oscillators in \eqref{eq: oscillator algebra} and \eqref{eq: oscillator algebra extension} according to the same rules, we obtain the representation of the centrally extended $\mathfrak{psu}(2,2|4)$ on on-shell superfields, which is well known from scattering amplitudes:\footnote{For vanishing central charge, the helicity $h=-\frac{1}{2}\lambda^\alpha\frac{\partial}{\partial \lambda^\alpha}+\frac{1}{2}\lambdat^\alphadot\frac{\partial}{\partial \lambdat^\alphadot}$ is connected to the hypercharge $\mathfrak{B}$ as $h=1-\frac{1}{2}\mathfrak{B}$.}
\begin{equation}\label{eq: onshell algebra}
 \begin{aligned}
  \mathfrak{L}^\alpha_{i,\beta}&=\lambda^{ \alpha}_i\partial_{i,\beta}-\frac{1}{2}\delta^\alpha_\beta \lambda^{ \gamma}_i\partial_{i,\gamma} \eqncom &
  \mathfrak{Q}^{\alpha A}_i&=\lambda^{\alpha}_i\eta^{ A}_i \eqncom \\
  \dot{\mathfrak{L}}^{\alphadot}_{i,\betadot}&=\lambdat^{ \alphadot}_i\partial_{i,\betadot}-\frac{1}{2}\delta^{\alphadot}_{\betadot} \lambdat^{ \gammadot}_i\partial_{i,\gammadot} \eqncom &
  \mathfrak{S}_{i,\alpha A}&=\partial_{i,\alpha}\partial_{i,A}  \eqncom\\
  \mathfrak{R}^A_{i,B}&=\eta^{ A}_i\partial_{i,B}-\frac{1}{4}\delta^A_B \eta^{ C}_i\partial_{i,C} \eqncom&
  \dot{\mathfrak{Q}}^{\alphadot}_{i,A}&=\lambdat^{\alphadot}_i\partial_{i,A}  \eqncom\\
  \mathfrak{D}_i&=\frac{1}{2}(\lambda^{ \gamma}_i\partial_{i,\gamma}+\lambdat^{ \gammadot}_i\partial_{i,\gammadot}+2) \eqncom&
  \dot{\mathfrak{S}}^{ A}_{i,\alphadot}&=\partial_{i,\alphadot}\eta^{ A}_i  \eqncom\\
  \mathfrak{C}_i&=\frac{1}{2}(\lambda^{ \gamma}_i\partial_{i,\gamma}-\lambdat^{ \gammadot}_i\partial_{i,\gammadot}-\eta^{ C}_i\partial_{i,C}+2) \eqncom&
  \mathfrak{P}^{\alpha \alphadot}_i&=\lambda^{\alpha}_i\lambdat^{ \alphadot}_i \eqncom \\
  \mathfrak{B}&=\eta^{ C}_i\partial_{i,C} \eqncom&
  \mathfrak{K}_{i,\alpha \alphadot}&=\partial_{i,\alpha}\partial_{i,\alphadot} \eqnsem
 \end{aligned}
\end{equation}
cf.\ \cite{Witten:2003nn}.

The action of this algebra on the on-shell part of the form factor \eqref{eq: form factor from spin chain}
is given by the sum of the respective terms on each external on-shell field:
\begin{equation}
 \sum_{i=1}^n\mathfrak{J}_i \hat{\cF}_\cO(1,\dots,n;q)\eqndot
\end{equation}
Comparing this with the action in \eqref{eq: oscillator algebra}, \eqref{eq: oscillator algebra extension}, \eqref{eq: oscillator action on composite operators}, we find that 
\begin{equation}\label{eq: action on form factor}
 \sum_{i=1}^L\mathfrak{J}_i\hat{\cF}_\cO(1,\dots,L;q)=\hat{\cF}_{\mathfrak{J}\cO}(1,\dots,L;q)
\end{equation}
for any generator $\mathfrak{J}$ of the centrally extended $\mathfrak{psu}(2,2|4)$.

We have seen that the replacement \eqref{eq: oscillator replacements} of super oscillators by super spinor helicity variables translates the action of $\mathfrak{psu}(2,2|4)$ on spin chains to the one on amplitudes and furthermore translates the spin-chain states to minimal tree-level form factors.\footnote{Apart from the aforementioned momentum-conserving delta function and normalisation factor.}
Hence, the minimal tree-level form factors exactly realise the spin chain of free \NfSYMt in the language of on-shell super fields.

In fact, \eqref{eq: action on form factor} is a special case of a Ward identity for form factors which was derived in \cite{Brandhuber:2011tv}.
In principle, this Ward identity is also valid in the interacting theory.
However, as we have already mentioned in the previous subsection, the algebra of the spin chain is deformed in the interacting theory. 
Also the action of \eqref{eq: onshell algebra} on scattering amplitudes is known to be deformed in the interacting theory; see \cite{Beisert:2010jq,Bargheer:2011mm} for reviews.%
\footnote{A comparison between both deformations and the occurring representations can be found in \cite{Beisert:2010jq}.}
For form factors, both deformations occur.
In the remainder of this paper, we find in particular the one-loop correction to the action of the dilatation operator on composite operators from form factors.
We leave an exact study of the relations between both deformations via form factors for future work.

\section{One-loop form factors}
\label{sec: one-loop form factors}

\enlargethispage{0.5\baselineskip}

In this section, we use generalised unitarity to obtain the cut-constructible part of the one-loop correction to the $L$-point form factor of a generic length-$L$ single-trace operator.

Generalised unitarity \cite{Bern:1994zx,Bern:1994cg,Britto:2004nc} exists in numerous variations and is well understood at one loop; see \cite{Bern:2007dw,Elvang:2013cua} for reviews.
The variant we use works as follows.%
\footnote{We give additional details on this methods for the case of the minimal form factor below.}
In strictly four dimensions, every Feynman integral can be written as a sum of box integrals, triangle integrals, bubble integrals, massive tadpoles integrals and rational terms \cite{Passarino:1978jh}.%
\footnote{%
In $D=4-2\varepsilon$ dimensions, also the pentagon integral occurs; see e.g.\ \cite{Henn:2014yza}.
} 
In the case of massless theories such as \NfSYMt, the tadpoles vanish.
Hence, a general ansatz can be made, which is depicted in figure \ref{fig: ansatz for generalised unitarity} for a form factor.%
\footnote{In the case of one-loop amplitudes in \NfSYMt, the triangle integrals, bubble integrals and rational terms do not contribute \cite{Bern:1994zx,Bern:1994cg}. This is no longer the case for one-loop form factors.}
\begin{figure}[htbp]
 \centering
\begin{equation*}
 \begin{aligned}
\settoheight{\eqoff}{$\times$}%
\setlength{\eqoff}{0.5\eqoff}%
\addtolength{\eqoff}{-14.5\unitlength}%
\raisebox{\eqoff}{%
\fmfframe(2,2)(6,2){%
\begin{fmfchar*}(25,25)
\fmfsurround{vp3,vp2,vp1,vq,vpL,vp}
\fmf{dbl_plain_arrow,tension=1}{vq,vqa}
\fmf{plain_arrow,tension=1}{vpLa,vpL}
\fmf{plain_arrow,tension=1}{vp3a,vp3}
\fmf{phantom,tension=1}{vpa,vp}
\fmf{plain_arrow}{vp1a,vp1}
\fmf{plain_arrow}{vp2a,vp2}
\fmf{dbl_plain_arrow,tension=2}{vqa,v1}
\fmf{plain_arrow,tension=2}{v1,vpLa}
\fmf{plain_arrow,tension=2}{v1,vp3a}
\fmf{phantom,tension=2}{v1,vpa}
\fmf{plain_arrow,tension=2}{v1,vp1a}
\fmf{plain_arrow,tension=2}{v1,vp2a}
\fmffreeze
\fmfdraw
 \fmfcmd{pair vertq, vertpone, vertptwo, vertpthree, vertpL, vertone, verttwo, vertp; vertone = vloc(__v1); verttwo = vloc(__v2); vertq = vloc(__vq); vertpone = vloc(__vp1); vertptwo = vloc(__vp2); vertpthree = vloc(__vp3);vertp = vloc(__vp);vertpL = vloc(__vpL);}
 \fmfiv{decor.shape=circle,decor.filled=30,decor.size=30}{vertone}
 \fmfiv{decor.shape=circle,decor.filled=0,decor.size=20}{vertone}
 \fmfiv{label=$ \hat{\cF}_{\cO}$,l.d=0,l.a=0}{vertone}
 \fmfiv{label=$\scriptstyle q$}{vertq}
 \fmfiv{label=$\scriptstyle p_1$}{vertpone}
 \fmfiv{label=$\scriptstyle p_2$}{vertptwo}
 \fmfiv{label=$\scriptstyle p_3$}{vertpthree}
 \fmfiv{label=$\scriptstyle p_n$}{vertpL}
 \fmfiv{label=$\cdot$,l.d=20,l.a=-45}{vertone}
 \fmfiv{label=$\cdot$,l.d=20,l.a=-60}{vertone}
 \fmfiv{label=$\cdot$,l.d=20,l.a=-75}{vertone}
\end{fmfchar*}%
}}%
&=
\sum_{i,j,k,l}c_{\text{box}}^{(i,j,k,l)}
\quad
\settoheight{\eqoff}{$\times$}%
\setlength{\eqoff}{0.5\eqoff}%
\addtolength{\eqoff}{-12\unitlength}%
\raisebox{\eqoff}{%
\fmfframe(2,2)(6,2){%
\begin{fmfchar*}(20,20)
\fmfsurround{vpjjj,vpj1,vpjj,vpj,vpiii,vpi1,vpii,vpi,vq,vpl1,vpll,vpl,vpkkk,vpk1,vpkk,vpk}
\fmf{dbl_plain_arrow,tension=0}{vq,vi}
\fmf{plain_arrow,tension=1}{vi,vpi}
\fmf{plain_arrow,tension=1}{vi,vpl1}
\fmf{plain_arrow,tension=1}{vj,vpj}
\fmf{plain_arrow,tension=1}{vj,vpi1}
\fmf{plain_arrow,tension=1}{vk,vpk}
\fmf{plain_arrow,tension=1}{vk,vpj1}
\fmf{plain_arrow,tension=1}{vl,vpl}
\fmf{plain_arrow,tension=1}{vl,vpk1}
\fmf{plain_arrow,tension=1}{vi,vj}
\fmf{plain_arrow,tension=1}{vi,vl}
\fmf{plain_arrow,tension=1}{vj,vk}
\fmf{plain_arrow,tension=1}{vl,vk}
\fmffreeze
\fmfdraw
 \fmfcmd{pair vertq, verti, vertj, vertk, vertl, vertpi, vertpj, vertpk, vertpl, vertpii, vertpjj, vertpkk, vertpll; vertq = vloc(__vq); verti = vloc(__vi); vertj = vloc(__vj); vertk = vloc(__vk); vertl = vloc(__vl);vertpi = vloc(__vpi); vertpj = vloc(__vpj); vertpk = vloc(__vpk); vertpl = vloc(__vpl);vertpii = vloc(__vpi1); vertpjj = vloc(__vpj1); vertpkk = vloc(__vpk1); vertpll = vloc(__vpl1);}
 \fmfiv{label=$\scriptstyle q$}{vertq}
 \fmfiv{label=$\scriptstyle p_i$}{vertpi}
 \fmfiv{label=$\scriptstyle p_{i+1}$}{vertpii}
 \fmfiv{label=$\scriptstyle p_j$}{vertpj}
 \fmfiv{label=$\scriptstyle p_{j+1}$}{vertpjj}
 \fmfiv{label=$\scriptstyle p_k$}{vertpk}
 \fmfiv{label=$\scriptstyle p_{k+1}$}{vertpkk}
 \fmfiv{label=$\scriptstyle p_l$}{vertpl}
 \fmfiv{label=$\scriptstyle p_{l+1}$}{vertpll}
 \fmfiv{label=$\cdot$,l.d=0,l.a=0}{verti+(12*sind(243),12*cosd(243))}
 \fmfiv{label=$\cdot$,l.d=0,l.a=0}{verti+(12*sind(-67),12*cosd(-67))}
 \fmfiv{label=$\cdot$,l.d=10,l.a=105}{vertj}
 \fmfiv{label=$\cdot$,l.d=10,l.a=90}{vertj}
 \fmfiv{label=$\cdot$,l.d=10,l.a=75}{vertj}
 \fmfiv{label=$\cdot$,l.d=10,l.a=-15}{vertk}
 \fmfiv{label=$\cdot$,l.d=10,l.a=-0}{vertk}
 \fmfiv{label=$\cdot$,l.d=10,l.a=+15}{vertk}
 \fmfiv{label=$\cdot$,l.d=10,l.a=-105}{vertl}
 \fmfiv{label=$\cdot$,l.d=10,l.a=-90}{vertl}
 \fmfiv{label=$\cdot$,l.d=10,l.a=-75}{vertl}
\end{fmfchar*}%
}}%
+\sum_{i,j,k}c_{\text{triangle}}^{(i,j,k)}
\quad
\settoheight{\eqoff}{$\times$}%
\setlength{\eqoff}{0.5\eqoff}%
\addtolength{\eqoff}{-12\unitlength}%
\raisebox{\eqoff}{%
\fmfframe(2,2)(6,2){%
\begin{fmfchar*}(20,20)
\fmfsurround{vpjj,vpj,vpiii,vpi1,vpii,vpiiii,vpi,vq,vpk1,vpkk,vpkkkk,vpk,vpjjj,vpj1}
\fmf{dbl_plain_arrow,tension=0}{vq,vi}
\fmf{plain_arrow,tension=1}{vi,vpi}
\fmf{plain_arrow,tension=1}{vi,vpk1}
\fmf{plain_arrow,tension=1}{vj,vpj}
\fmf{plain_arrow,tension=1}{vj,vpi1}
\fmf{plain_arrow,tension=1}{vk,vpk}
\fmf{plain_arrow,tension=1}{vk,vpj1}
\fmf{plain_arrow,tension=0.66}{vi,vj}
\fmf{plain_arrow,tension=0.66}{vi,vk}
\fmf{plain_arrow,tension=0.66}{vj,vk}
\fmffreeze
\fmfdraw
 \fmfcmd{pair vertq, verti, vertj, vertk, vertpi, vertpj, vertpk, vertpii, vertpjj, vertpkk; vertq = vloc(__vq); verti = vloc(__vi); vertj = vloc(__vj); vertk = vloc(__vk); vertpi = vloc(__vpi); vertpj = vloc(__vpj); vertpk = vloc(__vpk); vertpii = vloc(__vpi1); vertpjj = vloc(__vpj1); vertpkk = vloc(__vpk1); }
 \fmfiv{label=$\scriptstyle q$}{vertq}
 \fmfiv{label=$\scriptstyle p_i$}{vertpi}
 \fmfiv{label=$\scriptstyle p_{i+1}$}{vertpii}
 \fmfiv{label=$\scriptstyle p_j$}{vertpj}
 \fmfiv{label=$\scriptstyle p_{j+1}$}{vertpjj}
 \fmfiv{label=$\scriptstyle p_k$}{vertpk}
 \fmfiv{label=$\scriptstyle p_{k+1}$}{vertpkk}
 \fmfiv{label=$\cdot$,l.d=0,l.a=0}{verti+(12*sind(243),12*cosd(243))}
 \fmfiv{label=$\cdot$,l.d=0,l.a=0}{verti+(12*sind(-67),12*cosd(-67))}
 \fmfiv{label=$\cdot$,l.d=10,l.a=35}{vertj}
 \fmfiv{label=$\cdot$,l.d=10,l.a=50}{vertj}
 \fmfiv{label=$\cdot$,l.d=10,l.a=65}{vertj}
 \fmfiv{label=$\cdot$,l.d=10,l.a=-35}{vertk}
 \fmfiv{label=$\cdot$,l.d=10,l.a=-50}{vertk}
 \fmfiv{label=$\cdot$,l.d=10,l.a=-65}{vertk}
\end{fmfchar*}%
}}%
\\ &\phaneq
+\sum_{i,j}c_{\text{bubble}}^{(i,j)}
\quad
\settoheight{\eqoff}{$\times$}%
\setlength{\eqoff}{0.5\eqoff}%
\addtolength{\eqoff}{-12\unitlength}%
\raisebox{\eqoff}{%
\fmfframe(2,2)(6,2){%
\begin{fmfchar*}(20,20)
\fmfsurround{vpiii,vpi1,vpii,vpiiii,vpiiiii,vpiiiiii,vpi,vq,vpj1,vpjj,vpjjjj,vpjjjjj,vpjjjjjj,vpj}
\fmf{dbl_plain_arrow,tension=0}{vq,vi}
\fmf{plain_arrow,tension=1}{vi,vpi}
\fmf{plain_arrow,tension=1}{vi,vpj1}
\fmf{plain_arrow,tension=1}{vj,vpj}
\fmf{plain_arrow,tension=1}{vj,vpi1}
\fmf{plain_arrow,tension=0.5,right=0.66}{vi,vj}
\fmf{plain_arrow,tension=0.5,left=0.66}{vi,vj}
\fmffreeze
\fmfdraw
 \fmfcmd{pair vertq, verti, vertj,  vertpi, vertpj, vertpii, vertpjj; vertq = vloc(__vq); verti = vloc(__vi); vertj = vloc(__vj);  vertpi = vloc(__vpi); vertpj = vloc(__vpj);  vertpii = vloc(__vpi1); vertpjj = vloc(__vpj1);  }
 \fmfiv{label=$\scriptstyle q$}{vertq}
 \fmfiv{label=$\scriptstyle p_i$}{vertpi}
 \fmfiv{label=$\scriptstyle p_{i+1}$}{vertpii}
 \fmfiv{label=$\scriptstyle p_j$}{vertpj}
 \fmfiv{label=$\scriptstyle p_{j+1}$}{vertpjj}
 \fmfiv{label=$\cdot$,l.d=0,l.a=0}{verti+(12*sind(243),12*cosd(243))}
 \fmfiv{label=$\cdot$,l.d=0,l.a=0}{verti+(12*sind(-67),12*cosd(-67))}
 \fmfiv{label=$\cdot$,l.d=10,l.a=15}{vertj}
 \fmfiv{label=$\cdot$,l.d=10,l.a=0}{vertj}
 \fmfiv{label=$\cdot$,l.d=10,l.a=-15}{vertj}
\end{fmfchar*}%
}}%
+\text{rational terms}
\end{aligned}
\end{equation*}
\caption{The one-loop correction to the colour-ordered form factor $\hat{\cF}_\cO$ can be written as a linear combination of box integrals, triangle integrals, bubble integrals and rational terms. The coefficients are labelled by the different combinations of momenta flowing out of the corners of the integrals. While the on-shell momenta $p_i$ are colour ordered, the off-shell momentum $q$ of the operator does not take part in the colour ordering as the operator is a colour singlet.}
\label{fig: ansatz for generalised unitarity}
\end{figure}
The first three coefficients of this ansatz can be fixed by applying cuts to both sides of the equation in figure \ref{fig: ansatz for generalised unitarity} and integrating over all remaining free components of the loop momentum.
Here, a cut denotes replacing one or more propagators according to%
\footnote{Some authors define the cut procedure as replacing the propagator $\frac{1}{p^2}$ by $i$ and / or $2\pi$ times $\delta(p^2)\delta^+(p)$. Note that our result does not depend on the choice of this prefactor, as we are applying the cut to both sides of figure \ref{fig: ansatz for generalised unitarity}.}
\begin{equation}
\frac{1}{p^2}\to\delta_+(p^2)=\delta(p^2)\Theta(p_0)\eqncom 
\end{equation}
where the delta function sets the propagating particle on-shell and the Heaviside step function ensures positive energy.\footnote{For a detailed discussion of the differences between cuts and discontinuities across corresponding branch cuts in generalised unitarity, see \cite{Abreu:2014cla}.}  
The integral over the remaining free components of the loop momentum is also called phase space integral, and it is performed in strictly four dimensions.
On the cut, the one-loop form factor factorises into the product of the tree-level form factor and one or several tree-level amplitudes.
The general procedure is to first apply the quadruple cut, which isolates the box integral and allows to determine the box coefficient. 
Second, the triple cut is taken, which has contributions from the triangle integral and the box integral. Knowing the box coefficient, the triangle coefficient can be extracted from the triple cut. From the subsequent double cut, which has contributions from the box integral, the triangle integral and the bubble integral, the bubble coefficient can be extracted via the known box and triangle coefficients. 
This procedure determines the complete one-loop form factor except for the rational terms, which vanish in all cuts and are hence harder to obtain.\footnote{See \cite{Bern:2007dw} for a review of different strategies for obtaining rational terms in the context of amplitudes.}

The above procedure is similar to many approaches in the literature on amplitudes, but there are also important differences which are designed to make it applicable for form factors of generic operators. With the method of Ossola, Papadopoulos and Pittau (OPP) \cite{Ossola:2006us}, the above procedure shares the strategy of first fixing coefficients associated with a higher number of propagators in the loop and then using these to fix coefficients associated with a lower number of propagators; but in contrast to OPP, the above procedure works on the level of the integrals and not the integrands. The integral over the remaining unconstrained components of the loop momentum is also performed in several approaches, including some for the direct extraction of integral coefficients \cite{Forde:2007mi,ArkaniHamed:2008gz}. However, the parametrisation we use to perform this integral differs from the method of integration used in those approaches.

In this paper, we are considering the one-loop minimal form factor, i.e.\ the $L$-point form factor for an operator of length $L$.
This simplifies the general procedure described above considerably.
By looking at figure \ref{fig: ansatz for generalised unitarity}, it is easy to see that the box coefficient vanishes in this case; the box integral transforms two fields of the operator, which enter at its left corner, to at least three fields, which exit its top, right and bottom corner. It can hence only contribute to form factors with $n\geq L+1$ points.
Moreover, only one on-shell field can leave each of the two right corners of the triangle integral and only two on-shell fields can leave the right hand side of the bubble integral.
The resulting simplified ansatz is depicted in figure \ref{fig: simplified ansatz for generalised unitarity}.
\begin{figure}[t]
 \centering
\begin{equation*}
 \begin{aligned}
\settoheight{\eqoff}{$\times$}%
\setlength{\eqoff}{0.5\eqoff}%
\addtolength{\eqoff}{-14.5\unitlength}%
\raisebox{\eqoff}{%
\fmfframe(2,2)(6,2){%
\begin{fmfchar*}(25,25)
\fmfsurround{vp3,vp2,vp1,vq,vpL,vp}
\fmf{dbl_plain_arrow,tension=1}{vq,vqa}
\fmf{plain_arrow,tension=1}{vpLa,vpL}
\fmf{plain_arrow,tension=1}{vp3a,vp3}
\fmf{phantom,tension=1}{vpa,vp}
\fmf{plain_arrow}{vp1a,vp1}
\fmf{plain_arrow}{vp2a,vp2}
\fmf{dbl_plain_arrow,tension=2}{vqa,v1}
\fmf{plain_arrow,tension=2}{v1,vpLa}
\fmf{plain_arrow,tension=2}{v1,vp3a}
\fmf{phantom,tension=2}{v1,vpa}
\fmf{plain_arrow,tension=2}{v1,vp1a}
\fmf{plain_arrow,tension=2}{v1,vp2a}
\fmffreeze
\fmfdraw
 \fmfcmd{pair vertq, vertpone, vertptwo, vertpthree, vertpL, vertone, verttwo, vertp; vertone = vloc(__v1); verttwo = vloc(__v2); vertq = vloc(__vq); vertpone = vloc(__vp1); vertptwo = vloc(__vp2); vertpthree = vloc(__vp3);vertp = vloc(__vp);vertpL = vloc(__vpL);}
 \fmfiv{decor.shape=circle,decor.filled=30,decor.size=30}{vertone}
 \fmfiv{decor.shape=circle,decor.filled=0,decor.size=20}{vertone}
 \fmfiv{label=$ \hat{\cF}_{\cO}$,l.d=0,l.a=0}{vertone}
 \fmfiv{label=$\scriptstyle q$}{vertq}
 \fmfiv{label=$\scriptstyle p_1$}{vertpone}
 \fmfiv{label=$\scriptstyle p_2$}{vertptwo}
 \fmfiv{label=$\scriptstyle p_3$}{vertpthree}
 \fmfiv{label=$\scriptstyle p_L$}{vertpL}
 \fmfiv{label=$\cdot$,l.d=20,l.a=-45}{vertone}
 \fmfiv{label=$\cdot$,l.d=20,l.a=-60}{vertone}
 \fmfiv{label=$\cdot$,l.d=20,l.a=-75}{vertone}
\end{fmfchar*}%
}}%
&=
\sum_{i}c_{\text{triangle}}^{i,i+1}
\quad \,\,
\settoheight{\eqoff}{$\times$}%
\setlength{\eqoff}{0.5\eqoff}%
\addtolength{\eqoff}{-12\unitlength}%
\raisebox{\eqoff}{%
\fmfframe(2,2)(6,2){%
\begin{fmfchar*}(20,20)
\fmfsurround{vpjj,vpiii,vpj,vpi1,vpii,vpiiii,vpi,vq,vpk1,vpkk,vpkkkk,vpjjj,vpk,vpj1}
\fmf{dbl_plain_arrow,tension=0}{vq,vi}
\fmf{plain_arrow,tension=1}{vi,vpi}
\fmf{plain_arrow,tension=1}{vi,vpk1}
 \fmf{plain_arrow,tension=2}{vj,vpj}
 \fmf{plain_arrow,tension=2}{vk,vpk}
\fmf{plain_arrow,tension=0.66}{vi,vj}
\fmf{plain_arrow,tension=0.66}{vi,vk}
\fmf{plain_arrow,tension=0.66}{vj,vk}
\fmffreeze
\fmfdraw
 \fmfcmd{pair vertq, verti, vertj, vertk, vertpi, vertpj, vertpk, vertpii, vertpjj, vertpkk; vertq = vloc(__vq); verti = vloc(__vi); vertj = vloc(__vj); vertk = vloc(__vk); vertpi = vloc(__vpi); vertpj = vloc(__vpj); vertpk = vloc(__vpk); vertpii = vloc(__vpi1); vertpjj = vloc(__vpj1); vertpkk = vloc(__vpk1); }
 \fmfiv{label=$\scriptstyle q$}{vertq}
 \fmfiv{label=$\scriptstyle p_{i-1}$}{vertpi}
 \fmfiv{label=$\scriptstyle p_i$}{vertpj}
 \fmfiv{label=$\scriptstyle p_{i+1}$}{vertpk}
 \fmfiv{label=$\scriptstyle p_{i+2}$}{vertpkk}
 \fmfiv{label=$\cdot$,l.d=0,l.a=0}{verti+(12*sind(243),12*cosd(243))}
 \fmfiv{label=$\cdot$,l.d=0,l.a=0}{verti+(12*sind(-67),12*cosd(-67))}
\end{fmfchar*}%
}}%
\!\!\!\!+\sum_{i}c_{\text{bubble}}^{i,i+1}
\quad \,\,
\settoheight{\eqoff}{$\times$}%
\setlength{\eqoff}{0.5\eqoff}%
\addtolength{\eqoff}{-12\unitlength}%
\raisebox{\eqoff}{%
\fmfframe(2,2)(6,2){%
\begin{fmfchar*}(20,20)
\fmfsurround{vpiii,vpi1,vpii,vpiiii,vpiiiii,vpiiiiii,vpi,vq,vpj1,vpjj,vpjjjj,vpjjjjj,vpjjjjjj,vpj}
\fmf{dbl_plain_arrow,tension=0}{vq,vi}
\fmf{plain_arrow,tension=1}{vi,vpi}
\fmf{plain_arrow,tension=1}{vi,vpj1}
\fmf{plain_arrow,tension=1}{vj,vpj}
\fmf{plain_arrow,tension=1}{vj,vpi1}
\fmf{plain_arrow,tension=0.5,right=0.66}{vi,vj}
\fmf{plain_arrow,tension=0.5,left=0.66}{vi,vj}
\fmffreeze
\fmfdraw
 \fmfcmd{pair vertq, verti, vertj,  vertpi, vertpj, vertpii, vertpjj; vertq = vloc(__vq); verti = vloc(__vi); vertj = vloc(__vj);  vertpi = vloc(__vpi); vertpj = vloc(__vpj);  vertpii = vloc(__vpi1); vertpjj = vloc(__vpj1);  }
 \fmfiv{label=$\scriptstyle q$}{vertq}
 \fmfiv{label=$\scriptstyle p_{i-1}$}{vertpi}
 \fmfiv{label=$\scriptstyle p_{i}$}{vertpii}
 \fmfiv{label=$\scriptstyle p_{i+1}$}{vertpj}
 \fmfiv{label=$\scriptstyle p_{i+2}$}{vertpjj}
 \fmfiv{label=$\cdot$,l.d=0,l.a=0}{verti+(12*sind(243),12*cosd(243))}
 \fmfiv{label=$\cdot$,l.d=0,l.a=0}{verti+(12*sind(-67),12*cosd(-67))}
\end{fmfchar*}%
}}%
\\ &\phaneq
+\text{rational terms}
\end{aligned}
\end{equation*}
\caption{The one-loop correction to the form factor $\hat{\cF}_\cO$ for a minimal number of external fields can be written as a linear combination of one-mass triangle integrals, bubble integrals and rational terms. Here, the coefficients are labelled by the two momenta that leave the integrals on the right.}
\label{fig: simplified ansatz for generalised unitarity}
\end{figure}
As a consequence, the triple cut uniquely determines the triangle coefficient. This is depicted in figure \ref{fig: triple cut of simplified ansatz for generalised unitarity}. 
\begin{figure}[t]
 \centering
\begin{equation*}
 \begin{aligned}
\settoheight{\eqoff}{$\times$}%
\setlength{\eqoff}{0.5\eqoff}%
\addtolength{\eqoff}{-14.5\unitlength}%
\raisebox{\eqoff}{%
\fmfframe(2,2)(6,2){%
\begin{fmfchar*}(25,25)
\fmfsurround{vp3,vp2,vp1,vq,vpL,vp}
\fmf{dbl_plain_arrow,tension=1}{vq,vqa}
\fmf{plain_arrow,tension=1}{vpLa,vpL}
\fmf{plain_arrow,tension=1}{vp3a,vp3}
\fmf{phantom,tension=1}{vpa,vp}
\fmf{plain_arrow}{vp1a,vp1}
\fmf{plain_arrow}{vp2a,vp2}
\fmf{dbl_plain_arrow,tension=2}{vqa,v1}
\fmf{plain_arrow,tension=2}{v1,vpLa}
\fmf{plain_arrow,tension=2}{v1,vp3a}
\fmf{phantom,tension=2}{v1,vpa}
\fmf{plain_arrow,tension=2}{v1,vp1a}
\fmf{plain_arrow,tension=2}{v1,vp2a}
\fmffreeze
\fmfdraw
 \fmfcmd{pair vertq, vertpone, vertptwo, vertpthree, vertpL, vertone, verttwo, vertp; vertone = vloc(__v1); verttwo = vloc(__v2); vertq = vloc(__vq); vertpone = vloc(__vp1); vertptwo = vloc(__vp2); vertpthree = vloc(__vp3);vertp = vloc(__vp);vertpL = vloc(__vpL);}
 \fmfiv{decor.shape=circle,decor.filled=30,decor.size=30}{vertone}
 \fmfiv{decor.shape=circle,decor.filled=0,decor.size=20}{vertone}
 \fmfiv{label=$ \hat{\cF}_{\cO}$,l.d=0,l.a=0}{vertone}
 \fmfdraw
 \fmfi{dashes}{(vertone+(-18*sqrt(0.75),18*sqrt(0.25)))--(vertone+(-7*sqrt(0.75),7*sqrt(0.25)))}
 \fmfi{dashes}{(vertone+(0,18))--(vertone+(0,7))}
 \fmfi{dashes}{(vertone+(+18*sqrt(0.75),18*sqrt(0.25)))--(vertone+(+7*sqrt(0.75),7*sqrt(0.25)))}
 \fmfiv{label=$\scriptstyle q$}{vertq}
 \fmfiv{label=$\scriptstyle p_1$}{vertpone}
 \fmfiv{label=$\scriptstyle p_2$}{vertptwo}
 \fmfiv{label=$\scriptstyle p_3$}{vertpthree}
 \fmfiv{label=$\scriptstyle p_L$}{vertpL}
 \fmfiv{label=$\cdot$,l.d=20,l.a=-45}{vertone}
 \fmfiv{label=$\cdot$,l.d=20,l.a=-60}{vertone}
 \fmfiv{label=$\cdot$,l.d=20,l.a=-75}{vertone}
\end{fmfchar*}%
}}%
&=
c_{\text{triangle}}^{1,2}
\quad
\settoheight{\eqoff}{$\times$}%
\setlength{\eqoff}{0.5\eqoff}%
\addtolength{\eqoff}{-12\unitlength}%
\raisebox{\eqoff}{%
\fmfframe(2,2)(6,2){%
\begin{fmfchar*}(20,20)
\fmfsurround{vpjj,vpiii,vpj,vpi1,vpii,vpiiii,vpi,vq,vpk1,vpkk,vpkkkk,vpjjj,vpk,vpj1}
\fmf{dbl_plain_arrow,tension=0}{vq,vi}
\fmf{plain_arrow,tension=1}{vi,vpi}
\fmf{plain_arrow,tension=1}{vi,vpk1}
 \fmf{plain_arrow,tension=2}{vj,vpj}
 \fmf{plain_arrow,tension=2}{vk,vpk}
\fmf{plain_arrow,tension=0.66}{vi,vj}
\fmf{plain_arrow,tension=0.66}{vi,vk}
\fmf{plain_arrow,tension=0.66}{vj,vk}
\fmf{phantom_smallcut,tension=0}{vi,vj}
\fmf{phantom_smallcut,tension=0}{vi,vk}
\fmf{phantom_smallcut,tension=0}{vj,vk}
\fmffreeze
\fmfdraw
 \fmfcmd{pair vertq, verti, vertj, vertk, vertpi, vertpj, vertpk, vertpii, vertpjj, vertpkk; vertq = vloc(__vq); verti = vloc(__vi); vertj = vloc(__vj); vertk = vloc(__vk); vertpi = vloc(__vpi); vertpj = vloc(__vpj); vertpk = vloc(__vpk); vertpii = vloc(__vpi1); vertpjj = vloc(__vpj1); vertpkk = vloc(__vpk1); }
 \fmfiv{label=$\scriptstyle q$}{vertq}
 \fmfiv{label=$\scriptstyle p_{L}$}{vertpi}
 \fmfiv{label=$\scriptstyle p_1$}{vertpj}
 \fmfiv{label=$\scriptstyle p_{2}$}{vertpk}
 \fmfiv{label=$\scriptstyle p_{3}$}{vertpkk}
 \fmfiv{label=$\cdot$,l.d=0,l.a=0}{verti+(12*sind(243),12*cosd(243))}
 \fmfiv{label=$\cdot$,l.d=0,l.a=0}{verti+(12*sind(-67),12*cosd(-67))}
\end{fmfchar*}%
}}%
\end{aligned}
\end{equation*}
\caption{The triple cut between $p_1$, $p_2$ and the rest of the diagram isolates the triangle integral with external on-shell legs $p_1$ and $p_2$. To obtain its coefficient $c_{\text{triangle}}^{1,2}$, we integrate on both sides of the equation over all components of the loop momentum that are not fixed by the cut.}
\label{fig: triple cut of simplified ansatz for generalised unitarity}
\end{figure}
In fact, the triangle coefficient turns out to be universal, as shown in the next subsection. 
The cut triangle integral can then be subtracted from the double cut, which is depicted in figure \ref{fig: double cut of simplified ansatz for generalised unitarity}, to obtain the bubble coefficient. 
\begin{figure}[tbp]
 \centering
\begin{equation*}
 \begin{aligned}
\settoheight{\eqoff}{$\times$}%
\setlength{\eqoff}{0.5\eqoff}%
\addtolength{\eqoff}{-14.5\unitlength}%
\raisebox{\eqoff}{%
\fmfframe(2,2)(6,2){%
\begin{fmfchar*}(25,25)
\fmfsurround{vp3,vp2,vp1,vq,vpL,vp}
\fmf{dbl_plain_arrow,tension=1}{vq,vqa}
\fmf{plain_arrow,tension=1}{vpLa,vpL}
\fmf{plain_arrow,tension=1}{vp3a,vp3}
\fmf{phantom,tension=1}{vpa,vp}
\fmf{plain_arrow}{vp1a,vp1}
\fmf{plain_arrow}{vp2a,vp2}
\fmf{dbl_plain_arrow,tension=2}{vqa,v1}
\fmf{plain_arrow,tension=2}{v1,vpLa}
\fmf{plain_arrow,tension=2}{v1,vp3a}
\fmf{phantom,tension=2}{v1,vpa}
\fmf{plain_arrow,tension=2}{v1,vp1a}
\fmf{plain_arrow,tension=2}{v1,vp2a}
\fmffreeze
\fmfdraw
 \fmfcmd{pair vertq, vertpone, vertptwo, vertpthree, vertpL, vertone, verttwo, vertp; vertone = vloc(__v1); verttwo = vloc(__v2); vertq = vloc(__vq); vertpone = vloc(__vp1); vertptwo = vloc(__vp2); vertpthree = vloc(__vp3);vertp = vloc(__vp);vertpL = vloc(__vpL);}
 \fmfiv{decor.shape=circle,decor.filled=30,decor.size=30}{vertone}
 \fmfiv{decor.shape=circle,decor.filled=0,decor.size=20}{vertone}
 \fmfiv{label=$ \hat{\cF}_{\cO}$,l.d=0,l.a=0}{vertone}
 \fmfdraw
 \fmfi{dashes}{(vertone+(-18*sqrt(0.75),18*sqrt(0.25)))--(vertone+(-7*sqrt(0.75),7*sqrt(0.25)))}
 \fmfi{dashes}{(vertone+(+18*sqrt(0.75),18*sqrt(0.25)))--(vertone+(+7*sqrt(0.75),7*sqrt(0.25)))}
 \fmfiv{label=$\scriptstyle q$}{vertq}
 \fmfiv{label=$\scriptstyle p_1$}{vertpone}
 \fmfiv{label=$\scriptstyle p_2$}{vertptwo}
 \fmfiv{label=$\scriptstyle p_3$}{vertpthree}
 \fmfiv{label=$\scriptstyle p_L$}{vertpL}
 \fmfiv{label=$\cdot$,l.d=20,l.a=-45}{vertone}
 \fmfiv{label=$\cdot$,l.d=20,l.a=-60}{vertone}
 \fmfiv{label=$\cdot$,l.d=20,l.a=-75}{vertone}
\end{fmfchar*}%
}}%
&=
c_{\text{triangle}}^{1,2}
\quad
\settoheight{\eqoff}{$\times$}%
\setlength{\eqoff}{0.5\eqoff}%
\addtolength{\eqoff}{-12\unitlength}%
\raisebox{\eqoff}{%
\fmfframe(2,2)(6,2){%
\begin{fmfchar*}(20,20)
\fmfsurround{vpjj,vpiii,vpj,vpi1,vpii,vpiiii,vpi,vq,vpk1,vpkk,vpkkkk,vpjjj,vpk,vpj1}
\fmf{dbl_plain_arrow,tension=0}{vq,vi}
\fmf{plain_arrow,tension=1}{vi,vpi}
\fmf{plain_arrow,tension=1}{vi,vpk1}
 \fmf{plain_arrow,tension=2}{vj,vpj}
 \fmf{plain_arrow,tension=2}{vk,vpk}
\fmf{plain_arrow,tension=0.66}{vi,vj}
\fmf{plain_arrow,tension=0.66}{vi,vk}
\fmf{plain_arrow,tension=0.66}{vj,vk}
\fmf{phantom_smallcut,tension=0}{vi,vj}
\fmf{phantom_smallcut,tension=0}{vi,vk}
\fmffreeze
\fmfdraw
 \fmfcmd{pair vertq, verti, vertj, vertk, vertpi, vertpj, vertpk, vertpii, vertpjj, vertpkk; vertq = vloc(__vq); verti = vloc(__vi); vertj = vloc(__vj); vertk = vloc(__vk); vertpi = vloc(__vpi); vertpj = vloc(__vpj); vertpk = vloc(__vpk); vertpii = vloc(__vpi1); vertpjj = vloc(__vpj1); vertpkk = vloc(__vpk1); }
 \fmfiv{label=$\scriptstyle q$}{vertq}
 \fmfiv{label=$\scriptstyle p_{L}$}{vertpi}
 \fmfiv{label=$\scriptstyle p_1$}{vertpj}
 \fmfiv{label=$\scriptstyle p_{2}$}{vertpk}
 \fmfiv{label=$\scriptstyle p_{3}$}{vertpkk}
 \fmfiv{label=$\cdot$,l.d=0,l.a=0}{verti+(12*sind(243),12*cosd(243))}
 \fmfiv{label=$\cdot$,l.d=0,l.a=0}{verti+(12*sind(-67),12*cosd(-67))}
\end{fmfchar*}%
}}%
+
c_{\text{bubble}}^{1,2}
\quad
\settoheight{\eqoff}{$\times$}%
\setlength{\eqoff}{0.5\eqoff}%
\addtolength{\eqoff}{-12\unitlength}%
\raisebox{\eqoff}{%
\fmfframe(2,2)(6,2){%
\begin{fmfchar*}(20,20)
\fmfsurround{vpiii,vpi1,vpii,vpiiii,vpiiiii,vpiiiiii,vpi,vq,vpj1,vpjj,vpjjjj,vpjjjjj,vpjjjjjj,vpj}
\fmf{dbl_plain_arrow,tension=0}{vq,vi}
\fmf{plain_arrow,tension=1}{vi,vpi}
\fmf{plain_arrow,tension=1}{vi,vpj1}
\fmf{plain_arrow,tension=1}{vj,vpj}
\fmf{plain_arrow,tension=1}{vj,vpi1}
\fmf{plain_arrow,tension=0.5,right=0.66}{vi,vj}
\fmf{plain_arrow,tension=0.5,left=0.66}{vi,vj}
\fmf{phantom_smallcut,tension=0,right=0.66}{vi,vj}
\fmf{phantom_smallcut,tension=0,left=0.66}{vi,vj}
\fmffreeze
\fmfdraw
 \fmfcmd{pair vertq, verti, vertj,  vertpi, vertpj, vertpii, vertpjj; vertq = vloc(__vq); verti = vloc(__vi); vertj = vloc(__vj);  vertpi = vloc(__vpi); vertpj = vloc(__vpj);  vertpii = vloc(__vpi1); vertpjj = vloc(__vpj1);  }
 \fmfiv{label=$\scriptstyle q$}{vertq}
 \fmfiv{label=$\scriptstyle p_{L}$}{vertpi}
 \fmfiv{label=$\scriptstyle p_{1}$}{vertpii}
 \fmfiv{label=$\scriptstyle p_{2}$}{vertpj}
 \fmfiv{label=$\scriptstyle p_{3}$}{vertpjj}
 \fmfiv{label=$\cdot$,l.d=0,l.a=0}{verti+(12*sind(243),12*cosd(243))}
 \fmfiv{label=$\cdot$,l.d=0,l.a=0}{verti+(12*sind(-67),12*cosd(-67))}
\end{fmfchar*}%
}}%
\end{aligned}
\end{equation*}
\caption{The double cut between $p_1$, $p_2$ and the rest of the diagram isolates the triangle integral and bubble integral with external on-shell legs $p_1$ and $p_2$.
To obtain the bubble coefficient $c_{\text{bubble}}^{1,2}$, we bring the cut triangle integral and its known coefficient $c_{\text{triangle}}^{1,2}$ to the l.h.s.\ of the equation. 
We then integrate on both sides of the equation over all components of the loop momentum that are not fixed by the cut.}
\label{fig: double cut of simplified ansatz for generalised unitarity}
\end{figure}
We calculate it in subsection \ref{subsec: bubble coefficient}. 
Subsection \ref{subsec: summary} contains a summary of our results as well as a comment on rational terms.
Explicit expressions for the tree-level amplitudes occurring in the calculation are summarised in appendix \ref{app: amplitudes}.
For the convenience of the reader, we calculate the one-loop corrections to some examples of minimal form factors in appendix \ref{app: examples} via the more conventional approach of unitarity at the level of the integrand.

\FloatBarrier
\subsection{The triangle coefficient from the triple cut}

\begin{figure}[h]
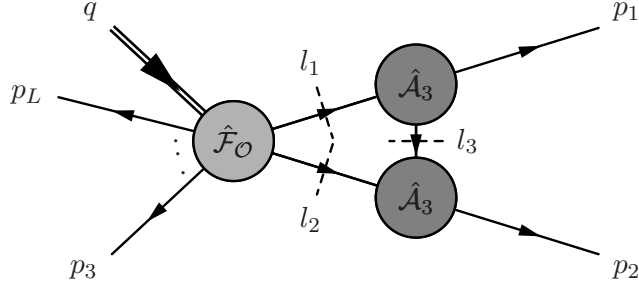

 \centering
$
\settoheight{\eqoff}{$\times$}%
\setlength{\eqoff}{0.5\eqoff}%
\addtolength{\eqoff}{-12.0\unitlength}%
\raisebox{\eqoff}{%
\fmfframe(2,2)(2,2){%
\begin{fmfchar*}(80,30)
\fmfleft{vp3,vp,vpL,vq}
\fmfright{vp2,vp1}
\fmf{dbl_plain_arrow,tension=1.5}{vq,v1}
\fmf{plain_arrow,tension=0}{v1,vpL}
\fmf{plain_arrow,tension=1.5}{v1,vp3}
\fmf{plain_arrow,left=0,l.s=left,label=$l_1\,,$,l.d=15}{v1,v2}
\fmf{plain_arrow,right=0,l.s=right,label=$l_2\,,$,l.d=15}{v1,v3}
\fmf{plain_arrow,right=0,tension=0,label=$l_3\,,$,l.d=-25}{v2,v3}
\fmf{phantom_cut,tension=0}{v1,v2}
\fmf{phantom_cut,tension=0}{v1,v3}
\fmf{phantom_cut,tension=0}{v2,v3}
\fmf{plain_arrow}{v2,vp1}
\fmf{plain_arrow}{v3,vp2}
\fmfv{decor.shape=circle,decor.filled=30,decor.size=30,label=$\hat{\cF}_{\cO}$,label.dist=0}{v1}
\fmfv{decor.shape=circle,decor.filled=50,decor.size=30,label=$\hat{\cA}_{3}$,label.dist=0}{v2}
\fmfv{decor.shape=circle,decor.filled=50,decor.size=30,label=$\hat{\cA}_{3}$,label.dist=0}{v3}
\fmffreeze
\fmfdraw
 \fmfcmd{pair vertq, vertpone, vertptwo, vertpthree, vertpL, vertone, verttwo; vertone = vloc(__v1); verttwo = vloc(__v2); vertq = vloc(__vq); vertpone = vloc(__vp1); vertptwo = vloc(__vp2); vertpthree = vloc(__vp3);vertpL = vloc(__vpL);}
 \fmfiv{label=$q$}{vertq}
 \fmfiv{label=$p_1$}{vertpone}
 \fmfiv{label=$p_2$}{vertptwo}
 \fmfiv{label=$p_3$}{vertpthree}
 \fmfiv{label=$p_L$}{vertpL}
 \fmfiv{label=$\cdot$,l.d=20,l.a=-150}{vertone}
 \fmfiv{label=$\cdot$,l.d=20,l.a=-165}{vertone}
 \fmfiv{label=$\cdot$,l.d=20,l.a=-180}{vertone}
\end{fmfchar*}%
}}%
$
\caption{The triple cut of the colour-ordered one-loop form factor in the channel of $p_1$ and $p_2$. On this cut, the r.h.s.\ of figure \ref{fig: triple cut of simplified ansatz for generalised unitarity} factorises into a product of two colour-ordered three-point tree-level amplitudes and the colour-ordered tree-level form factor.}
\label{fig: triple cut}
\end{figure}

As discussed above, the triangle coefficient can be extracted directly from the triple cut, which is depicted on the r.h.s.\ of 
figure \ref{fig: triple cut of simplified ansatz for generalised unitarity} and again in figure \ref{fig: triple cut}. To simplify the notation, we assume that this cut is taken between $p_1$, $p_2$ and the rest of the diagram.

The occurring phase space integral, i.e.\ the integral over the remaining free components of the loop momentum, reads
\begin{equation}\label{eq: triple cut integral}
 \begin{aligned}
\int \de \Lambda_{l_1}\de \Lambda_{l_2}\de \Lambda_{l_3} \hat{\cF}^{(0)}_\cO(\Lambda_{l_1},\Lambda_{l_2},\Lambda_{p_3},\dots,\Lambda_{p_L};q) \hat{\cA}^{(0)}(\Lambda_{l_1}^-,\Lambda_{p_1},\Lambda_{l_3})\hat{\cA}^{(0)}(\Lambda_{l_2}^-,\Lambda_{l_3}^-,\Lambda_{p_2})\eqncom  
 \end{aligned}
\end{equation}
where
\begin{equation}\label{eq: dLambda}
 \de\Lambda_{l_i}=\frac{\de^2\lambda_{l_i}\de^2\lambdat_{l_i}}{U(1)}\de^4\eta_{l_i} 
\end{equation}
with $i=1,2,3$. The superscript `$-$' on $\Lambda_{l_i}^-$ in \eqref{eq: triple cut integral} denotes a negative energy representation, i.e.\ that the respective momentum flows into the colour-ordered amplitude $\hat{\cA}_3$ and not out of it. In our conventions,
\begin{equation}
 \lambda_{-l_i}=-\lambda_{l_i}\eqncom \quad \lambdat_{-l_i}=\lambdat_{l_i}\eqndot
\end{equation}
The $U(1)$ in \eqref{eq: dLambda} accounts for the gauge freedom in the definition of the spinors $\lambda^\alpha$ and $\lambdat^\alphadot$. Multiplying $\lambda^\alpha$ by a phase factor $\e^{i\phi}\in U(1)$ and $\lambdat^\alphadot$ by its conjugate results in the same momentum $p^{\alpha\alphadot}=\lambda^\alpha\lambdat^\alphadot$.\footnote{The transformation behaviour under this multiplication is called little group scaling.} 
The final integral \eqref{eq: triple cut integral} has to be independent of the phase of the spinor-valued integration variables. Hence, we can integrate over the phase as well and use 
\begin{equation}\label{eq: dLambda without U(1)}
 \de\Lambda_{l_i}= \de^2\lambda_{l_i}\de^2\lambdat_{l_i} \de^4\eta_{l_i} \eqndot
\end{equation}
This leads to a total factor of $(2\pi)^3$ on both sides of figure \ref{fig: triple cut of simplified ansatz for generalised unitarity}, which hence drops out.

Choosing the loop momentum $l^\mu$ as in \eqref{eq: triangle integral}, the conditions imposed by the cut are 
\begin{equation}
\begin{aligned}
  l_1^2&=l^2=0 \eqncom\\
  l_2^2&=(p_1+p_2+l)^2=l^2+(p_1+p_2)^2+2(p_1+p_2)\cdot l=0\eqncom\\
  l_3^2&=(p_1+l)^2=l^2+2l\cdot p_1=0 \eqndot
\end{aligned}
\end{equation}
These are three conditions on the four components of $l^\mu$. 
Naively, one could hence expect a real one-parameter solution. 
In contrast to this expectation, the real solution for $l_1$ and $l_2$ is in fact unique: 
\begin{equation}\label{eq: condition}
 l_1=p_1\eqncom \quad l_2=p_2 \eqncom
\end{equation}
which is a consequence of $p_1^2=p_2^2=0$;  cf.\ for example \cite{ArkaniHamed:2008gz}.
As a massless momentum fixes the corresponding spinors except for an arbitrary phase, we can also write the condition \eqref{eq: condition} as\footnote{As before, we abbreviate $\lambda_{p_i}$ as $\lambda_{i}$ etc.}
\begin{equation}\label{eq: condition in spinors}
 \begin{aligned}
  \lambda_{l_1}^\alpha&=\e^{i\phi_1} \lambda_{1}^\alpha\eqncom & \lambda_{l_2}^\alpha&=\e^{i\phi_2} \lambda_{2}^\alpha\eqncom \\
  \lambdat_{l_1}^{\alphadot}&=\e^{-i\phi_1} \lambdat_{1}^{\alphadot}\eqncom & \lambdat_{l_2}^{\alphadot}&=\e^{-i\phi_2} \lambdat_{2}^{\alphadot}\eqndot
 \end{aligned}
\end{equation}

So far, we have neglected $l_3$. The combination of momentum conservation and the on-shell condition for $p_1$ and $p_2$ requires
\begin{equation}
\begin{aligned}
  p_1^2&=(l_1-l_3)^2=0 \eqncom\\
  p_2^2&=(l_2+l_3)^2=0 \eqndot
\end{aligned}
\end{equation}
In spinor helicity variables, these conditions read
\begin{equation}
\begin{aligned}
  \ab{l_1 l_3}\sb{l_1 l_3}&=0 \eqncom\\
  \ab{l_2 l_3}\sb{l_2 l_3}&=0 \eqndot
\end{aligned}
\end{equation}
At this point, we have to relax the constraint that $\lambda$ and $\lambdat$ are complex conjugates of each other, i.e.\ allow complex momenta, as is usual when considering massless three-particle kinematics.
The first equation then implies $\lambda_{l_3} \propto \lambda_{l_1}$ or $\lambdat_{l_3} \propto \lambdat_{l_1}$, while the second equation implies
$\lambda_{l_3} \propto \lambda_{l_2}$ or $\lambdat_{l_3} \propto \lambdat_{l_2}$. 
For generic $p_1$ and $p_2$, not all $\lambda_{l_i}$ or $\lambdat_{l_i}$ can be proportional to each other. 
This leaves us with either $\lambdat_{l_3} \propto \lambdat_{l_1}$ and $\lambda_{l_3} \propto \lambda_{l_2}$ or $\lambda_{l_3} \propto \lambda_{l_1}$ and $\lambdat_{l_3} \propto \lambdat_{l_2}$. 
The contributions of \eqref{eq: triple cut integral} for both complex solutions have to be summed with a prefactor of $\frac{1}{2}$, cf.\ \cite{Britto:2004nc,Kosower:2011ty}.

The two colour-ordered tree-level three-point amplitudes in the triple cut \eqref{eq: triple cut integral} shown in figure \ref{fig: triple cut} can be either of maximally-helicity-violating ($\MHV$) type or of the conjugate $\MHVb$ type. The respective expressions for the superamplitudes are shown in appendix \ref{app: amplitudes}.
On the first solution, only the combination of an upper $\MHV$ amplitude with a lower $\MHVb$ amplitude is nonvanishing. On the second solution, only the opposite combination gives a nonzero result.
In fact, the contributions from both solutions are equal, which cancels the prefactor of $\frac{1}{2}$ mentioned before. We hence focus on the first one.
It yields
\begin{equation}
 \begin{aligned}
 &\hat{\cA}^{\MHV\,(0)}(\Lambda_1,\Lambda_{l_3},\Lambda_{l_1}^-) 
 \hat{\cA}^{\MHVb\,(0)}(\Lambda_{l_3}^-,\Lambda_{2},\Lambda_{l_2}^-) \\
 &=\frac{\delta^4(p_1+l_3-l_1)}{\ab{1 l_3}\ab{l_3 l_1}\ab{l_1 1}}\prod_{A=1}^4(\ab{1l_3}\eta_1^A\eta_{l_3}^A-\ab{l_1l_3}\eta_{l_1}^A\eta_{l_3}^A-\ab{1l_1}\eta_1^A\eta_{l_1}^A)\\
 &\phaneq
 \frac{\delta^4(p_2-l_2-l_3)}{\sb{ 2l_2}\sb{l_2 l_3}\sb{l_3 2}}\prod_{A=1}^4(\sb{l_2l_3}\eta_2^A+\sb{2l_2}\eta_{l_3}^A+\sb{l_3 2}\eta_{l_2}^A)
 \end{aligned}
\end{equation}
\begin{equation}
 \begin{aligned}
&=\delta^4(p_1+l_3-l_1)\delta^4(p_2-l_2-l_3)\frac{\ab{12}\e^{2i(\phi_1+\phi_2)}}{\sb{12}^3\ab{1l_1}^4}\\
&\phaneq\prod_{A=1}^4 \Big(\ab{1l_3}(\eta_1^A-\e^{i\phi_1}\eta_{l_1}^A)\eta_{l_3}^A-\ab{1l_1}\eta_1^A\eta_{l_1}^A\Big)
\prod_{A=1}^4 \Big(\sb{2l_3}(\e^{-i\phi_2}\eta_2^A-\eta_{l_2}^A)+\sb{2l_2}\eta_{l_3}^A\Big) \eqncom
\label{eq: product of three-point amplitudes}
 \end{aligned}
\end{equation}
where we have used \eqref{eq: condition in spinors} and the relations
\begin{equation}\label{eq: identities}
 \begin{aligned}
 \ab{1l_3}\sb{l_3 2}&= \asb{1}{l_3}{2}= \asb{1}{l_1}{2}=\ab{1 l_1}\sb{l_12}=\ab{1l_1}\sb{12}\e^{-i\phi_1} \eqncom\\
 \ab{l_3l_1}\sb{l_2l_3}&=\sab{l_2}{l_3}{l_1}=-\sab{l_2}{p_1}{l_1}=-\ab{1l_1}\sb{l_21}=\ab{1l_1}\sb{12}\e^{-i\phi_2}\eqncom\\
 \ab{12}\sb{2l_2}&=\asb{1}{p_2}{l_2}=\asb{1}{l_1}{l_2}=\ab{1l_1}\sb{l_1l_2}=\ab{1l_1}\sb{12}\e^{-i(\phi_1+\phi_2)}\eqncom
\end{aligned}
\end{equation}
which are a consequence of momentum conservation and \eqref{eq: condition in spinors}. 
Performing the fermionic integral of \eqref{eq: product of three-point amplitudes} over $\eta_{l_3}$ yields
\begin{multline}
\delta^4(p_1+l_3-l_1)\delta^4(p_2-l_2-l_3)\frac{\ab{12}\e^{2i(\phi_1+\phi_2)}}{\sb{12}^3\ab{1l_1}^4}\\
\prod_{A=1}^4 \Big(\ab{1l_3}(\eta_1^A-\e^{i\phi_1}\eta_{l_1}^A)\sb{2l_3}(\e^{-i\phi_2}\eta_2^A-\eta_{l_2}^A)+\ab{1l_1}\eta_1^A\eta_{l_1}^A\sb{2l_2}\Big)
\eqndot
 \end{multline}
Applying \eqref{eq: identities} another time, we can eliminate all spinors in the denominator and find that the second term in the parenthesis vanishes.\footnote{Note that we can only use the condition $\sb{2l_2}=0$ now that we have eliminated all potentially singular terms in the denominator.} We are left with
\begin{equation}
 \begin{aligned}
-(p_1+p_2)^2 \delta^4(p_1+l_3-l_1)\delta^4(p_2-l_2-l_3) \e^{2i(\phi_1+\phi_2)} \prod_{A=1}^4\!\Big((\e^{-i\phi_1}\eta_1^A-\eta_{l_1}^A)(\e^{-i\phi_2}\eta_2^A-\eta_{l_2}^A)\Big)\eqndot
\end{aligned}
\end{equation}
It is easy to see that the fermionic integral $\int \de^4\eta_{l_1}\de^4\eta_{l_2}$ of the product of this expression with the tree-level form factor simply replaces
\begin{equation}
 \eta_{l_1}^A\to\e^{-i\phi_1}\eta_1^A\eqncom \quad \eta_{l_2}^A\to \e^{-i\phi_2} \eta_2^A 
\end{equation}
in the tree-level form factor.
Via \eqref{eq: condition in spinors}, the phase space integral leads to similar replacements of the bosonic variables in the tree-level form factor. 
Assembling all phase factors, we find $\e^{2 i\phi_1 C_{l_1}}\e^{2 i\phi_2 C_{l_2}}$, where $C_{l_1}$ and $C_{l_2}$ are the central charges of the respective first two legs of the tree-level form factor $\hat{\cF}_{\cO}^{(0)}$. 
The integrals over $\phi_1$, $\phi_2$ and the analogous phase in $\Lambda_{l_3}$ yield $(2\pi)\delta_{C_{l_1}}$, $(2\pi)\delta_{C_{l_2}}$ and $(2\pi)$, respectively. 
As all fields in $\hat{\cF}_{\cO}^{(0)}$ obey the central charge constraint, the total factor is $(2\pi)^3$.

Hence, the complete phase space integral \eqref{eq: triple cut integral} of the triple-cut in figure \ref{fig: triple cut} reads
\begin{equation}
 -(p_1+p_2)^2 (2\pi)^3\hat{\cF}^{(0)}_{\cO}(\Lambda_1,\Lambda_2,\Lambda_3,\dots,\Lambda_L;q)\eqndot
\end{equation}
To obtain the triangle coefficient, we have to compare this to the phase space integral of the triple cut of the triangle integral. The latter simply gives $(2\pi)^3$ from the phase integrations, showing that the triangle coefficient is universally given by\footnote{In particular, this result is consistent with the so-called rung rule \cite{Bern:1997nh,Bern:1998ug}.}
\begin{equation}
 c^{1,2}_{\text{triangle}}=-(p_1+p_2)^2 \hat{\cF}^{(0)}_{\cO}(\Lambda_1,\Lambda_2,\Lambda_3,\dots,\Lambda_L;q)\eqndot
\end{equation}
Note that in the calculation we have neglected a factor of the modified effective planar coupling constant 
\begin{equation}\label{eq: modified effective planar coupling constant}
 \tilde{g}^2=\left(4\pi\e^{-\gamma_{\text{E}}}\right)^\varepsilon g^2=\left(4\pi\e^{-\gamma_{\text{E}}}\right)^\varepsilon \frac{g_\YM^2 N}{(4\pi)^2}\eqncom
\end{equation}
where $\gamma_{\text{E}}$ is the Euler-Mascheroni constant, on both sides of figure \ref{fig: triple cut of simplified ansatz for generalised unitarity}. 
It arises on the l.h.s.\ as loop expansion parameter and on the r.h.s.\ as combination of factors of $g_\YM$ from each three-point tree-level amplitude, $N$ from the colour contractions and $\left(4\pi\e^{-\gamma_{\text{E}}}\right)^\varepsilon (4\pi)^{-2}$ from the loop integral.

\subsection{The bubble coefficient from the double cut}
\label{subsec: bubble coefficient}

\begin{figure}[h]
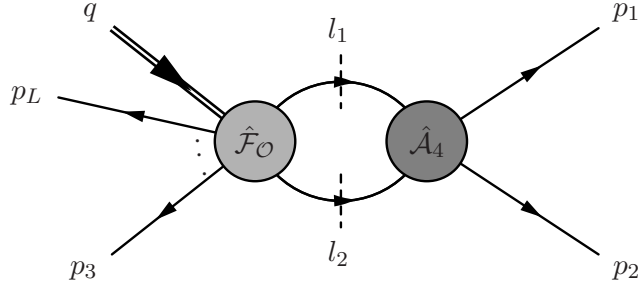

 \centering
$
\settoheight{\eqoff}{$\times$}%
\setlength{\eqoff}{0.5\eqoff}%
\addtolength{\eqoff}{-12.0\unitlength}%
\raisebox{\eqoff}{%
\fmfframe(2,2)(2,2){%
\begin{fmfchar*}(80,30)
\fmfleft{vp3,vp,vpL,vq}
\fmfright{vp2,vp1}
\fmf{dbl_plain_arrow,tension=1.2}{vq,v1}
\fmf{plain_arrow,tension=0}{v1,vpL}
\fmf{plain_arrow,tension=1.2}{v1,vp3}
\fmf{plain_arrow,left=0.7,label=$l_1\,,$,l.d=15}{v1,v2}
\fmf{plain_arrow,right=0.7,label=$l_2\,,$,l.d=15}{v1,v2}
\fmf{phantom_cut,left=0.7,tension=0}{v1,v2}
\fmf{phantom_cut,right=0.7,tension=0}{v1,v2}
\fmf{plain_arrow}{v2,vp1}
\fmf{plain_arrow}{v2,vp2}
\fmfv{decor.shape=circle,decor.filled=30,decor.size=30,label=$\hat{\cF}_{\cO}$,label.dist=0}{v1}
\fmfv{decor.shape=circle,decor.filled=50,decor.size=30,label=$\hat{\cA}_{4}$,label.dist=0}{v2}
\fmffreeze
 \fmfcmd{pair vertq, vertpone, vertptwo, vertpthree, vertpL, vertone, verttwo; vertone = vloc(__v1); verttwo = vloc(__v2); vertq = vloc(__vq); vertpone = vloc(__vp1); vertptwo = vloc(__vp2); vertpthree = vloc(__vp3);vertpL = vloc(__vpL);}
 \fmfiv{label=$q$}{vertq}
 \fmfiv{label=$p_1$}{vertpone}
 \fmfiv{label=$p_2$}{vertptwo}
 \fmfiv{label=$p_3$}{vertpthree}
 \fmfiv{label=$p_L$}{vertpL}
 \fmfiv{label=$\cdot$,l.d=20,l.a=-150}{vertone}
 \fmfiv{label=$\cdot$,l.d=20,l.a=-165}{vertone}
 \fmfiv{label=$\cdot$,l.d=20,l.a=-180}{vertone}
\end{fmfchar*}%
}}%
$
\caption{The double cut of the colour-ordered one-loop form factor in the $(p_1+p_2)^2$ channel. On this cut, the r.h.s.\ of figure \ref{fig: double cut of simplified ansatz for generalised unitarity} factorises into the product of the colour-ordered four-point tree-level amplitude and the colour-ordered tree-level form factor.}
\label{fig: double cut}
\end{figure}

The double cut of the one-loop minimal form factor is shown on the r.h.s.\ of figure \ref{fig: double cut of simplified ansatz for generalised unitarity} and again in figure \ref{fig: double cut}.
Both the triangle and the bubble integral contribute to it, as shown in figure \ref{fig: double cut of simplified ansatz for generalised unitarity}. From the last subsection, we know that the contribution from the triangle integral is given by 
\begin{equation}
 c^{1,2}_{\text{triangle}} \FDinline[triangle,cut,topleglabel=${\scriptscriptstyle 1}$,bottomleglabel=${\scriptscriptstyle 2}$]= -\frac{(p_1+p_2)^2}{(l_1-p_1)^2}\hat{\cF}^{(0)}_\cO(\Lambda_1,\Lambda_2,\Lambda_3,\dots,\Lambda_L;q)\eqncom
\end{equation}
where the graph on the l.h.s.\ denotes the integrand of the depicted integral \eqref{eq: triangle integral} excluding the cut propagators and the factor $\frac{\e^{\varepsilon\gamma_{\text{E}}}}{i\pi^{D/2}}$.
We can subtract this contribution from the double cut to obtain the bubble coefficient from  
\begin{equation}\label{eq: bubble coefficient}
 \begin{aligned}
  \int \de \Lambda_{l_1}\de \Lambda_{l_2}  \Big(&\hat{\cF}_\cO^{(0)}(\Lambda_{l_1},\Lambda_{l_2},\Lambda_{p_3},\dots,\Lambda_{p_L};q) \hat{\cA}_4^{(0)}(\Lambda_{l_2}^-,\Lambda_{l_1}^-,\Lambda_{p_1},\Lambda_{p_2})\\
  &+\frac{(p_1+p_2)^2}{(l_1-p_1)^2} \delta^{4}(P)(\eta_{l_1})^4(\eta_{l_2})^4\hat{\cF}^{(0)}_\cO(\Lambda_{p_1},\Lambda_{p_2},\Lambda_{p_3},\dots,\Lambda_{p_L};q)
  \Big)
  \eqncom
 \end{aligned}
\end{equation}
where we have added $(\eta_{l_1})^4(\eta_{l_2})^4$ in order to write both lines inside of the full integral over $\Lambda_{l_1}$ and $\Lambda_{l_2}$.
An explicit expression for the colour-ordered four-point tree-level amplitude $\hat{\cA}_4^{(0)}$ is given in \eqref{eq: colour-ordered MHV super amplitude}.

The phase space integral in strictly four dimensions can be performed using a specific parametrisation. 
To make contact with the observation of \cite{Zwiebel:2011bx} on the connection between amplitudes and the dilatation operator, we use the same parametrisation as that paper: 
\begin{equation}\label{eq: parametrisation}
\begin{aligned}
 \left( \begin{array}{c}
\lambda_{l_1}^1  \\
\lambda_{l_2}^1    \end{array} \right)
&=r_1 \e^{i\sigma_1} U  \left( \begin{array}{c}
\lambda_{1}^1  \\
\lambda_{2}^1    \end{array} \right)\eqncom
\qquad
 \left( \begin{array}{c}
\lambda_{l_1}^2  \\
\lambda_{l_2}^2    \end{array} \right)
&=r_2  U  V(\sigma_2) \left( \begin{array}{c}
\lambda_{1}^2  \\
\lambda_{2}^2    \end{array} \right) \eqncom
\end{aligned} 
\end{equation}
where
\begin{equation}
U=\text{diag}(\e^{i\phi_2},\e^{i\phi_3})V(\theta)\text{diag}(1,\e^{i\phi_1})\eqncom\quad 
V(\theta)=
\left( \begin{array}{cc}
\cos\theta & -\sin\theta \\
\sin\theta & \cos\theta \end{array} \right) \eqndot
\end{equation}
The $\lambdat_{l_i}^\alphadot$ are given by the complex conjugates of \eqref{eq: parametrisation} and we have $r_1,r_2\in(0,\infty)$, $\theta,\sigma_2\in(0,\frac{\pi}{2})$, $\sigma_1,\phi_1,\phi_2,\phi_3\in(0,2\pi)$. 
Note that the phase factors involving $\phi_2$ and $\phi_3$ change the spinor helicity variables but not the physical momenta $l_1$ and $l_2$. 
Hence, the result has to be independent of $\phi_2$ and $\phi_3$, as before.

In \cite{Zwiebel:2011bx}, this parametrisation was used to arrive at a compact expression for the first summand in \eqref{eq: bubble coefficient}.
In the following, we briefly review this calculation.

The momentum-conserving delta function $\delta^4(P)$ transforms under the above change of variables according to 
\begin{equation}\label{eq: delta function change}
\begin{aligned}
 \delta^4(P)&=\delta^4(p_1+p_2-l_1-l_2)=\prod_{\alpha=1}^2\prod_{\alphadot=\DOT1}^{\DOT2}\delta(\lambda_1^\alpha\lambdat_1^{\alphadot}+\lambda_2^\alpha\lambdat_2^{\alphadot}-\lambda_{l_1}^\alpha\lambdat_{l_1}^{\alphadot}-\lambda_{l_2}^\alpha\lambdat_{l_2}^{\alphadot})\\
 &=\frac{i \delta(1-r_1)\delta(1-r_2)\delta(\sigma_1)\delta(\sigma_2)}{4(\lambda_1^1\lambdat_1^{\DOT1}+\lambda_2^1\lambdat_2^{\DOT1})(\lambda_1^2\lambdat_1^{\DOT2}+\lambda_2^2\lambdat_2^{\DOT2})\left(-\ab{12}(\lambdat_1^{\DOT1}\lambdat_1^{\DOT2}+\lambdat_2^{\DOT1}\lambdat_2^{\DOT2})+\sb{12}(\lambda_1^1\lambda_1^{2}+\lambda_2^1\lambda_2^{2})\right)} \eqncom
\end{aligned}
\end{equation}
i.e.\ the integrals over $r_1, r_2$ and $\sigma_1, \sigma_2$ localise at $1$ and $0$, respectively.
At these values, the Jacobian from the change of variables is $2 \cos\theta \sin\theta$ times the denominator of the second line in \eqref{eq: delta function change}. 
Hence, the combination of measure and delta function changes according to 
\begin{equation}
 \de^2\lambda_{l_1}\de^2\lambdat_{l_1}\de^2\lambda_{l_2}\de^2\lambdat_{l_2} \, \delta^4(P) \to \de\phi_1\de\phi_2 \de\phi_3\de\theta \, 2i\cos\theta\sin\theta\eqndot
\end{equation}
The $\MHV$ denominator of the four-point amplitude gives
\begin{equation}
 \ab{12}\ab{2l_2}\ab{l_2 l_1}\ab{l_1 1}=   \ab{12}^4 \e^{2 i (\phi_1+\phi_2+\phi_3)} \sin^2\theta \eqndot
\end{equation}
Finally, the supercharge-conserving delta function is 
\begin{equation}\label{eq: expansion of super momentum conserving delta function}
\begin{aligned}
 \delta^8(Q)
 &=\prod_{A=1}^4 \Big( \ab{12}\eta_{1}^A\eta_{2}^A-\ab{1l_1}\eta_{1}^A\eta_{l_1}^A-\ab{1l_2}\eta_{1}^A\eta_{l_2}^A
 -\ab{2l_1}\eta_{2}^A\eta_{l_1}^A-\ab{2l_2}\eta_{2}^A\eta_{l_2}^A+\ab{l_1l_2}\eta_{l_1}^A\eta_{l_2}^A\Big)\\
 &=\ab{12}^4 \e^{4 i (\phi_1+\phi_2+\phi_3)} \prod_{A=1}^4 \Big(\e^{-i (\phi_1+\phi_2+\phi_3)} \, \eta_{1}^A\eta_{2}^A +\eta_{l_1}^A\eta_{l_2}^A \\
 &\phaneq
 +\e^{-i\phi_3}(\sin\theta   \, \eta_{1}^A+\e^{-i \phi_1} \cos \theta \, \eta_{2}^A)\eta_{l_1}^A +\e^{-i \phi_2}(\e^{-i \phi_1} \sin \theta \, \eta_{2}^A-  \cos \theta \, \eta_{1}^A)\eta_{l_2}^A \Big)\eqndot
 \end{aligned}
\end{equation}
The subsequent integration of the fermionic loop variables replaces the fermionic variables in $\hat{\cF}_\cO^{(0)}$ according to 
\begin{equation}
\begin{aligned}
   \left( \begin{array}{c}
\eta_{l_1}^{A}  \\
\eta_{l_2}^{A}  \end{array} \right)
= U^*
\left( \begin{array}{c}
\eta^A_{1}  \\
\eta^A_{2}    \end{array} \right) \eqndot
\end{aligned} 
\end{equation}
In comparison, the bosonic loop variables are 
\begin{equation}\label{eq: bosonic loop spinor variables}
\begin{aligned}
  \left( \begin{array}{c}
\lambda_{l_1}^{\alpha}  \\
\lambda_{l_2}^{\alpha}  \end{array} \right)=U
\left( \begin{array}{c}
\lambda_{1}^\alpha  \\
\lambda_{2}^\alpha    \end{array} \right)\eqncom\quad
  \left( \begin{array}{c}
\lambdat_{l_1}^{\alphadot}  \\
\lambdat_{l_2}^{\alphadot}  \end{array} \right)=U^*
\left( \begin{array}{c}
\lambdat^{\alphadot}_{1}  \\
\lambdat^{\alphadot}_{2}    \end{array} \right)\eqndot
\end{aligned} 
\end{equation}
Combining the previous steps and in particular all phase factors, we find 
\begin{equation}
\begin{aligned}
 2i \int_0^{2\pi}\de \phi_1 \e^{2i\phi_1 C_{p_2}}\int_0^{2\pi}\de \phi_2  \e^{2i\phi_2 C_{l_1}} \int_0^{2\pi}\de \phi_3  \e^{2i\phi_3 C_{l_2}} \int_0^{\frac{\pi}{2}} \de \theta \cot\theta 
 \\ \hat{\cF}^{(0)}_\cO(\Lambda_{1}^\prime,\Lambda_{2}^\prime,\Lambda_{3},\dots,\Lambda_{L};q)\eqncom
\end{aligned}
\end{equation}
where $C_{p_2}$, $C_{l_1}$ and $C_{l_2}$ are the central charges of the respective fields and  
\begin{equation}
 \begin{aligned}
  \left( \begin{array}{c}
\lambda_{1}^{\prime \alpha}  \\
\lambda_{2}^{\prime \alpha}  \end{array} \right)=V(\theta)
\left( \begin{array}{c}
\lambda_{1}^\alpha  \\
\lambda_{2}^\alpha    \end{array} \right)\eqncom\quad
  \left( \begin{array}{c}
\lambdat_{1}^{\prime \alphadot}  \\
\lambdat_{2}^{\prime \alphadot}  \end{array} \right)=V(\theta)
\left( \begin{array}{c}
\lambdat^{\alphadot}_{1}  \\
\lambdat^{\alphadot}_{2}    \end{array} \right)\eqncom\quad
   \left( \begin{array}{c}
\eta_{1}^{\prime A}  \\
\eta_{2}^{\prime A}  \end{array} \right)
=V(\theta)
\left( \begin{array}{c}
\eta^A_{1}  \\
\eta^A_{2}    \end{array} \right)\eqndot
 \end{aligned}
\end{equation}
Performing the integration over $\phi_1$, $\phi_2$ and $\phi_3$ yields
\begin{equation}
\int_0^{2\pi}\de \phi_1 \e^{2i\phi_1 C_{p_2}}\int_0^{2\pi}\de \phi_2  \e^{2i\phi_2 C_{l_1}} \int_0^{2\pi}\de \phi_3  \e^{2i\phi_3 C_{l_2}} = (2\pi)^3 \delta_{C_{p_2},0}\,\delta_{C_{l_1},0}\,\delta_{C_{l_2},0} \eqndot
\end{equation}
The central charge condition is automatically fulfilled at $l_1$ and $l_2$, as these variables correspond to the tree-level form factor. 
The remaining Kronecker delta $\delta_{C_{p_2},0}$ ensures that the central charge condition is also fulfilled for the one-loop correction.

In total, we have
\begin{equation}
\begin{aligned}
&\int \de \Lambda_{l_1}\de \Lambda_{l_2}  \Big(\hat{\cF}^{(0)}_\cO(\Lambda_{l_1},\Lambda_{l_2},\Lambda_{3},\dots,\Lambda_{L};q) \hat{\cA}^{(0)}(\Lambda_{l_2}^-,\Lambda_{l_1}^-,\Lambda_{1},\Lambda_{2})\Big)\\ 
&=2i (2\pi)^3 \delta_{C_{p_2},0} \int_0^{\frac{\pi}{2}} \de \theta \cot\theta \hat{\cF}^{(0)}_\cO(\Lambda_{1}^\prime,\Lambda_{2}^\prime,\Lambda_{3},\dots,\Lambda_{L};q)\eqndot
\end{aligned}
\end{equation}

The remaining task is to compute the second summand in \eqref{eq: bubble coefficient}.
The only term that depends on the integration variables is
\begin{equation}
 (l_1-p_1)^2=\ab{l_1 1}\sb{l_1 1}\eqndot
\end{equation}
From \eqref{eq: bosonic loop spinor variables}, we have 
\begin{equation}
 \ab{l_1 1}= \ab{ 1 2} \e^{i(\phi_1+\phi_2)} \sin\theta \eqndot
\end{equation}
Thus, 
\begin{equation}
 (l_1-p_1)^2=\ab{ 1 2}\sb{ 1 2}\sin^2\theta=-(p_1+p_2)^2\sin^2\theta\eqndot
\end{equation}
This expression is free of $\phi_1,\phi_2,\phi_3$, and the corresponding integrations hence yield a factor of $(2\pi)^3$. 
The central charge automatically vanishes for all fields. Hence, we can include $\delta_{C_{p_2},0}$ and write the complete second summand in \eqref{eq: bubble coefficient} as 
\begin{equation}
 \begin{aligned}
 -2i (2\pi)^3 \delta_{C_{p_2},0} \int_0^{\frac{\pi}{2}} \de\theta \cot\theta \hat{\cF}^{(0)}_\cO(\Lambda_{1},\Lambda_{2},\Lambda_{3},\dots,\Lambda_{L};q)
  \eqndot
 \end{aligned}
\end{equation}

To obtain the bubble coefficient, we have to compare the above to the phase space integral of the double cut of the bubble integral. It is given by
\begin{equation}\label{eq: cut of bubble}
\begin{aligned}
\int \de \Lambda_{l_1}\de \Lambda_{l_2} \delta^{4}(P) (\eta_{l_1})^4(\eta_{l_2})^4 \FDinline[bubble,cut,topleglabel=$\scriptscriptstyle 1$,bottomleglabel=$\scriptscriptstyle 2$] 
&=2i (2\pi)^3  \int_0^{\frac{\pi}{2}} \de \theta \sin\theta \cos\theta = i (2\pi)^3 \eqncom
\end{aligned}
\end{equation}
where the depicted cut integrand is simply 1.

The bubble coefficient is given by the ratio of \eqref{eq: bubble coefficient} and \eqref{eq: cut of bubble}:\footnote{As before, we have neglected a factor of $\tilde{g}^2$ on both sides of figure \ref{fig: ansatz for generalised unitarity}.}
\begin{equation}\label{eq: bubble coefficient final}
\begin{aligned}
c_{\text{bubble}}^{1,2}=-2 \delta_{C_{p_2},0} \int_0^{\frac{\pi}{2}} \de \theta \cot\theta \Big(& \hat{\cF}^{(0)}_\cO(\Lambda_{1},\Lambda_{2},\Lambda_{3},\dots,\Lambda_{L};q)\\&-\hat{\cF}^{(0)}_\cO(\Lambda_{1}^\prime,\Lambda_{2}^\prime,\Lambda_{3},\dots,\Lambda_{L};q) \Big)\eqndot
\end{aligned}
\end{equation}
Apart from a factor of $-2$, this is precisely the result obtained by replacing all oscillators in the one-loop dilatation operator \eqref{eq: one-loop dila in oscillators} by on-shell variables via \eqref{eq: oscillator replacements}. We come back to this point in the next section.

\subsection{Summary of the result}
\label{subsec: summary}

In the previous subsections, we have explicitly calculated the coefficients of the integrals in figure \ref{fig: simplified ansatz for generalised unitarity} that involve the legs $p_1$ and $p_2$. Analogous results are valid for any other pair of neighbouring legs $p_i$ and $p_{i+1}$.

Hence, the final result for the cut-constructible part of the planar colour-ordered minimal one-loop form factor of a generic single-trace operator $\cO$ is 
 \begin{equation}\label{eq: summary}
 \begin{aligned}
  \hat{\cF}^{(1)}_\cO(1,\dots,L;q)&= -  \sum_{i=1}^L s_{i\, i+1} \hat{\cF}^{(0)}_\cO(1,\dots,L;q) \FDinline[triangle,topleglabel=$\scriptscriptstyle i$,bottomleglabel=$\scriptscriptstyle i+1$]
  \\   &\phaneq
  +  \sum_{i=1}^L B_{i\,i+1}\hat{\cF}^{(0)}_\cO(1,\dots,L;q) \FDinline[bubble,topleglabel=$\scriptscriptstyle i$,bottomleglabel=$\scriptscriptstyle i+1$] \\
  &\phaneq+\text{rational terms} \eqncom
 \end{aligned}
 \end{equation}
where $s_{i\,i+1}=(p_i+p_{i+1})^2$ and 
\begin{equation}\label{eq: bubble coefficient summary}
\begin{aligned}
B_{i\,i+1}\hat{\cF}^{(0)}_\cO(\Lambda_{1},\dots,\Lambda_{L};q)=-2 \delta_{C_{i+1},0} \int_0^{\frac{\pi}{2}} \de \theta \cot\theta \Big( &\hat{\cF}^{(0)}_\cO(\Lambda_{1},\dots,\Lambda_{i},\Lambda_{i+1},\dots,\Lambda_{L};q)\\
&-\hat{\cF}^{(0)}_\cO(\Lambda_{1},\dots,\Lambda_{i}^\prime,\Lambda_{i+1}^\prime,\dots,\Lambda_{L};q) \Big)\eqndot
\end{aligned}
\end{equation}
with
\begin{equation}
 \begin{aligned}
   \left( \begin{array}{c}
\Lambda_{i\phantom{+1}}^{\prime}  \\
\Lambda_{i+1}^{\prime }  \end{array} \right)
=V(\theta)
\left( \begin{array}{c}
\Lambda_{i\phantom{+1}}  \\
\Lambda_{i+1}    \end{array} \right)\eqncom \qquad V(\theta)=
\left( \begin{array}{cc}
\cos\theta & -\sin\theta \\
\sin\theta & \cos\theta \end{array} \right)\eqndot
 \end{aligned}
\end{equation} 
We give explicit expressions for the integrals occurring in \eqref{eq: summary} in \eqref{eq: bubble integral} and \eqref{eq: triangle integral}.
The above one-loop correction occurs as second term in the expansion of the minimal form factor in the modified effective planar coupling constant \eqref{eq: modified effective planar coupling constant}:
\begin{equation}
\hat{\cF}_\cO(1,\dots,L;q)=\hat{\cF}^{(0)}_\cO(1,\dots,L;q)+\tilde{g}^2\hat{\cF}^{(1)}_\cO(1,\dots,L;q)+\cO(\tilde{g}^3)
 \eqndot
\end{equation}

Moreover, the above result is not limited to the planar theory; it can be straightforwardly generalised to finite $N$. 
To this end, one can either work out the colour factor which arises from \eqref{eq: def colour-ordered form factor} and \eqref{eq: def colour-ordered amplitude} for a cut in non-adjacent legs by hand or modify the prescription of \cite{Beisert:2003jj} for applying the complete one-loop dilatation operator at finite $N$.\footnote{This is similar to the situation for the one-loop amplitude, where the non-planar double-trace contributions can be reconstructed from the knowledge of the planar single-trace contribution as well \cite{Bern:1994zx}.}

Finally, let us comment on the finite rational terms in \eqref{eq: summary}, which cannot be detected by unitary cuts in four dimensions.
By comparing with the known one-loop minimal form factors in the literature \cite{Brandhuber:2010ad,Bork:2010wf,Penante:2014sza}, these are found to be absent for the BPS operators $\tr[\phi_{12}\phi_{12}]$ and $\tr[(\phi_{12})^k]$.
In \cite{Bern:1994cg}, it was found that rational terms can only occur if three or more powers of numerator momenta appear in a box integral or if two or more powers of numerator momenta appear in a triangle or bubble integral.
For amplitudes, this could be used to severely constrain the appearance of rational terms \cite{Bern:1994cg}.
For form factor, however, we have an unlimited number of numerator momenta due to the insertion of the composite operator. For example, for an operator in the $\mathfrak{sl}(2)$ subsector with $k$ covariant derivatives, we have at least $k$ numerator momenta in the loop.
Indeed, a small example calculation using unitarity at the level of the integrand and a subsequent Passarino-Veltman reduction \cite{Passarino:1978jh} shows that rational terms in general occur.
This calculation is presented in appendix \ref{app: SL2 sector}.
For amplitudes, several methods to obtain the rational terms are reviewed in \cite{Bern:2007dw}. These appear to be applicable for form factors as well. We leave this for future investigations.

Moreover, an important subtlety arises for operators with contracted flavour or vector indices, as we discuss in full detail for the example of the Konishi operator in \cite{Nandan:2014oga}. At one-loop level, this subtlety gives rise to additional rational terms \cite{Nandan:2014oga}.

\section{Divergences, renormalisation and the dilatation operator}
\label{sec: one-loop dila}

In this section, we discuss the divergences occurring in the previously derived one-loop result.
In particular, the UV divergences require renormalisation and yield the complete one-loop dilatation operator.

At one loop, the only UV divergent integral is the bubble integral. The triangle integral is IR divergent but UV finite, and the possible rational terms are UV and IR finite.\footnote{The box integral is IR divergent and UV finite as well but does not occur here.} Both UV and IR divergences are regularised by continuing the dimension of spacetime to $D=4-2\varepsilon$.

For any operator $\cO$, the IR divergent part of the planar one-loop form factor \eqref{eq: summary} is given by 
\begin{equation}
 \left.\hat{\cF}^{(1)}_\cO(1,\dots,L;q)\right|_{\text{IR}}= -\left(\sum_{i=1}^L s_{i\,i+1} \FDinline[triangle,topleglabel=$\scriptscriptstyle i$,bottomleglabel=$\scriptscriptstyle i+1$] \,\,\,\,\, \bigg|_{\text{IR}} \right) \hat{\cF}^{(0)}_\cO(1,\dots,L;q)\eqncom
\end{equation}
where the triangle integral is given in appendix \ref{app: integrals}.
Hence,
\begin{equation}
 \left.\frac{\hat{\cF}^{(1)}_\cO(1,\dots,L;q)}{\hat{\cF}^{(0)}_\cO(1,\dots,L;q)}\right|_{\text{IR}}=-\frac{1}{\varepsilon^2}
 \sum_{i=1}^L  (-s_{i\,i+1})^{-\varepsilon}\eqndot
\end{equation} 
This agrees with the universal IR behaviour of form factors \cite{vanNeerven:1985ja,Henn:2011by,Penante:2014sza},
as well as the conjectured duality between form factors and periodic Wilson loops \cite{Brandhuber:2010ad}.
Moreover, it agrees with a straightforward generalisation of the BDS-ansatz-type form \cite{Bern:2005iz} from amplitudes to form factors \cite{Brandhuber:2010ad,Bork:2010wf}:%
\footnote{This form was checked for the minimal form factor of the BPS operator $\tr[\phi_{12}\phi_{12}]$ up to the third loop order \cite{Gehrmann:2011xn}, for the minimal form factor of the BPS operator $\tr[(\phi_{12})^k]$ up to the second loop order \cite{Brandhuber:2014ica}, for the $n$-point MHV form factor of the BPS operator $\tr[\phi_{12}\phi_{12}]$ up to the first loop order \cite{Brandhuber:2010ad} and for the three-point MHV form factor of the BPS operator $\tr[\phi_{12}\phi_{12}]$ up to the second loop order \cite{Brandhuber:2012vm}.}
\begin{equation}\label{eq: universal IR divergences}
  \log\left(\frac{\hat{\cF}_\cO(1,\dots,n;q)}{\hat{\cF}^{(0)}_\cO(1,\dots,n;q)}\right)= \sum_{l=1}^\infty \tilde{g}^{2l}\left[ -\frac{\gamma^{(l)}_{\text{cusp}}}{8(l\varepsilon)^2}-\frac{\cG_0^{(l)}}{4l\varepsilon}\right]\sum_{i=1}^n(-s_{i\,i+1})^{-l\varepsilon} + \text{Fin}(\tilde{g}^2) +\cO(\varepsilon)
  \eqncom
\end{equation}
where 
\begin{equation}
 \gamma_{\text{cusp}}(\tilde{g}^2)=\sum_{l=1}^\infty \tilde{g}^{2l}\gamma^{(l)}_{\text{cusp}}= 8 \tilde{g}^2-\frac{8 \pi ^2 }{3}\tilde{g}^4+\frac{88 \pi^4 }{45}\tilde{g}^6+\cO(\tilde{g}^8)
\end{equation}
is the cusp anomalous dimension,
\begin{equation}
 \cG_0(\tilde{g}^2)=\sum_{l=1}^\infty \tilde{g}^{2l}\cG^{(l)}_0= -4  \zeta_3\tilde{g}^4+8  \left(\frac{5 \pi ^2 \zeta_3}{9}+4 \zeta_5\right)\tilde{g}^6+\cO(\tilde{g}^8)
\end{equation}
is the collinear anomalous dimension and $\text{Fin}(\tilde{g}^2)$ is a function that is finite in the limit $\varepsilon \to 0$.

The IR divergences are constrained by the Kinoshita-Lee-Nauenberg theorem \cite{Kinoshita:1962ur,Lee:1964is}.
In all observables, like cross sections or correlation functions, they cancel among contributions with the same order of the coupling constant but different numbers of loops and external legs.

The UV divergences, on the other hand, require renormalisation.
In \NfSYMt, the appropriate combinations of the self-energies of the elementary fields%
\footnote{In the formulation in $\cN=1$ superspace, the self-energies of the superfields themselves are already UV finite. 
In the formulation in terms of component fields, however, the self-energies of the elementary fields have nonvanishing UV divergences; see e.g.\ \cite{Minahan:2002ve}.} 
 and the one-particle-irreducible (1PI) corrections to the elementary vertices are UV finite, ensuring the vanishing of the $\beta$-function.
The only sources for UV divergences are the insertions of composite operators as external states, which hence need to be renormalised.
The renormalised operators can be written in terms of the bare operators and the renormalisation constant $\cZ$ as
\begin{equation}\label{eq: def renomalisation constant}
 \cO_{\text{ren}}^a=\cZ^a{}_b \cO_{\text{bare}}^b \eqncom
\end{equation}
where the renormalisation constant is a matrix due to the possibility of operator mixing. In terms of the modified effective planar coupling constant $\tilde{g}$  \eqref{eq: modified effective planar coupling constant}, $\cZ$ can be expanded as 
\begin{equation}\label{eq: expansion of Z}
 \cZ^a{}_b=\delta^a{}_b+\tilde{g}^2 (\cZ^{(1)})^a{}_b+\cO(\tilde{g}^3) \eqndot
\end{equation}
The renormalisation constant $\cZ$ is universal; it has to cancel the UV divergences associated to the respective operator in all gauge-invariant 
correlation functions,
in particular those defining the form factor.
Hence, we can use the UV divergences in the form factor to determine $\cZ$.\footnote{In \cite{Engelund:2012re}, three-point correlation functions were computed via generalised unitarity and the cancellation of the UV divergences therein was used to determine the one-loop dilatation operator in the $\mathfrak{sl}(2)$ subsector.}
In particular, \eqref{eq: universal IR divergences} is expected to be valid for the form factor of the renormalised operator.
So far, we have worked with bare operators. Inserting \eqref{eq: def renomalisation constant} into the form factor yields
\begin{equation}
 \hat{\cF}_{\cO_{\text{ren}}^a}(1,\dots,L;q)
 =\cZ^a{}_b \hat{\cF}_{ \cO_{\text{bare}}^b}(1,\dots,L;q)\eqncom
\end{equation}
which can be expanded to the first loop order to arrive at  
\begin{equation}\label{eq: form factor of renormalised operator at one-loop}
 \hat{\cF}^{(1)}_{\cO_{\text{ren}}^a}(1,\dots,L;q)= \hat{\cF}^{(1)}_{ \cO_{\text{bare}}^a}(1,\dots,L;q)
 +(\cZ^{(1)})^a{}_b\hat{\cF}^{(0)}_{ \cO_{\text{bare}}^b}(1,\dots,L;q)\eqndot
\end{equation}
This has to be UV finite.
Accordingly, the second term in \eqref{eq: form factor of renormalised operator at one-loop} has to cancel the UV divergent part of the first term, i.e.\ the UV divergent part of the bubble integral times the bubble coefficient \eqref{eq: bubble coefficient summary}:
\begin{equation}
\begin{aligned}
 (\cZ^{(1)})^a{}_b\hat{\cF}^{(0)}_{ \cO_{\text{bare}}^b}(1,\dots,L;q)
 &=-\left.\hat{\cF}^{(1)}_{ \cO_{\text{bare}}^a}(1,\dots,L;q)\right|_{\text{UV}}\\
 &=-\sum_{i=1}^L B_{i\,i+1}\hat{\cF}^{(0)}_{\cO_{\text{bare}}^a}(1,\dots,L;q) \times \Kop\left[ \FDinline[bubble,topleglabel=$\scriptscriptstyle i$,bottomleglabel=$\scriptscriptstyle i+1$] \,\,\,\,\, \right]\eqncom
\end{aligned}
\end{equation}
where the operator $\Kop$ extract the divergence.
The bubble integral is given in \eqref{eq: bubble integral}, and its UV divergence is 
\begin{equation}
\Kop\left[ \FDinline[bubble,topleglabel=$\scriptscriptstyle i$,bottomleglabel=$\scriptscriptstyle i+1$] \,\,\,\,\, \right]=\frac{1}{\varepsilon} \eqndot 
\end{equation} 

The connection between $\cZ$ and the anomalous part of the dilatation operator $\delta\mathfrak{D}$ is, cf.\ e.g.\ \cite{Sieg:2010jt},
\begin{equation}
 \delta\mathfrak{D}=\lim_{\varepsilon\to0} \varepsilon \tilde{g} \frac{\partial}{\partial \tilde{g}} \ln \cZ \eqncom
\end{equation}
where the logarithm is understood as series in the loop corrections to $\cZ$, cf.\ \eqref{eq: expansion of Z}.
The derivative with respect to $\tilde{g}$ yields a factor of $2K$ for loop order $K$. 
In the one-loop case, we have 
\begin{equation}
 \mathfrak{D}_2=\frac{1}{\tilde{g}^2}\lim_{\varepsilon\to0} \varepsilon \tilde{g} \frac{\partial}{\partial \tilde{g}}\left(\tilde{g}^2 \cZ^{(1)}\right)=-2 \sum_{i=1}^L B_{i\,i+1} \eqncom
\end{equation}
or, for its density,
\begin{equation}
 (\mathfrak{D}_2)_{i\,i+1}= - 2 B_{i\,i+1} \eqndot
\end{equation}
This exactly yields the known expression \eqref{eq: one-loop dila in oscillators} for the dilatation-operator density once we replace the oscillators by spinor helicity variables according to \eqref{eq: oscillator replacements}.\footnote{The choice to use $\tilde{g}$ instead of $g$ as expansion parameter, i.e.\ to use the modified minimal subtraction scheme instead of the minimal subtraction scheme, does not affect the one-loop dilatation operator.}

Some remarks are in order.
As we are working in $D=4-2\varepsilon$ dimensions, massless bubble integrals with on-shell external momentum vanish, in particular those potentially occurring at the massless external legs. Massless bubble integrals with on-shell external momentum are, however, UV and IR divergent, and only vanish in $D=4-2\varepsilon$ due to a cancellation between the UV and IR poles. 
In particular, their UV divergence has to be taken into account in the calculation of renormalisation group coefficients. For example when calculating the $\beta$-function of pure Yang-Mills theory, neglecting them 
leads to the wrong sign \cite{ArkaniHamed:2008gz}.
Because of the aforementioned cancellation,
the UV divergences from the massless bubble integrals with on-shell external momentum can be reconstructed from the IR divergences. Since the IR divergences in these bubble integrals are cancelled by the UV divergences, they can no longer cancel the IR divergences from the contribution with one loop less and one external leg more as predicted by the Kinoshita-Lee-Nauenberg theorem. Hence, the UV divergences can be reconstructed from those uncancelled IR divergences \cite{ArkaniHamed:2008gz}.\footnote{This observation in \cite{ArkaniHamed:2008gz} goes back to Lance Dixon; see the respective remark in \cite{ArkaniHamed:2008gz}.}    
In our case, all UV divergences of massless bubbles with on-shell external momentum are correctly accounted for. This can be seen as follows.\footnote{Of course, we also know the expression for the complete one-loop dilatation operator and have seen that it is correctly reproduced by the UV divergence of our result.} 
As discussed above, the IR divergences of the one-loop form factor agree with the expected universal form \eqref{eq: universal IR divergences}, see also \cite{vanNeerven:1985ja,Henn:2011by,Penante:2014sza}. 
But as the IR divergences are correctly accounted for, so are the UV divergences, as massless bubbles with on-shell external momentum affect both at once.

Finally, let us comment on a completely different kind of divergence that occurs in the calculation.
Both the integrals of the first and the second summand in \eqref{eq: bubble coefficient}, and hence also in \eqref{eq: bubble coefficient final}, diverge when taken individually. 
The divergence occurs in the integral region where the uncut propagator in the triangle integral goes on-shell, i.e.\ for $\theta=0$.
It is hence a collinear divergence of the tree-level four-point amplitude.\footnote{Recall that the phase space integral is taken to be strictly four dimensional.} 
In \cite{Zwiebel:2011bx}, the author only obtains the second term in \eqref{eq: bubble coefficient final} from an amplitude-like expression and adds the first term in \eqref{eq: bubble coefficient final} as a regularisation.
He (numerically) finds that the commutation relations among the length-changing generators of the superconformal algebra fix this regularisation uniquely, but states that a (more) physical argument would be desirable.  
In this work, we have found such an argument and moreover derived the complete result from field theory.

\section{Conclusion and outlook}
\label{sec: conclusion and outlook}

In this paper, we have initialised the study of form factors for generic gauge-invariant local composite operators in \NfSYMt. 
In the free theory, the form factors exactly realise the spin-chain picture for composite operators in the language of on-shell superfields.
The loop corrections to the form factor require operator renormalisation and hence yield the dilatation operator, a central object of integrability.
Form factors thus allow us to revisit and extend the study of integrability for composite operators via on-shell methods.

At tree-level, we have shown that the minimal form factor of an irreducible single-trace operator is obtained by replacing the spin-chain oscillators $(\aosc_i^{\dagger\alpha},\bosc_i^{\dagger\alphadot},\dosc_i^{\dagger A})$ by the on-shell super momenta $(\lambda_i^{\alpha},\lambdat_i^{\alphadot},\eta_i^{A})$.
We then used generalised unitarity to obtain the cut-constructible part of the one-loop correction to the minimal form factor.
Its IR divergence is independent of the precise operator and agrees with the universal prediction by a BDS-ansatz-type form and the conjectured duality to periodic Wilson loops. 
Its UV divergence, on the other hand, delicately depends on the operator. It requires operator renormalisation and yields the complete one-loop dilatation operator of \NfSYMt in the formulation of \cite{Zwiebel:2007cpa}.  
To our knowledge, this constitutes the first derivation of the complete $\mathfrak{D}_2$ entirely from field theory and without exploiting symmetries to lift results from closed subsectors to the full theory as in \cite{Beisert:2003jj,Beisert:2004ry}.
In the progress, we have in particular derived and explained a connection between $\mathfrak{D}_2$ and the four-point amplitude, which was observed in \cite{Zwiebel:2011bx} but could up to now not be derived via the methods of field theory.  
The minimal one-loop form factor also contains rational terms, which are not detectable by unitary cuts in four dimensions. These are, however, UV and IR finite.
\newline

Our results suggest many interesting routes for further investigation.\footnote{All three directions below are currently under active investigation.}  
One direction is towards more loops and legs and a field-theoretic understanding of integrability.  
To complete the result for the minimal one-loop form factor, the rational terms need to be calculated. On-shell techniques as reviewed in \cite{Bern:2007dw} seem promising for this purpose. Moreover, the one-loop method used in this paper is not limited to the minimal form factor; it can also be applied for a higher number of points.
A next important step is to go to the second loop order and calculate the minimal two-loop form factor for a generic operator.\footnote{The minimal two-loop form factor for the operator $\tr[(\phi_{12})^k]$ was recently calculated in \cite{Brandhuber:2014ica}.}
In particular, this would yield the complete two-loop dilatation operator of \NfSYMt. 
The complete two-loop dilatation operator is currently unknown and would be highly interesting from a conceptual point of view. It would show which quantity the Bethe ansatz diagonalises beyond one loop and could hence further the field-theoretic understanding of integrability, possibly even leading to a field-theoretic proof.
Furthermore, the two-loop dilatation operator would also be interesting from a practical point of view, as the current approach of integrability for three-point functions requires its eigenstates and not only its eigenvalues; see \cite{Escobedo:2010xs,Bajnok:2014sza} and references therein.
If one is only interested in the two-loop dilatation operator, determining the UV divergent contributions to the two-loop form factor suffices. One could hence avoid the detour of calculating the coefficients of all diagrams by generalising the methods for the direct extraction of integral coefficient developed at one-loop in \cite{Forde:2007mi,ArkaniHamed:2008gz} to the second loop order.
A step towards the two-loop dilatation operator would be to derive the expressions for the leading length-changing contributions of the dilatation operator, which are completely fixed by symmetry and were obtained in \cite{Zwiebel:2011bx}.
Finally, the on-shell methods for the calculation of the dilatation operator are not limited to \NfSYMt and could be applied in other interesting theories.%
\footnote{A different interesting approach to the dilatation operator via two-point functions and the twistor action is presented in the contemporaneously appearing paper \cite{Koster:2014fva}, where the one-loop dilatation operator in the $\mathfrak{so}(6)$ subsector is obtained. In the forthcoming publication \cite{KMSW}, we will show that the twistor action can also be employed for the calculation of form factors.}

A second direction is to better understand the connection between the action of the complete $\mathfrak{psu}(2,2|4)$ algebra on amplitudes and composite operators via form factors. In both cases, the generators obtain corrections in the interacting theory; apart from $\mathfrak{D}$ also $\mathfrak{Q}$, $\dot{\mathfrak{Q}}$, $\mathfrak{S}$, $\dot{\mathfrak{S}}$, $\mathfrak{P}$ and $\mathfrak{K}$. 
In particular, these corrections can be length-changing.
On amplitudes, they occur due to the holomorphic anomaly. On composite operators, they occur due to the interaction terms in the equations of motion that are used to reduce the fields in the spin-chain picture. A connection between the two corrections exists and was already used in the algebraic derivation of \cite{Zwiebel:2011bx}. On form factors, this connection becomes manifest.

The third and perhaps most interesting direction is to compute form factors by a modification of the approach of integrability for amplitudes at weak coupling.\footnote{At strong coupling, it is already known how to compute form factors via a modification of the respective approach of integrability for amplitudes \cite{Maldacena:2010kp,Gao:2013dza}. It would also be interesting to generalise the approach of \cite{Basso:2013vsa}, which works for finite coupling in certain kinematic regimes, from amplitudes to form factors.}
The current approach to integrability for composite operators yields the eigenvalues of the dilatation operator, which are the anomalous dimensions, but not the corresponding eigenstates.
Integrability for amplitudes yields eigenstates, namely Yangian invariants, but the corresponding eigenvalues are trivial. Unifying both approaches promises to yield not only the spectrum of anomalous dimensions but also the corresponding eigenstates --- both these pieces of information are contained in the UV divergent part of the form factor. Moreover, the form factor has a finite and an IR divergent part, which should be accessible by integrability as well.\footnote{To obtain \emph{normalised} eigenstates, these parts are required as well.}${}^{\text{,}}$\footnote{Note that also the finite-size effect of wrapping \cite{Sieg:2005kd,Ambjorn:2005wa} occurs for form factors. While $\cN=4$ supersymmetry dictates that it can only affect the UV divergent part starting at the fourth loop order, wrapping affects the IR divergent part of $\cF_{\tr[\phi_{12}\phi_{12}]}$ already at the second loop order \cite{vanNeerven:1985ja,Henn:2011by}.} 
Form factors form a bridge between amplitudes and correlation functions. The latter can be constructed from (generalised) form factors and amplitudes via generalised unitarity \cite{Engelund:2012re}. Hence, an understanding of integrability for form factors should also provide valuable information for the computation of correlation functions via integrability and could hence be an important step towards a final solution of (planar) \NfSYMt.

\section*{Acknowledgements}

It is a pleasure to thank Dhritiman Nandan, my coadviser Christoph Sieg and Gang Yang for collaboration on the related work \cite{Nandan:2014oga}, numerous discussions and comments on the manuscript.
I thank my adviser, Matthias Staudacher, for suggesting to find a field-theoretic derivation of the observation of \cite{Zwiebel:2011bx}, for many discussions as well as for remarks on the manuscript.
I am grateful to Zvi Bern, Lance Dixon, Nils Kanning, Kasper Larsen, Florian Loebbert, Brenda Penante, Jan Plefka and Radu Roiban for discussions on various aspects of this work. 
I thank the Simons Center for Geometry and Physics, Stony Brook University, as well as the C.N.\ Yang Institute for Theoretical Physics for warm hospitality during the final stage of this project.
This work was supported by the Marie Curie International Research Staff Exchange Network UNIFY of the European Union's Seventh Framework Programme [FP7-People-2010-IRSES] under grant agreement number 269217, which allowed me to visit Stony Brook University.
It was also supported in part by the SFB 647 \emph{``Raum-Zeit-Materie. Analytische und Geometrische Strukturen''} and the Marie Curie network GATIS (\texttt{\href{http://gatis.desy.eu}{gatis.desy.eu}}) of the European Union’s Seventh Framework Programme FP7/2007-2013/ under REA Grant Agreement No 317089. 
Ich danke der Studienstiftung des deutschen Volkes für ein Promotionsförderstipendium.

\appendix

\section{Some facts about amplitudes}
\label{app: amplitudes}

In this appendix, we give some basic expressions for scattering amplitudes which we are using in section \ref{sec: one-loop form factors}. 
We refer the reader to the review \cite{Elvang:2013cua} for details.

The $n$-point tree-level amplitudes $\cA^{(0)}_n$ of \NfSYMt can be expressed in terms of a trace over the generators $\T^a$ of the gauge group $SU(N)$ and a colour-ordered amplitude $\hat{\cA}^{(0)}_n$:
\begin{equation}\label{eq: def colour-ordered amplitude}
 \begin{aligned}
 \cA^{(0)}_n(1,\dots,n)= g_\YM^{n-2}\sum_{\sigma\in S_n/Z_n} \Tr[\T^{a_{\sigma(1)}}\cdots\T^{a_{\sigma(n)}}] \hat{\cA}^{(0)}_n(\sigma(1),\dots,\sigma(n))
 \eqncom
 \end{aligned}
\end{equation}
where the sum is over all non-cyclic permutations. Moreover, we have written the explicit dependence on the Yang-Mills coupling constant $g_\YM$ in front of the colour-ordered amplitude.

The colour-ordered $n$-point tree-level \MHV 
superamplitude of \NfSYMt is given by
\begin{equation}\label{eq: colour-ordered MHV super amplitude}
 \begin{aligned}
  \hat{\cA}^{\MHV\,(0)}_n(1,\dots,n)=\frac{\delta^4(P)\delta^8(Q)}{\ab{12}\ab{23}\cdots\ab{n1}} \eqncom
 \end{aligned}
\end{equation}
where all fields are assumed to be outgoing and \MHV refers to the (minimal) fermionic degree of $\cA^\MHV_n$, which is $8$.
The bosonic momentum-conserving delta function explicitly reads
\begin{equation}\label{eq: bosonic delta function}
 \delta^4(P)=\prod_{\alpha=1}^2\prod_{\alphadot=\DOT1}^{\DOT2} \delta\left(\sum_{i=1}^n\lambda_i^\alpha\lambdat_i^\alphadot\right)\eqncom
\end{equation}
while the fermionic supermomentum-conserving delta function reads 
\begin{equation}
 \delta^8(Q)=\frac{1}{2^4}\prod_{A=1}^4 \sum_{i,j=1}^{n} \epsilon_{\alpha\beta}\mathfrak{Q}_i^{\alpha A}\mathfrak{Q}^{\beta A}_j=\prod_{A=1}^4 \sum_{1\leq i< j\leq n} \ab{ij}\eta_i^A\eta_j^A\eqndot
\end{equation}

The colour-ordered $n$-point tree-level \MHVb superamplitude of \NfSYMt can be obtained from its \MHV counterpart by applying the conjugation rule \eqref{eq: conjugation in on-shell variables}. 
Here, \MHVb refers to the (maximal) fermionic degree of $\cA^\MHVb_n$, which is $4n-8$.\footnote{Note that the statement about the minimal or maximal fermionic degree applies for $n\geq 4$. For $n=3$, $\cA^\MHV$ has a higher fermionic degree than $\cA^\MHVb$.}
For example,
the colour-ordered three-point tree-level \MHVb superamplitude is 
\begin{equation}
 \begin{aligned}
  \hat{\cA}^{\MHVb\,(0)}_3(1,2,3)=\frac{\delta^4(P)\delta^4(\bar{Q})}{\sb{12}\sb{23}\sb{31}} \eqncom
 \end{aligned}
\end{equation}
where the fermionic delta function reads
\begin{equation}
 \delta^4(\bar{Q})=\prod_{A=1}^4  \left(\sb{12}\eta_3^A+\sb{23}\eta_1^A+\sb{31}\eta_2^A\right) \eqndot
\end{equation}

\section{Example calculations}
\label{app: examples}

In this appendix, we use a more pedestrian approach, namely ordinary unitarity at the level of the integrand, to compute the one-loop form factor for some example operators. In particular, we derive the one-loop dilatation operator in the $\mathfrak{su}(2)$ subsector, which is given by the Hamiltonian of the Heisenberg XXX spin chain.\footnote{The dilatation operator of the $\mathfrak{su}(2)$ subsector is currently known from direct field-theory calculations \cite{Sieg:2010tz} up to three loops.}
Moreover, we demonstrate that rational terms can in general occur.

\subsection{\texorpdfstring{$\mathfrak{su}(2)$}{su(2)} subsector}

The $\mathfrak{su}(2)$ subsector of \NfSYMt can be built from two kinds of different scalars $\phi_{AC}$ and $\phi_{BC}$ with one common $SU(4)$ index, say $\phi_{24}$ and $\phi_{34}$. These are identified with spin up $(\uparrow)$ and spin down $(\downarrow)$, respectively.
Single-trace operators in this subsector are built as traces of $L$ of these scalars.

As in the main text, we consider the cut in the legs $p_1$ and $p_2$. This leads to four different possibilities: the combinations of fields we can encounter at these positions in the operator or tree-level form factor are $\uparrow\uparrow$, $\downarrow\uparrow$, $\uparrow\downarrow$ and $\downarrow\downarrow$. In fact, it is sufficient to consider $\uparrow\downarrow$ and $\downarrow\downarrow$, as the remaining two combinations are related to these two via the $SU(4)_R$ symmetry.  

Let us first consider the case $\downarrow\downarrow$ resp. $\displaystyle \cO=\tr[\phi_{34}\phi_{34}\cdots]+\dots$, where the ellipses denote additional fields and terms that are not relevant for the present discussion. According to \eqref{eq: oscillator replacements}, this leads to the colour-ordered minimal tree-level form factor 
\begin{equation}
 \hat{\cF}^{(0)}_\cO(\Lambda_1,\Lambda_2,\dots,\Lambda_L;q)=\delta^4\left(\sum_{i=1}^L\lambda_i\lambdat_i-q\right)(\eta_1^3\eta_1^4\eta_2^3\eta_2^4 \dots+\dots)\eqndot
\end{equation}
The double cut depicted in figure \ref{fig: double cut} reads
\begin{equation}\label{eq: su2 zz 1}
 \begin{aligned}
 &\int \de \Lambda_{l_1}\de \Lambda_{l_2}  \hat{\cF}^{(0)}_\cO(\Lambda_{l_1},\Lambda_{l_2},\Lambda_{3},\dots,\Lambda_{L};q) \hat{\cA}(\Lambda_{l_2}^-,\Lambda_{l_1}^-,\Lambda_{1},\Lambda_{2})\\
 = &\int \de \Lambda_{l_1}\de \Lambda_{l_2} \delta^4\left(\lambda_{l_1}\lambdat_{l_1}+\lambda_{l_2}\lambdat_{l_2}+\sum_{i=3}^L\lambda_i\lambdat_i-q\right)(\eta_{l_1}^3\eta_{l_1}^4\eta_{l_2}^3\eta_{l_2}^4\dots+\dots)\\
 & \phantom{\int \de \Lambda_{l_1}\de \Lambda_{l_2} }
 \frac{\delta^4(\lambda_{1}\lambdat_{1}+\lambda_{2}\lambdat_{2}-\lambda_{l_1}\lambdat_{l_1}-\lambda_{l_2}\lambdat_{l_2})\delta^8(Q)}{\ab{12}\ab{2l_2}\ab{l_2l_1}\ab{l_1 1}}\eqndot
 \end{aligned}
\end{equation}
Expanding the supermomentum-conserving delta function as in the first line of \eqref{eq: expansion of super momentum conserving delta function} and performing the Grassmann integral over $\eta^A_{l_1}$ and $\eta^A_{l_2}$, we find
\begin{equation}\label{eq: su2 zz 2}
 \begin{aligned}
  \int \de^2 \lambda_{l_1}\de^2 \lambdat_{l_1}\de^2 \lambda_{l_2}\de^2 \lambdat_{l_2} &\delta^4\left(\lambda_{1}\lambdat_{1}+\lambda_{2}\lambdat_{2}-\lambda_{l_1}\lambdat_{l_1}-\lambda_{l_2}\lambdat_{l_2}\right)\\ &\delta^4\left(\sum_{i=1}^L\lambda_i\lambdat_i-q\right)\left(\frac{\ab{1 2}\ab{ l_1l_2}}{\ab{1 l_1}\ab{2 l_2}}\eta_1^3\eta_1^4\eta_2^3\eta_2^4\dots +\dots \right)\eqndot
 \end{aligned}
\end{equation}  
Using momentum conservation and some spinor identities yields the relation
\begin{equation}\label{eq: triangle propagator identity}
 \frac{\ab{ 1 2}\ab{ l_1 l_2}}{\ab{1 l_1}\ab{2 l_2}}=-\frac{(p_1+p_2)^2}{(p_1-l_1)^2}\eqndot
\end{equation}
Hence, the cut of the one-loop form factor is proportional to the cut of the triangle integral:
\begin{equation}\label{eq: su2 zz 3}
 \begin{aligned}  
  \int \de^2 \lambda_{l_1}\de^2 \lambdat_{l_1}\de^2 \lambda_{l_2}\de^2 \lambdat_{l_2} &\delta^4\left(\lambda_{1}\lambdat_{1}+\lambda_{2}\lambdat_{2}-\lambda_{l_1}\lambdat_{l_1}-\lambda_{l_2}\lambdat_{l_2}\right)\\ &\delta^4\left(\sum_{i=1}^L\lambda_i\lambdat_i-q\right)\left(-(p_1+p_2)^2\FDinline[cut,triangle,topleglabel=${\scriptscriptstyle 1}$,bottomleglabel=${\scriptscriptstyle 2}$]\eta_1^3\eta_1^4\eta_2^3\eta_2^4\dots+\dots \right)
\eqncom
 \end{aligned}
\end{equation}
where the graph denotes the integrand of the depicted integral \eqref{eq: triangle integral} excluding the cut propagators and the factor $\frac{\e^{\varepsilon\gamma_{\text{E}}}}{i\pi^{D/2}}$.
In contrast to the approach in the main text, which consists of applying unitarity at the level of the integral, we do not perform the integration to fix coefficients etc. Instead, we work with unitarity at the level of the integrand.
It involves lifting the result for the cut to the full integral, i.e.\ concluding that the uncut one-loop form factor is proportional to the uncut triangle integral --- apart from terms that are invisible in this cut. 
This result, as well as the corresponding discussion, agrees with the analysis of \cite{Brandhuber:2010ad}.

Let us now turn to the case $\uparrow\downarrow$ resp. $\cO=\tr[\phi_{24}\phi_{34}\cdots]+\dots$, which leads to the colour-ordered minimal tree-level form factor
\begin{equation}
 \hat{\cF}^{(0)}_\cO(\Lambda_1,\Lambda_2,\dots,\Lambda_L;q)=\delta^4\left(\sum_{i=1}^L\lambda_i\lambdat_i-q\right)(\eta_1^2\eta_1^4\eta_2^3\eta_2^4 \dots+\dots)\eqndot
\end{equation}
After the Grassmann integration of the analogue of \eqref{eq: su2 zz 1}, we find
\begin{equation}\label{eq: su2 yz}
 \begin{aligned}
 &\int \de^2 \lambda_{l_1}\de^2 \lambdat_{l_1}\de^2 \lambda_{l_2}\de^2 \lambdat_{l_2} \delta^4\left(\lambda_{1}\lambdat_{1}+\lambda_{2}\lambdat_{2}-\lambda_{l_1}\lambdat_{l_1}-\lambda_{l_2}\lambdat_{l_2}\right)\delta^4\left(\sum_{i=1}^L\lambda_i\lambdat_i-q\right) \\
 &\hphantom{\int}\left(+\eta_1^3\eta_1^4\eta_2^2\eta_2^4 -\frac{\ab{1 l_2}\ab{2 l_1}}{\ab{1 l_1}\ab{2 l_2}}\eta_1^2\eta_1^4\eta_2^3\eta_2^4
+\frac{\ab{1 l_2}}{\ab{2 l_2}}\eta_1^2\eta_1^3\eta_1^4\eta_2^4+\frac{\ab{2 l_1}}{\ab{1 l_1}}\eta_1^4\eta_2^2\eta_2^3\eta_2^4\right)\dots+\dots\\
= &{}\int \de^2 \lambda_{l_1}\de^2 \lambdat_{l_1}\de^2 \lambda_{l_2}\de^2 \lambdat_{l_2} \delta^4\left(\lambda_{1}\lambdat_{1}+\lambda_{2}\lambdat_{2}-\lambda_{l_1}\lambdat_{l_1}-\lambda_{l_2}\lambdat_{l_2}\right)\delta^4\left(\sum_{i=1}^L\lambda_i\lambdat_i-q\right) \\ &\hphantom{\int}\Bigg(+\FDinline[cut,bubble,topleglabel=${\scriptscriptstyle 1}$,bottomleglabel=${\scriptscriptstyle 2}$]\eta_1^3\eta_1^4\eta_2^2\eta_2^4 -\left(\FDinline[cut,bubble,topleglabel=${\scriptscriptstyle 1}$,bottomleglabel=${\scriptscriptstyle 2}$]+(p_1+p_2)^2\FDinline[cut,triangle,topleglabel=${\scriptscriptstyle 1}$,bottomleglabel=${\scriptscriptstyle 2}$]\right) \eta_1^2\eta_1^4\eta_2^3\eta_2^4\\
&\hphantom{\int}\phantom{\Bigg(}{}-\langle1|\gamma_\mu | 2]l_2^\mu\FDinline[cut,triangle,topleglabel=${\scriptscriptstyle 1}$,bottomleglabel=${\scriptscriptstyle 2}$] \eta_1^2\eta_1^3\eta_1^4\eta_2^4-\langle2|\gamma_\mu | 1]l_1^\mu\FDinline[cut,triangle,topleglabel=${\scriptscriptstyle 1}$,bottomleglabel=${\scriptscriptstyle 2}$]\eta_1^4\eta_2^2\eta_2^3\eta_2^4\Bigg)\dots+\dots
\eqncom
 \end{aligned}
\end{equation}
where we have used a Schouten identity, \eqref{eq: triangle propagator identity} and expansion with $\sb{1l_1}$ and $\sb{2l_2}$ to arrive at the second line.
As before, we can lift this result to the full integral, apart from possible terms that vanish in the cut. 
Note that the two terms with fermionic final states drop out of the final result since the linear tensor triangle integral results in a linear combination of the momenta $p_1$ and $p_2$, which vanishes when inserted into $\langle p_1|\gamma_\mu|p_2]$ or $\langle p_2|\gamma_\mu|p_1]$. 

Combining the results of both cases, we find that the one-loop correction to the planar colour-ordered minimal form factor of an operator from the $\mathfrak{su}(2)$ subsector is given by 
 \begin{equation}\label{eq: summary SU(2)}
 \begin{aligned}
  \hat{\cF}^{(1)}_\cO(1,\dots,L;q)&= - \sum_{i=1}^L s_{i\, i+1} \hat{\cF}^{(0)}_\cO(1,\dots,L;q) \FDinline[triangle,topleglabel=$\scriptscriptstyle i$,bottomleglabel=$\scriptscriptstyle i+1$]\\
  &\phaneq-  \sum_{i=1}^L (\idm-\PP)_{i\,i+1}\hat{\cF}^{(0)}_\cO(1,\dots,L;q) \FDinline[bubble,topleglabel=$\scriptscriptstyle i$,bottomleglabel=$\scriptscriptstyle i+1$] \quad \eqncom
 \end{aligned}
 \end{equation}
where $\PP_{i\,i+1}$ is the permutation operator acting on the flavours of the legs $i$ and $i+1$ and $\idm$ is the identity matrix.
The coefficients of the triangle and bubble integral agree with the analysis of section \ref{sec: one-loop form factors} and in particular with \eqref{eq: summary}.
Moreover, it is easy to show via Feynman diagrams and the criterion of \cite{Bern:1994cg} that rational terms are absent in this case, such that \eqref{eq: summary SU(2)} is indeed the complete result.

As discussed in section \ref{sec: one-loop dila}, the one-loop dilatation operator is given by $-2$ times the coefficients of the bubble integral with the tree-level form factor stripped off. In the $\mathfrak{su}(2)$ subsector, this yields
\begin{equation}
 (\mathfrak{D}_2)_{i\,i+1} = 2(\idm-\PP)_{i\,i+1}\eqndot
\end{equation}
This is precisely the well-known Hamiltonian density of the Heisenberg XXX spin chain.\footnote{The factor of $2$ depends on the convention for the modified effective planar coupling constant \eqref{eq: modified effective planar coupling constant}.}

\subsection{\texorpdfstring{$\mathfrak{sl}(2)$}{sl(2)} subsector}
\label{app: SL2 sector}

Next, we come to the $\mathfrak{sl}(2)$ subsector. Operators from this subsector are built from an arbitrary number of covariant derivatives of one type, say $\cder_{1\DOT1}$, acting on scalars of one type, say $\phi_{34}$.
In contrast to the $\mathfrak{su}(2)$ subsector, the $\mathfrak{sl}(2)$ subsector is non-compact, which leads to several complications. 

Let us consider the operator $\cO=\tr[(\cder_{1\DOT1})^k\phi_{34}(\cder_{1\DOT1})^{n-k}\phi_{34}\cdots]+\dots$, whose colour-ordered minimal tree-level form factor is given by
\begin{equation}
 \hat{\cF}^{(0)}_\cO(\Lambda_1,\Lambda_2,\dots,\Lambda_L;q)=\delta^4\left(\sum_{i=1}^L\lambda_i\lambdat_i-q\right)\Big((\lambda_1^1\lambdat_1^{\DOT1})^k\eta_1^3\eta_1^4(\lambda_2^1\lambdat_2^{\DOT1})^{n-k}\eta_2^3\eta_2^4\dots+\dots \Big) \eqndot
\end{equation}
After the Grassmann integration of the analogue of \eqref{eq: su2 zz 1}, this gives 
\begin{equation}\label{eq: sl2 zz}
 \begin{aligned}
  &\int \de^2 \lambda_{l_1}\de^2 \lambdat_{l_1}\de^2 \lambda_{l_2}\de^2 \lambdat_{l_2}  \delta^4\left(\lambda_{1}\lambdat_{1}+\lambda_{2}\lambdat_{2}-\lambda_{l_1}\lambdat_{l_1}-\lambda_{l_2}\lambdat_{l_2}\right)\delta^4\left(\sum_{i=1}^L\lambda_i\lambdat_i-q\right) \\
 &\hphantom{\int}\left(\frac{\langle1 2\rangle\langle l_1l_2\rangle (\lambda_{l_1}^1\lambdat_{l_1}^{\DOT1})^k(\lambda_{l_2}^1\lambdat_{l_2}^{\DOT1})^{n-k}}{\langle1 l_1\rangle\langle2 l_2\rangle}\eta_1^3\eta_1^4\eta_2^3\eta_2^4\dots+\dots \right)
  \end{aligned}
\end{equation}
\begin{equation}\label{eq: sl2 zz 2}
 \begin{aligned}
=&\int \de^2 \lambda_{l_1}\de^2 \lambdat_{l_1}\de^2 \lambda_{l_2}\de^2 \lambdat_{l_2} \delta^4\left(\lambda_{1}\lambdat_{1}+\lambda_{2}\lambdat_{2}-\lambda_{l_1}\lambdat_{l_1}-\lambda_{l_2}\lambdat_{l_2}\right)\delta^4\left(\sum_{i=1}^L\lambda_i\lambdat_i-q\right) \\
 &\hphantom{\int}\left(-(\lambda_{l_1}^1\lambdat_{l_1}^{\DOT1})^k(\lambda_{l_2}^1\lambdat_{l_2}^{\DOT1})^{n-k}(p_1+p_2)^2\FDinline[cut,triangle,topleglabel=$\scriptscriptstyle 1$,bottomleglabel=$\scriptscriptstyle 2$]\eta_1^3\eta_1^4\eta_2^3\eta_2^4\dots+\dots \right)
\eqndot
 \end{aligned}
\end{equation}
Again, we can lift this result to the full integral. 
In contrast to the $\mathfrak{su}(2)$ case, we obtain tensor triangle integrals of arbitrary rank. 
They can be reduced to scalar integrals via Passarino-Veltman (PV) reduction \cite{Passarino:1978jh}.

In the case of $k=1,n=1$, this gives
\begin{equation}\label{eq: k1n1}
\begin{aligned}
  &-(p_1+p_2)^2(\sigma_\mu)^{1\DOT1} \,\text{Tri}[l_1^\mu]\,\eta_1^3\eta_1^4\eta_2^3\eta_2^4\dots+\dots \\
 &=\left(-\lambda_1^1\lambdat_1^{\DOT1}\FDinline[bubble,topleglabel=$\scriptscriptstyle 1$,bottomleglabel=$\scriptscriptstyle 2$] + \lambda_2^1\lambdat_2^{\DOT1}\FDinline[bubble,topleglabel=$\scriptscriptstyle 1$,bottomleglabel=$\scriptscriptstyle 2$]
  -\lambda_1^1\lambdat_1^{\DOT1}(p_1+p_2)^2\FDinline[triangle,topleglabel=$\scriptscriptstyle 1$,bottomleglabel=$\scriptscriptstyle 2$]\right)\eta_1^3\eta_1^4\eta_2^3\eta_2^4\dots+\dots \eqncom
\end{aligned}
\end{equation}
where $\text{Tri}[l_1^\mu]$ denotes the linear triangle integral with numerator momentum $l_1^\mu$.
For $k=0,n=1$, we find
\begin{equation}\label{eq: k0n1}
\begin{aligned}
  &-(p_1+p_2)^2(\sigma_\mu)^{1\DOT1}\,\text{Tri}[l_2^\mu]\,\eta_1^3\eta_1^4\eta_2^3\eta_2^4\dots+\dots\\
&=  \left(\lambda_1^1\lambdat_1^{\DOT1}\FDinline[bubble,topleglabel=$\scriptscriptstyle 1$,bottomleglabel=$\scriptscriptstyle 2$] - \lambda_2^1\lambdat_2^{\DOT1}\FDinline[bubble,topleglabel=$\scriptscriptstyle 1$,bottomleglabel=$\scriptscriptstyle 2$]
  -\lambda_2^1\lambdat_2^{\DOT1}(p_1+p_2)^2\FDinline[triangle,topleglabel=$\scriptscriptstyle 1$,bottomleglabel=$\scriptscriptstyle 2$]\right)\eta_1^3\eta_1^4\eta_2^3\eta_2^4\dots+\dots \eqndot
 \end{aligned}
\end{equation}
For the next case of $k=2,n=2$, the PV reduction yields
\begin{equation}\label{eq: PV of two magnons}
\begin{aligned}
  &-(p_1+p_2)^2(\sigma_\mu)^{1\DOT1}(\sigma_\nu)^{1\DOT1}\,\text{Tri}[l_1^\mu l_1^\nu]\,\eta_1^3\eta_1^4\eta_2^3\eta_2^4\dots+\dots \\
 &= \Bigg(
 -\frac{3}{2} (\lambda_1^1\lambdat_1^{\DOT1})^2  \FDinline[bubble,topleglabel=$\scriptscriptstyle 1$,bottomleglabel=$\scriptscriptstyle 2$]
 + 2 (\lambda_1^1\lambdat_1^{\DOT1})^1(\lambda_2^1\lambdat_2^{\DOT1})^1\FDinline[bubble,topleglabel=$\scriptscriptstyle 1$,bottomleglabel=$\scriptscriptstyle 2$]
 + \frac{1}{2} (\lambda_2^1\lambdat_2^{\DOT1})^2 \FDinline[bubble,topleglabel=$\scriptscriptstyle 1$,bottomleglabel=$\scriptscriptstyle 2$] 
 \\ &\phaneq\phantom{\Big(}
 - (\lambda_1^1\lambdat_1^{\DOT1})^2 (p_1+p_2)^2 \FDinline[triangle,topleglabel=$\scriptscriptstyle 1$,bottomleglabel=$\scriptscriptstyle 2$]
 + 2 (\lambda_1^1\lambdat_1^{\DOT1})^1(\lambda_2^1\lambdat_2^{\DOT1})^1\,\text{Tri}[l_{\varepsilon}^2]\,
 \Bigg)\eta_1^3\eta_1^4\eta_2^3\eta_2^4\dots+\dots \eqncom
 \end{aligned}
\end{equation}
where $\text{Tri}[l_1^\mu l_1^\nu]$ denotes the tensor-two triangle integral with numerator momentum $l_1^\mu l_1^\nu$ and $l_{\varepsilon}$ in the numerator of $\text{Tri}[l_{\varepsilon}^2]$ denotes the $2\varepsilon$-dimensional part of the loop momentum.
The latter integral yields a rational term equal to \cite{Bern:1995ix}
\begin{equation}
 \text{Tri}[l_{\varepsilon}^2]=\frac{1}{2}+\cO(\varepsilon)\eqndot
\end{equation}

The results \eqref{eq: k1n1}, \eqref{eq: k0n1} and \eqref{eq: PV of two magnons} agree with the general analysis of section \ref{sec: one-loop form factors} except for the finite rational term, which cannot be detected by the method of section \ref{sec: one-loop form factors}.%
\footnote{In particular, the UV divergence of the above results is consistent with the known matrix elements of the dilatation operator in the $\mathfrak{sl}(2)$ subsector.} 
Moreover, they show that rational terms can indeed occur.%
\footnote{Rational terms are in general called not cut-constructible as they cannot be obtained via generalised unitarity alone, e.g.\ in the variant of OPP \cite{Ossola:2006us} or section \ref{sec: one-loop form factors}. In the example calculation of this subsection, we have obtained the rational term in \eqref{eq: PV of two magnons} via unitarity and subsequent PV reduction. }

The analysis in the $\mathfrak{sl}(2)$ subsector also demonstrates the advantages of generalised unitary at the level of the integral, which we have used in the main text, in comparison to unitarity at the level of the integrand.
At the level of the integral, far more relations are available, as nonzero integrands can integrate to zero.
In the non-compact $\mathfrak{sl}(2)$ subsector, this leads to an infinite number of basis tensor integrands before PV reduction, while the basis integrals are still the scalar bubble, triangle and box\footnote{As discussed previously, the box integral does not contribute in the case of the minimal form factor.} integral.

\section{Basis integrals}
\label{app: integrals}

In this appendix, we provide explicit expressions for the triangle and bubble integrals.

The bubble integral is
\begin{equation}\label{eq: bubble integral}
\begin{aligned}
 \text{Bub}(q)&=\FDinline[bubble,topleglabel=${\scriptscriptstyle 1}$,bottomleglabel=${\scriptscriptstyle 2}$]=
 \e^{\varepsilon\gamma_{\text{E}}}\int \frac{\de^D l}{i\pi^{D/2}} \, \frac{1}{l^2(q-l)^2}= \e^{\varepsilon\gamma_{\text{E}}}  \frac{\Gamma(\varepsilon)\Gamma(1-\varepsilon)^2}{\Gamma(2-2\varepsilon)(-q^2)^{\varepsilon}} \\
 &= (-q^2)^{-\varepsilon} \left(\frac{1}{\varepsilon}+2+\left(4-\frac{\pi ^2}{12}\right) \varepsilon+
 \left(-\frac{7 \zeta_3}{3}-\frac{\pi ^2}{6}+8\right) \varepsilon^2+
 \cO(\varepsilon^3)\right) \eqncom
\end{aligned} 
\end{equation}
where $\Gamma$ is the Euler gamma function and $\gamma_{\text{E}}$ the Euler-Mascheroni constant; cf.\ \cite{Smirnov:2004ym}.

The one-mass triangle integral is 
\begin{equation}\label{eq: triangle integral}
\begin{aligned}
 \text{Tri}(q)&=\FDinline[triangle,topleglabel=${\scriptscriptstyle 1}$,bottomleglabel=${\scriptscriptstyle 2}$]=\e^{\varepsilon\gamma_{\text{E}}}\int \frac{\de^D l}{i\pi^{D/2}} \, \frac{1}{l^2(l+p_1)^2(l+p_1+p_2)^2}= - \e^{\varepsilon\gamma_{\text{E}}}
 \frac{\Gamma(1+\varepsilon)\Gamma(-\varepsilon)^2}{\Gamma(1-2\varepsilon)(-q^2)^{1+\varepsilon}}\\
&=(-q^2)^{-1-\varepsilon} \left( -\frac{1}{\varepsilon^2}+\frac{\pi^2}{12}+\frac{7 \zeta_3}{3}\varepsilon+\frac{47 \pi^4}{1440}\varepsilon^2+\cO(\varepsilon^3)\right)
\eqncom
 \end{aligned}
\end{equation}
where $q=p_1+p_2$ and $p_1^2=p_2^2=0$; cf.\ \cite{Smirnov:2004ym}. 

\bibliographystyle{utcaps}
\bibliography{ThesisINSPIRE}

\end{fmffile}
\end{document}

%% file: Macros2.tex
\newcommand{\la}{\langle}
\newcommand{\ra}{\rangle}
\newcommand{\ab}[1]{\la #1 \ra}
\renewcommand{\sb}[1]{[ #1 ]}
\newcommand{\asb}[3]{\la #1 | #2 | #3 ]}
\newcommand{\sab}[3]{[ #1 | #2 | #3 \ra}
\newcommand{\MHV}{{\ensuremath{\text{MHV}} \xspace}}
\newcommand{\MHVb}{{\ensuremath{\overline{\text{MHV}}} \xspace}}

\newcommand{\lambdat}{\tilde{\lambda}}
\newcommand{\etat}{\tilde{\eta}}

\newcommand{\ZZ}{\ensuremath{\mathbb{Z}}}
\renewcommand{\SS}{\ensuremath{\mathbb{S}}}
\newcommand{\NN}{\ensuremath{\mathbb{N}}}
\newcommand{\PP}{\ensuremath{\mathbb{P}}}

\DeclareMathOperator{\idm}{\mathds{1}}

\newcommand{\comm}[3][]{{}[{}#2{}\,{\overset{#1}{,}}\,{}#3{}]{}}
\newcommand{\acomm}[3][]{{}\{{}#2{}\,{\overset{#1}{,}}\,{}#3{}\}{}}

\newcommand{\YM}{{\mathrm{\scriptscriptstyle YM}}}

\DeclareMathOperator{\tr}{tr}
\newcommand{\Tr}{\tr}

\DeclareMathOperator{\phaneq}{\phantom{{}=}}

\newcommand{\e}{\operatorname{e}}

\newcommand{\ket}[1]{\mathopen{\mid}{}#1{}\mathclose{\rangle}}

\newcommand{\bra}[1]{\mathopen{\langle}{}#1{}\mathclose{\mid}}
\newcommand{\braket}[2]{\mathopen{\langle}{}#1{}|{}#2{}\mathclose{\rangle}}

\newcommand{\vac}{\operatorname{\mid} 0 \, \operatorname{\rangle}}

\DeclareMathOperator{\Kop}{K}

\newcommand{\eqndot}{\, .}
\newcommand{\eqncom}{\, ,}
\newcommand{\eqnsem}{\, ;}

\newcommand{\de}{\operatorname{d}\!}

\newcommand{\cA}{\mathcal{A}}

\newcommand{\cF}{\mathcal{F}}
\newcommand{\cG}{\mathcal{G}}

\newcommand{\cN}{\mathcal{N}}
\newcommand{\cO}{\mathcal{O}}

\newcommand{\cV}{\mathcal{V}}

\newcommand{\cZ}{\mathcal{Z}}

\newcommand{\ba}{\mathbf{a}}
\newcommand{\bb}{\mathbf{b}}
\newcommand{\bc}{\mathbf{c}}
\newcommand{\bd}{\mathbf{d}}

\newcommand{\bA}{\mathbf{A}}

\newcommand{\alphadot}{{\dot{\alpha}}}
\newcommand{\betadot}{{\dot{\beta}}}
\newcommand{\gammadot}{{\dot{\gamma}}}

\newcommand{\Nfour}{$\mathcal{N}=4$\xspace}

\newcommand{\NfSYMt}{\Nfour SYM theory\xspace}

\newcommand{\ferm}{\psi}
\newcommand{\antiferm}{\bar{\psi}}

\newcommand{\cfstrength}{F}
\newcommand{\cantifstrength}{\bar{\cfstrength}}

\newcommand{\cder}{\D}

\newcommand{\aosc}{\ba}
\newcommand{\aoscdag}{{\aosc^\dagger}}
\newcommand{\bosc}{\bb}
\newcommand{\boscdag}{{\bosc^\dagger}}
\newcommand{\cosc}{\bc}
\newcommand{\coscdag}{{\cosc^\dagger}}
\newcommand{\dosc}{\bd}
\newcommand{\doscdag}{{\dosc^\dagger}}
\newcommand{\Aosc}{\bA}

\newcommand{\akind}[1][ ]{a^{#1}}
\newcommand{\bkind}[1][ ]{b^{#1}}

\newcommand{\dkind}[1][ ]{d^{#1}}
\newcommand{\akindsite}[2][ ]{a^{#1}_{#2}}
\newcommand{\bkindsite}[2][ ]{b^{#1}_{#2}}

\newcommand{\dkindsite}[2][ ]{d^{#1}_{#2}}

\DeclareMathOperator{\T}{T}

\newlength{\eqoff}

\DeclareMathOperator{\D}{D}